\documentclass[prd,onecolumn,preprintnumbers,amsmath,amssymb,superscriptaddress]{revtex4}

\usepackage{epsfig}
\usepackage{amsmath}
\usepackage{graphicx}  
\usepackage{dcolumn}   
\usepackage{bm}        

\newcommand\Eqn[1]     {Eq.\,(\ref{#1})}
\newcommand\Eqns[2]    {Eqs\,(\ref{#1}) and~(\ref{#2})}

\newcommand\eqn[1]     {Eq.\,(\ref{#1})}
\newcommand\eqns[2]    {Eqs\,(\ref{#1}) and~(\ref{#2})}
\newcommand\eqnss[2]   {Eqs\,(\ref{#1})--(\ref{#2})}

\newcommand\Fig[1]     {Fig.\,{\ref{#1}}}
\newcommand\Figs[2]    {Figs\,{\ref{#1}} and~{\ref{#2}}}
\newcommand\fig[1]     {Fig.\,{\ref{#1}}}
\newcommand\figs[2]    {Figs\,\ref{#1} and~{\ref{#2}}}

\newcommand\Sect[1]    {\S\ref{#1}}

\newcommand\app[1]     {Appendix~\ref{#1}}

\newcommand\nt         {\notag}
\newcommand\nn         {\nonumber}

\def\beq{\begin{equation}}
\def\eeq{\end{equation}}
\def\bsp#1\esp{\begin{split}#1\end{split}}
\def\bal#1\eal{\begin{align}#1\end{align}}

\newcommand\bom[1]     {{\mbox{\boldmath $#1$}}}

\newcommand\pd     {\partial}
\newcommand\del     {\delta}
\newcommand\tht     {\theta}
\newcommand\Tht     {\Theta}
\newcommand\Psij[1]     {\Psi^{(#1)}}

\newcommand\dGj[1]     {\delta G^{(#1)}}
\newcommand\Ddel     {\del^D}

\newcommand\kv     {\hbox{\bf k}}

\newcommand\vv     {\hbox{\bf v}}
\newcommand\xv     {\hbox{\bf x}}
\newcommand\ts     {\tau}

\newcommand\grad     {\nabla}
\newcommand\cH     {{\cal H}}
\newcommand\cP     {{\cal P}}
\newcommand\rd     {{\mathrm d}}
\newcommand\re     {{\mathrm e}}
\newcommand\ri     {{\mathrm i}}

\newcommand\al     {\alpha}
\newcommand\be     {\beta}

\newcommand\gas     {\gamma^{\mathrm{(s)}}}
\newcommand\bgas     {\bar{\gamma}^{\mathrm{(s)}}}

\newcommand\fb {f^{\rm b}}

\newcommand\rmm   {\mathrm{m}}
\newcommand\rc   {\mathrm{c}}
\newcommand\rb   {\mathrm{b}}
\newcommand\rr   {\mathrm{r}}
\newcommand\lin   {\mathrm{lin}}

\newcommand{\ba}{\begin{eqnarray}}
\newcommand{\ea}{\end{eqnarray}}

\def\bk{{\bf k}}

\def\kMpc{\, h \, {\rm Mpc}^{-1}}

\def\facOm{\left[1-\frac{\Omega_m}{2}+\Omega_{\Lambda}\right]}
\def\tot{-\frac{3}{2}}

\def\mnras{{MNRAS}}

\def\prd{{PRD}}

\def\apj{{ApJ}}
\def\apjs{{ApJS}}

\def\aap{{A\&A}}

\def\physrep{{Phys.~ Rep.~}}
\def\araa{Annual Reviews of Astronomy \& Astrophysics}

\begin{document}


\title{Cosmological perturbation theory for baryons and
dark matter I: \\ one-loop corrections in the RPT framework}


\date{\today}


\author{G\'abor Somogyi}
\email{Gabor.Somogyi@desy.de}
\affiliation{Deutsches Elektronensynchrotron DESY, Platanenallee 6, D-15738 Zeuthen, Germany} 
\affiliation{Institute for Theoretical Physics, University of Zurich, 
CH-8037 Zurich, Switzerland}

\author{Robert E. Smith} 
\email{res@physik.unizh.ch}
\affiliation{Institute for Theoretical Physics, University of Zurich, 
CH-8037 Zurich, Switzerland}


\begin{abstract}
We generalize the ``renormalized'' perturbation theory (RPT) formalism
of \citet{CrocceScoccimarro2006a} to deal with multiple fluids in the
Universe and here we present the complete calculations up to the
one-loop level in the RPT. We apply this approach to the problem of
following the non-linear evolution of baryon and cold dark matter
(CDM) perturbations, evolving from the distinct sets of initial
conditions, from the high redshift post-recombination Universe right
through to the present day.  In current theoretical and numerical
models of structure formation, it is standard practice to treat
baryons and CDM as an effective single matter fluid -- the so called
dark matter only modeling. In this approximation, one uses a weighed
sum of late time baryon and CDM transfer functions to set initial mass
fluctuations. In this paper we explore whether this approach can be
employed for high precision modeling of structure formation. We show
that, even if we only follow the linear evolution, there is a
large-scale scale-dependent bias between baryons and CDM for the
currently favored WMAP5 $\Lambda$CDM model. This time evolving bias
is significant $(>1\%)$ until the present day, when it is driven
towards unity through gravitational relaxation processes. Using the
RPT formalism we test this approximation in the non-linear regime. We
show that the non-linear CDM power spectrum in the 2-component fluid
differs from that obtained from an effective mean-mass 1-component
fluid by $\sim3\%$ on scales of order $k\sim0.05\kMpc$ at $z=10$, and
by $\sim0.5\%$ at $z=0$. However, for the case of the non-linear
evolution of the baryons the situation is worse and we find that the
power spectrum is suppressed, relative to the total matter, by
$\sim15\%$ on scales $k\sim0.05\kMpc$ at $z=10$, and by $\sim3-5\%$ at
$z=0$. Importantly, besides the suppression of the spectrum, the
Baryonic Acoustic Oscillation (BAO) features are amplified for baryon
and slightly damped for CDM spectra. If we compare the total matter
power spectra in the 2- and 1-component fluid approaches, then we find
excellent agreement, with deviations being $<0.5\%$ throughout the
evolution.  Consequences: high precision modeling of the large-scale
distribution of baryons in the Universe can not be achieved through an
effective mean-mass 1-component fluid approximation; detection
significance of BAO will be amplified in probes that study baryonic
matter, relative to probes that study the CDM or total mass only. The
CDM distribution can be modeled accurately at late times and the
total matter at all times. This is good news for probes that are
sensitive to the total mass, such as gravitational weak lensing as
existing modeling techniques are good enough. Lastly, we identify an
analytic approximation that greatly simplifies the evaluation of the
full PT expressions, and it is better than $<1\%$ over the full range
of scales and times considered.
\end{abstract}


\maketitle


\section{Introduction}\label{sec:intro}

In the current paradigm for structure formation in the Universe, it is
supposed that there was an initial period of inflation, during which,
quantum fluctuations were generated and inflated up to super-horizon
scales; producing a near scale-invariant and near Gaussian set of
primordial potential fluctuations. At the end of inflation the
Universe is reheated, and particles and radiation are synthesized. In
this hot early phase, cold thermal relics are also produced, these
particles interact gravitationally and possibly through the weak
interaction -- dubbed Cold Dark Matter (CDM).  Prior to recombination
photons are coupled to electrons through Thompson scattering, and in
turn electrons to baryonic nuclei through the Coulomb interaction.
Baryon fluctuations are pressure supported on scales smaller than the
sound horizon scale.  Subsequent to recombination photons free-stream
out of perturbations and baryons cool into the CDM potential
wells. Structure formation then proceeds in a hierarchical way with
small objects collapsing first and then merging to form larger
objects.  Eventually, sufficiently dense gas wells are accumulated and
galaxies form.  At late times the Universe switches from a
decelerating phase of expansion to an accelerated phase. This is
attributed to the non-zero constant energy-density of the vacuum. This
paradigm has been dubbed: $\Lambda$CDM
\citep[][]{PeeblesRatra2003,Spergeletal2007,Komatsuetal2009}.

The present day energy-density budget for the $\Lambda$CDM model is
distributed into several components, which in units of the critical
density are: vacuum energy $\Omega_{\Lambda,0}\approx0.73$, matter
$\Omega_{\rmm,0}\approx0.27$, neutrinos $\Omega_{\nu,0}\lesssim
10^{-2}$ and radiation $\Omega_{\rr,0}\approx 5\times 10^{-4}$. The
matter distribution can be subdivided further into contributions from
CDM and baryons, with present day values $\Omega_{\rc}\approx0.225$
and $\Omega_{\rb}\approx0.045$. The detailed physics for the evolution
of the radiation, CDM, baryon and neutrino fluctuations
($\delta^{\rr}, \delta^{\rc}, \delta^{\rb}, \delta^{\nu}$), from the
early Universe through to recombination can be obtained by solving the
Einstein--Boltzmann equations. These are a set of coupled non-linear
partial differential equations, however while the fluctuations are
small they may be linearized and solved \citep[for a review
see][]{MaBertschinger1995,SeljakZaldarriaga1996}.  At later times
these fluctuations enter a phase of non-linear growth. Their evolution
during this period can no longer be described accurately using the
linear Einstein--Boltzmann theory and must be followed using higher
order perturbation theory (hereafter PT) techniques
\cite{Bernardeauetal2002} or more directly through $N$-body
simulations \cite{Bertschinger2001}.


\begin{figure}[t!]
\centering{
  \includegraphics[width=8cm,clip=]{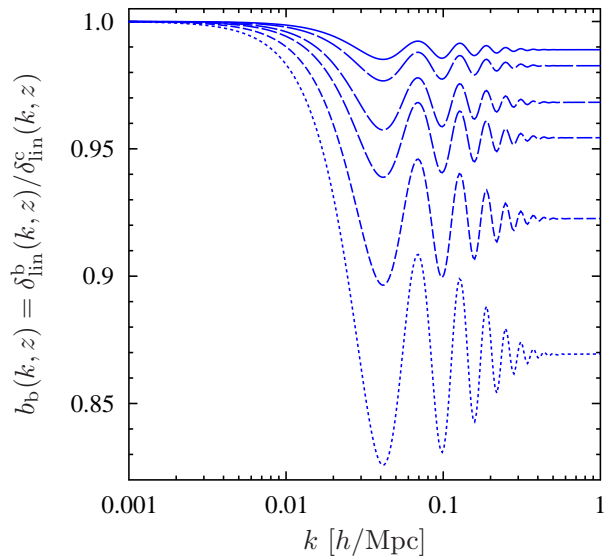}}
\caption{\small{Baryon bias as a function of inverse spatial scale,
    where we have defined the baryon bias $b_{\rb}(k,z)\equiv
    \delta^{\rb}_{\mathrm{lin}}(k,z)/\delta^{\rc}_{\mathrm{lin}}(k,z)$,
    and used \Eqns{eq:Lin_c}{eq:Lin_b} from \S\ref{ssec:application}.
    Solid through to dashed lines show results for redshifts $z=\{0,
    1.0, 3.0, 5.0, 10.0, 20.0\}$.}\label{fig:baryonbias}}
\end{figure}


The cross-over from the linear to the non-linear theory is, in general,
not straightforward.  Most of the theoretical and numerical approaches
to this latter task were developed during the previous two decades,
during which time the Standard CDM model was favored: essentially
Einstein-de Sitter spacetime, $\Omega_{\rmm}=1$, with energy-density
dominated by CDM $\gtrsim 97\%$, and baryons contributing
$\lesssim3\%$ to the total matter. For this case it is natural to
assume that the CDM fluctuations dominate the gravitational potentials
at nearly all times, with baryons playing little role in the formation
of large-scale structure \citep[see for
  example][]{Efstathiouetal1985,ThomasCouchman1992,Couchmanetal1995}.
Thus, one simply requires a transfer function for the CDM distribution
at some late time, to model the evolution of both components.

In the latter part of the 1990s, cosmological tests pointed towards
the $\Lambda$CDM paradigm, and it was recognized that baryons should
play some role in shaping the transfer function of matter fluctuations
\citep{MaBertschinger1995,Sugiyama1995,EisensteinHu1998,MeiksinWhitePeacock1999}.
However, rather than attempting to follow the CDM and baryons as
separate fluids evolving from distinct sets of initial conditions, a
simpler approximate scheme was adopted. This scheme is currently
standard practice for all studies of structure formation in the
Universe
\citep{Pearceetal2001,Teyssier2002,Springel2005,Springel2009}. It can
be summarized as follows:
\begin{enumerate}
\item Fix the cosmological model, specifying $\Omega_{\rc}$ and
  $\Omega_{\rb}$, and hence the fraction of baryons, $\fb$. Solve for the evolution of all
  perturbed species using the linearized coupled Einstein--Boltzmann
  equations. Obtain transfer functions at $z=0$, where CDM and baryons
  are fully relaxed: i.e.~$T^{\rc}(k,z=0)\approx T^{\rb}(k,z=0)$,
  where $T^{\rc}$ and $T^{\rb}$ are the transfer functions of the CDM
  and baryons, respectively.
\item Use the transfer functions to generate the linear matter power
  spectrum, normalized to the present day: i.e.
$P_{\bar{\delta}\bar{\delta}}(\bk,z=0)\approx \left[(1-\fb)
    T^{\rc}(k,z=0)+\fb T^{\rb}(k,z=0)\right]^2Ak^n \approx
  [T^{\rc}(k,z=0)]^2 Ak^n\ ,$
where the total matter fluctuation is
$\bar{\delta}=(1-\fb)\delta^{\rc}+\fb\delta^{\rb}$.
\item Scale this power spectrum back to the initial time $z_i$, using
  the linear growth factor for the total matter fluctuation
  $\bar{\delta}$, which obtains from solving the equations of motion
  for a single fluid.
\item Generate the initial CDM density field and assume that the
  baryons are perfect tracers of the CDM.
\item Evolve this effective CDM+baryon fluid under gravity using full
  non-linear equations of motion, either as a single fluid for dark
  matter only, or as a 2-component fluid if hydrodynamics are also to be
  followed.
\end{enumerate}

This effective description for the formation of structure becomes poor
as the baryon fraction $\fb=\Omega_{\rb}/\Omega_{\rmm}$ approaches
$\sim O(1)$, and for the currently favored WMAP5 cosmology
$\fb\approx0.17$. In fact, in linear theory the gravitational
relaxation between baryonic and dark matter components is only
achieved at the level of $<1\%$ by $z=0$.  Moreover, as can be seen in
\fig{fig:baryonbias} there is a non-trivial scale-dependent bias
between baryons and dark matter that exists right through to the
present day 
\footnote{Note that we have defined the baryon bias
    $b_{\rb}(k,z)\equiv\delta^{\rb}_{\mathrm{lin}}(k,z)/
    \delta^{\rc}_{\mathrm{lin}}(k,z)$, where these linear fluctuations
    are given by \Eqns{eq:Lin_c}{eq:Lin_b} from
    \S\ref{ssec:application}. In this analysis we have neglected any
    residual effect of the coupling between the photons and the
    baryons, and have assumed that the baryons are cold at $z=100$,
    hence $p_{\rm b}\propto \rho T\sim 0$. Furthermore, this result is
    sensitive to our choice for the initial $\delta_i$ and $\theta_i$,
    and indeed had we chosen the initial baryon and CDM velocity fields
    to be proportional to the {\em total matter} fluctuation, then
    these effects would be slightly reduced. However, the resultant
    bias will still remain non-negligible.}.
Failing, to take these biases into account may lead to non-negligible
systematic errors in studies that attempt to obtain cosmological
constraints from Large-Scale Structure (LSS) tests. With the next
generation of LSS tests aiming to achieving 1\% precision in
measurements of the matter power spectrum \cite{DETF2006,ESO2006}, it
seems timely to investigate whether such modeling assumptions may
impact inferences. Where we expect such systematic effects to play an
important role is for studies based on baryonic physics such as: using
21cm emission from neutral hydrogen to trace mass density at the epoch
of reionization $z_{\rm reion} \approx 10$
\citep{Zhangetal2004,Ilievetal2006,Furlanettoetal2006,PillepichPorciani2007};
and studies of the Lyman alpha forest to probe the matter power
spectrum \citep{Croftetal1998,McDonaldetal2006}.

In this, the first in a series of papers, we test whether this
effective procedure can indeed be employed to follow, at high
precision, the non-linear growth of structure formation in baryon and
CDM fluids, from recombination right through to the present day. We do
this by generalizing the matrix based ``renormalized'' perturbation theory (RPT) 
\footnote{It appears to us that what may be implemented in the RPT
  framework is not so much what would usually be understood as the
  ``perturbative renormalization'' of the theory, but rather a
  ``resummation'' of the perturbative series, in some specific
  approximation (the ``high-$k$'' limit). We will continue to use the
  acronym RPT to refer to the method, but would suggest that it is
  more properly read as {\em resummed}, and not {\em renormalized}
  perturbation theory.}
techniques of
\citet{CrocceScoccimarro2006a,CrocceScoccimarro2006b,CrocceScoccimarro2008}
to an arbitrary number of gravitating fluids. In this paper we content
our selves with calculating the observables up to the one-loop level
in the theory. In a future paper we consider the resummation to
arbitrary orders (Somogyi \& Smith 2009, in preparation).

The late-time linear theory evolution of coupled baryon and CDM
fluctuations, with sudden late time reionization of the inter-galactic
medium was studied by \citep{Nusser2000,MatarreseMohayaee2002}.
Recently, \citep{ShojiKomatsu2009} explored a similar system using
non-linear PT techniques. The work presented by these authors differs
from that presented here in a number of fundamental ways. Firstly,
these authors have assumed that, on large scales, baryons and CDM
fluctuations evolve with the same initial conditions and that the only
physical difference between the two fluids arises on small scales,
where galaxy formation processes are able to provide some pressure
support. They model this with a fixed comoving-scale Jeans
criterion. In this work we are not so much interested in following how
small-scale baryonic collapse feeds back on the mass distribution, but
are more interested in following the very large-scale baryon and CDM
fluctuations, initialized with their correct post-recombination
spatial distributions, into the non-linear regime. Secondly, their
work was developed within standard PT, the problems associated with
this formalism are well documented \citep{Bernardeauetal2002}. Here,
we develop the more successful RPT approach
\citep{CrocceScoccimarro2006a}.  In passing, we note that a number of
additional complimentary approaches and extensions to single-fluid RPT
have recently been developed
\citep{MatarresePietroni2007,MatarresePietroni2008,Bernardeauetal2008,Pietroni2008,Matsubara2008}.
In addition, there has recently been progress on realistic modeling of
massive neutrinos and CDM
\citep{Wong2008,Saitoetal2008,Ichikietal2009,Brandbygeetal2008,BrandbygeHannestad2009a,BrandbygeHannestad2009b,Lesgourguesetal2009},
for simplicity we shall assume that neutrinos play a negligible role.

The paper is broken down as follows: In \Sect{sec:EoM} we set up the
theoretical formalism: presenting the coupled equations of motion for
the $N$-component fluid. Then we show how, when written in matrix
form, the equations may be solved at linear order. We discuss
generalized initial conditions, and then discuss the specific case of
baryons and CDM. In \Sect{sec:PT} we develop the non-linear PT
expansion for the solutions, and show that they may be constructed to
arbitrary order using a set of Feynman rules.  In \Sect{sec:stats} we
discuss the two-point statistics of the fields, which we take as our
lowest order observables.  In \Sect{sec:OneLoopPower} we give specific
details of how we calculate the one-loop corrections in our theory.
In \Sect{sec:results} we present results for various quantities of
interest. In \Sect{sec:approx} we give details of an approximate RPT
scheme, which matches the exact calculation to within $<1\%$, for all
times of interest and on all scales where the perturbation theory is
valid. Finally, in \Sect{sec:conclusions} we draw our conclusions.


\section{Equations of motion}\label{sec:EoM}


\subsection{Standard form}

For an $N$-component fluid in a uniformly expanding spacetime, the
relevant equations of motion are, conservation of mass, momentum, and
the Poisson equation. These may be written ($i=1,\ldots,N$ labels the
$i$-th component) \citep{Bernardeauetal2002}:
\bal &
\frac{\pd\del_i(\xv,\ts)}{\pd\ts}+\grad\cdot[(1+\del_i(\xv,\ts))\vv_i(\xv,\ts)]
= 0\,;
\label{eq:conteq}\\
&
\frac{\pd\vv_i(\xv,\ts)}{\pd\ts}+\cH(\ts)\vv_i(\xv,\ts)
	+(\vv_i(\xv,\ts)\cdot\grad)\vv_i(\xv,\ts)
	= -\grad\Phi(\xv,\ts)\,;
\label{eq:eulereq} \\
& 
\grad^2\Phi(\xv,\ts) 
	= 4\pi Ga^2\sum_{i=1}^{N}\bar{\rho}_i(\ts)\del_i(\xv,\ts)=
\frac{3}{2}\Omega_{\rmm}(\ts)\cH^2(\ts)\sum_{i=1}^{N}w_i\del_i(\xv,\ts)\,.
\label{eq:poissoneq}
\eal
where $\ts$ is conformal time $\rd\ts\equiv \rd t/a$, with $a$ the scale
factor. Here we have introduced the density perturbations in each
component as
$\del_i(\xv,\ts)\equiv\rho_i(\xv,\ts)/\bar{\rho}_i-1
=\rho_i(\xv,\ts)/w_i\bar{\rho}_{\rm m}-1$,
where $\rho_i$ and $\bar{\rho}_i$ are the density and mean
density. $\vv_i(\xv,\ts)$ is the peculiar velocity. $\cH(\ts)=\rd\ln
a/\rd\ts$ is the conformal Hubble parameter, related to the usual
Hubble parameter, $H\equiv \dot{a}/a$, through $\cH(\ts)=a H$.  The
$w_i\equiv\Omega_i/\Omega_{\rmm}$ are the fractional contributions of
each fluid species to the total matter density. They are constant in
time, and normalized
\beq \sum_{i=1}^N w_i=1\,. 
\label{eq:sumwi}
\eeq
In the above we have also expressed the mean energy-density in each
species in units of the critical density,
\beq \Omega_{i}(\ts) \equiv \frac{\bar{\rho}_{i}(\ts)}{\rho_{\rm crit}(\ts)} = 
\frac{8\pi G\bar{\rho}_{i}(\ts)}{3a^2\cH^2(\ts)}
\ ,\eeq
and with subscript 0 denoting present day quantities,
$\Omega_{i,0}=\Omega_{i}(\ts=\ts_0)$.  Finally, the Hubble and density
parameters of each species are related through the Friedmann equation,
which can be written:
\beq 
\cH^2(\ts)=a^{-2}\cH_0^2\left[\Omega_{\rmm,0}a^{-3}+\Omega_{\Lambda,0}+\Omega_{K,0}
a^{-2}\right] \ ;\ \ \ \ \Omega_{K,0}\equiv1-\Omega_{\rmm,0}-\Omega_{\Lambda,0}\ .
\eeq

Taking the divergence of \eqn{eq:eulereq} and using \eqn{eq:poissoneq}
to eliminate $\Phi$, we obtain
\ba
& & \frac{\pd\Tht_i(\xv,\ts)}{\pd\ts}+\cH(\ts)\Tht_i(\xv,\ts)
	+\sum_{j,k}[\pd_j (\vv_i)_k(\xv,\ts)][\pd_k (\vv_i)_j(\xv,\ts)] 
	+ (\vv_i(\xv,\ts)\cdot\grad)\Tht_i(\xv,\ts)
	\nn \\
& & \qquad\qquad
= -\frac{3}{2}\Omega_{\rmm}(\ts)\cH^2(\ts)\sum_{j=1}^{N}w_j\del_j(\xv,\ts)\,,
\label{eq:euler_poisson}
\ea
where we have set $\grad\cdot\vv(\xv,\ts)=\Tht(\xv,\ts)$ and $j$ and
$k$ denote vector components. Next, we go to Fourier space. For a
quantity $A(\xv,\ts)$, we define the Fourier transform,
$\tilde{A}(\kv,\ts)$ and its inverse, as follows
\beq 
A(\xv,\ts) =
\int \rd^3 \kv\, \re^{\ri\kv\cdot\xv}\tilde{A}(\kv,\ts)
\quad
\Leftrightarrow 
\quad
\tilde{A}(\kv,\ts) =
\int\frac{\rd^3\xv}{(2\pi)^3} \re^{-\ri\kv\cdot\xv}A(\xv,\ts)
\label{eq:fouriertr}\ .
\eeq
Assuming $\vv_i(\xv,\ts)$ to be irrotational, there exists a scalar
field $u_i(\xv,\ts)$, such that $\vv_i(\xv,\ts) = -\grad u_i(\xv,\ts)$
and thus
\beq \tilde{\vv}_i(\kv,\ts) = -\ri \kv \tilde{u}_i(\kv,\ts)\,,\quad
\tilde{\Tht}_i(\kv,\ts) = \kv^2
\tilde{u}_i(\kv,\ts)\quad\Rightarrow\quad \tilde{\vv}_i(\kv,\ts) =
-\ri \kv \frac{\tilde{\Tht}_i(\kv,\ts)}{\kv^2}\,.  
\eeq 
On inserting this relation into the equations of motion we find:
\bal & \frac{\pd\tilde{\del}_i(\kv,\ts)}{\pd\ts} +
\tilde{\Tht}_i(\kv,\ts)
+\int\rd^3\kv_1\,\rd^3\kv_2\,\Ddel(\kv-\kv_1-\kv_2)
\left(1+\frac{\kv_1\cdot\kv_2}{\kv_2^2}\right)
\tilde{\del}_i(\kv_1,\ts)\tilde{\Tht}_i(\kv_2,\ts)=0\,; \\ 
&
\frac{\pd\tilde{\Tht}_i(\kv,\ts)}{\pd\ts}+\cH(\ts)\tilde{\Tht}_i(\kv,\ts)
+\frac{3}{2}\Omega_{\rmm}(\ts)\cH^2(\ts)\sum_{j=1}^{N}w_j\tilde{\del}_j(\kv,\ts)
\nt\\
&\qquad\qquad\qquad
+\int\rd^3\kv_1\,\rd^3\kv_2\,\Ddel(\kv-\kv_1-\kv_2)
\left[\frac{(\kv_1+\kv_2)^2 \kv_1\cdot\kv_2}{2\kv_1^2\kv_2^2}\right]
\tilde{\Tht}_i(\kv_1,\ts)\tilde{\Tht}_i(\kv_2,\ts)=0\,.  \eal 
Finally, we introduce the variable $\tilde{\tht}_i(\kv,\ts)\equiv
-\tilde{\Tht}_i(\kv,\ts)/\cH(\ts)$ and change the ``time'' variable to
$\eta = \ln a(\ts)$. Using 
\beq \frac{\rd}{\rd\ts} = \frac{\rd}{\rd\eta}\frac{\rd\eta}{\rd\ts} =
\frac{\rd}{\rd\eta}\frac{\rd\ln a}{\rd\ts}=\cH\frac{\rd}{\rd\eta}\,,
\eeq
and where $\rd \cH/\rd\ts = -\cH^2[\Omega_{\rmm}/2-\Omega_{\Lambda}]$,
(see \app{app:Hubble}), we obtain:
\beq
\frac{\pd\tilde{\del}_i(\kv,\eta)}{\pd\eta} -
\tilde{\tht}_i(\kv,\eta) = 
\int\rd^3\kv_1\,\rd^3\kv_2\,\Ddel(\kv-\kv_1-\kv_2) \al(\kv_2,\kv_1)
\tilde{\del}_i(\kv_1,\eta)\tilde{\tht}_i(\kv_2,\eta)\, ; \label{eq:eom1} 
\eeq
\ba
\frac{\pd\tilde{\tht}_i(\kv,\eta)}{\pd\eta}+\tilde{\tht}_i(\kv,\eta)
\left[1-\frac{\Omega_{\rmm}(\eta)}{2}+\Omega_{\Lambda}(\eta)\right]
	-\frac{3}{2}\Omega_{\rmm}(\eta)\sum_{j=1}^{N}w_j\tilde{\del}_j(\kv,\eta) &  & \nn \\
 & & \hspace{-6cm} = \int\rd^3\kv_1\,\rd^3\kv_2\,\Ddel(\kv-\kv_1-\kv_2)
\be(\kv_1,\kv_2)
\tilde{\tht}_i(\kv_1,\eta)\tilde{\tht}_i(\kv_2,\eta)\, ,
\label{eq:eom2}
\ea
where in the above expressions we introduced the mode-coupling
functions:
\beq 
\al(\kv_1,\kv_2) = 1+\frac{\kv_1\cdot\kv_2}{\kv_1^2}\,, \qquad
\be(\kv_1,\kv_2) = \frac{(\kv_1+\kv_2)^2
  \kv_1\cdot\kv_2}{2\kv_1^2\kv_2^2}\,.
\label{eq:albedef}
\eeq


\subsection{Matrix form}

Following \cite{CrocceScoccimarro2006a} we now rewrite the equations
of motion in matrix form. To do this, we first introduce a
$2N$-component ``vector'' of fields $\Psi_a$, the transpose of which is
given by
\beq
\Psi^{T}_a(\kv,\eta) = 
\left[\tilde{\del}_1(\kv,\eta),\ 
  \tilde{\tht}_1(\kv,\eta),\  
  \dots,\   
  \tilde{\del}_N(\kv,\eta),\   
  \tilde{\tht}_N(\kv,\eta)	
  \right] \label{eq:psi}\ ,
\eeq
where the index $a=1,2,\ldots,2N$. With this notation, the equations
of motion, \eqns{eq:eom1}{eq:eom2}, can be written in the compact form
(repeated indices are summed over)
\beq 
\pd_\eta \Psi_a(\kv,\eta) + \Omega_{ab}\Psi_b(\kv,\eta) =
\int\rd^3\kv_1\,\rd^3\kv_2\,
\gas_{abc}(\kv,\kv_1,\kv_2)\Psi_b(\kv_1,\eta)\Psi_c(\kv_2,\eta)\,.
\label{eq:eom}
\eeq
We shall also introduce the additional compactification that repeated
$k$-vectors are to be integrated over, hence we may rewrite
\eqn{eq:eom} as
\beq \rightarrow \pd_\eta \Psi_a(\kv,\eta) +
\Omega_{ab}\Psi_b(\kv,\eta) =
\gas_{abc}(\kv,\kv_1,\kv_2)\Psi_b(\kv_1,\eta)\Psi_c(\kv_2,\eta)
\label{eq:eomCompact} \ .
\eeq
The $2N\times 2N$ matrix $\Omega_{ab}$ appearing in \eqn{eq:eom} reads
\beq \Omega_{ab} = 
\left[\begin{array}{ccccccccc} 
0 & -1 & 0 & 0 & 0 & 0 &\ldots & 0 & 0 \\ 
\tot\Omega_{\rmm} w_1 & \facOm & \tot\Omega_{\rmm} w_2 & 0 & \tot\Omega_{\rmm} w_3 & 0 & \ldots & \tot\Omega_{\rmm} w_N & 0 \\ 
0 & 0 & 0 & -1 & 0 & 0 & \ldots & 0 & 0 \\ 
\tot\Omega_{\rmm} w_1 & 0 & 
          \tot\Omega_{\rmm} w_2 & \facOm & \tot\Omega_{\rmm} w_3 & 0 & \ldots & \tot\Omega_{\rmm} w_N & 0 \\ 
\vdots & \vdots & \vdots & \vdots & \vdots & \vdots & & \vdots & \vdots\\ 
0 & 0 & 0 & 0 & 0 & 0 & \ldots & 0 & -1 \\ 
\tot\Omega_{\rmm} w_1 & 0 & \tot\Omega_{\rmm} w_2 & 0 & \tot\Omega_{\rmm} w_3 & 0 & \ldots & \tot\Omega_{\rmm} w_N & \facOm
\end{array}\right]\,,
\label{eq:Omegaab}
\eeq
where for the cases of $N=1$ and $N=2$ we simply take the top left
$2\times2$ and $4\times4$ sub-matrices, respectively.

The symmetrized vertex matrix may be written for general $N$ as,
\ba
\gas_{(2j-1)(2j-1)(2j)}(\kv,\kv_1,\kv_2) 
	& = & \frac{\al(\kv_2,\kv_1)}{2}\Ddel(\kv-\kv_1-\kv_2) \nn\\
\gas_{(2j-1)(2j)(2j-1)}(\kv,\kv_1,\kv_2) 
	& = & \frac{\al(\kv_1,\kv_2)}{2}\Ddel(\kv-\kv_1-\kv_2) \nn\\
\gas_{(2j)(2j)(2j)}(\kv,\kv_1,\kv_2) 
	& = & \be(\kv_1,\kv_2)\Ddel(\kv-\kv_1-\kv_2)\,,
\ea
where $j=1,2,\ldots,N$, and all other matrix elements are zero. 
For the sake of clarity, let us write explicitly the vertex matrix for $N=1$ and $N=2$:
\begin{itemize}
\item
For $N=1$ the vertex matrix reads:
\bal
\bgas_{1bc}(\kv_1,\kv_2)& = 
	\left[\begin{array}{cc}
	0 & \al(\kv_2,\kv_1)/2 \\
	\al(\kv_1,\kv_2)/2 & 0 
	\end{array}\right]\,,
&
\bgas_{2bc}(\kv_1,\kv_2)& = 
	\left[\begin{array}{cc}
	0 & 0 \\
	0 & \be(\kv_1,\kv_2) 
	\end{array}\right]\,.
\eal 
\item
For $N=2$ the vertex matrix reads:
\bal
\bgas_{1bc}(\kv_1,\kv_2)& = 
	\left[\begin{array}{cccc}
	0 & \al(\kv_2,\kv_1)/2 & 0 & 0 \\
	\al(\kv_1,\kv_2)/2 & 0 & 0 & 0 \\
	0 & 0 & 0 & 0 \\
	0 & 0 & 0 & 0
	\end{array}\right]\,,
&
\bgas_{2bc}(\kv_1,\kv_2)& = 
	\left[\begin{array}{cccc}
	0 & 0 & 0 & 0 \\
	0 & \be(\kv_1,\kv_2) & 0 & 0 \\
	0 & 0 & 0 & 0 \\
	0 & 0 & 0 & 0
	\end{array}\right]\,,
\nonumber \\ 
\bgas_{3bc}(\kv_1,\kv_2)& =
	\left[\begin{array}{cccc}
	0 & 0 & 0 & 0 \\
	0 & 0 & 0 & 0 \\
	0 & 0 & 0 & \al(\kv_2,\kv_1)/2 \\
	0 & 0 & \al(\kv_1,\kv_2)/2 & 0
	\end{array}\right]\,,
&
\bgas_{4bc}(\kv_1,\kv_2)& = 
	\left[\begin{array}{cccc}
	0 & 0 & 0 & 0 \\
	0 & 0 & 0 & 0 \\
	0 & 0 & 0 & 0 \\
	0 & 0 & 0 & \be(\kv_1,\kv_2)
	\end{array}\right]\,.
\eal
\end{itemize}
In the above we defined $\bgas$ to be the vertex matrix {\em without} the
delta function for momentum conservation, i.e.~$\gas_{abc}(\kv,\kv_1,\kv_2) =
\bgas_{abc}(\kv_1,\kv_2)\Ddel(\kv-\kv_1-\kv_2)$.


\subsection{Solving for the propagator}

The main advantage of writing \eqn{eq:eom} in this compact form
is, as was pointed out in \citep{CrocceScoccimarro2006a}, that an
implicit integral solution can be obtained by Laplace transforming in
the variable $\eta$. This approach, however, requires that we set the
background cosmological model to be Einstein--de Sitter:
i.e.~$\{\Omega_{\rmm}(\eta)=1, \Omega_{\Lambda}(\eta)=0\}$ (we shall
discuss generalizations to other cosmological models in
\Sect{ssec:gencosmo}). This means that the matrix $\Omega_{ab}$ is
independent of time, hence
\beq
\sigma_{ab}^{-1}(s)\Psi_b(\bk,s)=\phi^{(0)}_a(\bk)+\gamma^{(s)}_{abc}(\bk,\bk_1,\bk_2)
\oint \frac{\rd s_1}{2\pi i} \Psi_b(\bk,s_1)\Psi_c(\bk,s-s_1) \ ,
\eeq
where $\phi^{(0)}_a(\bk)\equiv\Psi_a(\bk,\eta=0)$ denotes the initial conditions,
set when the growth factor is one and
$\sigma_{ab}^{-1}(s)=s\delta_{ab}+\Omega_{ab}$. Multiplying by the
matrix $\sigma_{ab}(s)$ and performing the inverse Laplace transform
gives the formal solution \citep{CrocceScoccimarro2006a},
\beq 
\Psi_a(\bk,a)=g_{ab}(\eta)\phi^{(0)}_b(\bk)+\int_0^{\eta} \rd\eta' g_{ab}(\eta-\eta')
\gamma^{(s)}_{bcd}(\bk,\bk_1\bk_2)\Psi_c(\bk_1,\eta')\Psi_d(\bk_1,\eta')\ , 
\label{eq:gensol}
\eeq
where $g_{ab}(\eta)$ is the {\em linear propagator} given by
\beq
g_{ab}(\eta)={\cal L}^{-1}\left[\sigma_{ab}(s),s,\eta\right]\,,
\label{eq:gabdef}
\eeq
where ${\cal L}^{-1}[f(s),s,\eta]$ denotes the inverse Laplace
transform of the function $f(s)$ from the variable $s$ to the variable
$\eta$. In \app{app:propagator_N}, we prove that the linear propagator has 
the general form ($j,k=1,2,\ldots,N$)
\ba
g_{(2j-1)(2k-1)} & = & \frac{1}{5}\left(3\re^\eta-5+2\re^{-3\eta/2}\right)w_{k}
	+\del_{jk}\,, 
\nn \\
g_{(2j)(2k-1)} & = & \frac{3}{5}\left(\re^\eta-\re^{-3\eta/2}\right)w_{k}\,,
\nn \\
g_{(2j-1)(2k)} & = & \frac{2}{5}\left(\re^\eta-5+5\re^{-\eta/2}-\re^{-3\eta/2}\right)
	w_{k} + (2-2\re^{-\eta/2})\delta_{jk}\,,
\nn \\
g_{(2j)(2k)} & = & \frac{1}{5}\left(2\re^\eta-5\re^{-\eta/2}+3\re^{-3\eta/2}\right)w_{k}
	+\re^{-\eta/2}\del_{jk}\,,
\label{eq:gall}
\ea
for $\eta\ge 0$, and $g_{ab}(\eta)=0$ for $\eta<0$ due to causality,
and furthermore $g_{ab}(\eta)\to \del^K_{ab}$ as $\eta\to 0^+$.
Clearly we can write the propagator as
\beq
g_{ab}(\eta) = \sum_l \re^{l\eta} g_{ab,l}\,,
\label{eq:gab}
\eeq
where the summation runs over the set $l\in\{1,0,-1/2,-3/2\}$ and the
$g_{ab,l}$ are constant matrices. Explicitly we find:
\begin{itemize} 
\item For $N=1$, we have $w_1=1$ and the propagator reads:
\beq
g_{ab}(\eta) = \frac{1}{5}
	\left[\begin{array}{rr}
	3\re^\eta+2\re^{-3\eta/2} & 2\re^\eta-2\re^{-3\eta/2} \\
	3\re^\eta-3\re^{-3\eta/2} & 2\re^\eta+3\re^{-3\eta/2}
	\end{array}\right]
=\frac{\re^\eta}{5}
	\left[\begin{array}{rr}
	3 & 2 \\
	3 & 2
	\end{array}\right]
-\frac{\re^{-3\eta/2}}{5}
	\left[\begin{array}{rr}
	-2 & 2 \\
	3 & -3
	\end{array}\right]\,,
\eeq
\end{itemize}
in agreement with the usual expression \cite{CrocceScoccimarro2006a},
where there are two solutions for the time evolution of the modes: a
growing $D_{+}\propto e^{\eta}$ and a decaying mode $D_{-}\propto
e^{-3\eta/2}$. Thus {\em for $N=1$ only}, $g_{ab,0}=g_{ab,-1/2}=0$ and
\bal
g_{ab,1}& = \frac{1}{5}
	\left[\begin{array}{rr}
	3 & 2 \\
	3 & 2
	\end{array}\right]\,,
&
g_{ab,-3/2}& = \frac{1}{5}
	\left[\begin{array}{rr}
	2 & -2 \\
	-3 & 3
	\end{array}\right]\,.
\eal
\begin{itemize}
\item 
For $N=2$, the propagator takes the general form of \eqn{eq:gab}, with
all the $g_{ab,l}$ non-zero:
\bal
g_{ab,1}& = \frac{1}{5}
	\left[\begin{array}{rrrr}
	3w_1 & 2w_1 & 3w_2 & 2w_2 \\
	3w_1 & 2w_1 & 3w_2 & 2w_2 \\
	3w_1 & 2w_1 & 3w_2 & 2w_2 \\
	3w_1 & 2w_1 & 3w_2 & 2w_2 \\
	\end{array}\right]\,,
&
g_{ab,0}& =
	\left[\begin{array}{rrrr}
	1-w_1 & 2(1-w_1) & -w_2 & -2w_2 \\
	0 & 0 & 0 & 0 \\
	-w_1 & -2w_1 & 1-w_2 & 2(1-w_2) \\
	0 & 0 & 0 & 0
	\end{array}\right]\,,
\nonumber \\ 
g_{ab,-1/2}& = 
	\left[\begin{array}{rrrr}
	0 & -2(1-w_1) & 0 & 2w_2 \\
	0 & 1-w_1 & 0 & -w_2 \\
	0 & 2w_1 & 0 & -2(1-w_2) \\
	0 & -w_1 & 0 & 1-w_2
	\end{array}\right]\,,
&
g_{ab,-3/2}& = \frac{1}{5}
	\left[\begin{array}{rrrr}
	2w_1 & -2w_1 & 2w_2 & -2w_2 \\
	-3w_1 & 3w_1 & -3w_2 & 3w_2 \\
	2w_1 & -2w_1 & 2w_2 & -2w_2 \\
	-3w_1 & 3w_1 & -3w_2 & 3w_2
	\end{array}\right]\,.
\eal
\end{itemize}

Interestingly, we now see that there are more solutions for the time
evolution of the modes, and besides the usual growing and decaying
modes, there is an additional decaying mode $g_{ab}\propto
e^{-\eta/2}$, and a static mode $g_{ab}\propto {\rm const}$. In 
\app{app:linpropN3} we give an explicit expression for the
linear propagator for the specific case of $N=3$. The interesting
point to note in going from $N=2$ to $N=3$, is that, whilst the
eigenvectors do change, no new eigenvalues are generated and this
holds for general $N$. This explicitly follows from \eqn{eq:gall}.


\subsection{Extension to general cosmological models}\label{ssec:gencosmo}

An approximate treatment of the dependence of the PT and RPT kernels
on cosmology was discussed by
\citep{Scoccimarroetal1998,CrocceScoccimarro2006b}, and we shall adopt
the same approach here. The necessary changes to be made are that we
use a new time variable,
\beq \eta \equiv \log D_{+}(\tau) \ ,\eeq
where $D_+(\tau)$ is the linear growth factor for the appropriate
cosmology. Also, instead of working with the solution vector of the
form \eqn{eq:psi}, we work with vectors
\beq
\Psi^{T}_a(\kv,\eta) = 
\left[\tilde{\del}_1(\kv,\eta),\ 
  \tilde{\tht}_1(\kv,\eta)/f(\eta),\  
  \dots,\   
  \tilde{\del}_N(\kv,\eta),\   
  \tilde{\tht}_N(\kv,\eta)/f(\eta)
  \right]\ ,
\eeq
where $f(\eta)\equiv \rd\log D_{+}/\rd\log a$ is the logarithmic growth
factor for peculiar velocity fields. On performing these
transformations \eqn{eq:eom} remains structurally the same, except for
the elements of the matrix $\Omega_{ab}$, which gain time
dependence. This means that it is no longer straightforward to perform
the Laplace transforms and so we do not obtain
\eqn{eq:gensol}. However, as was argued by
\citep[][]{CrocceScoccimarro2006b}, most of the cosmological
dependence is encoded in $D_+(\eta)$, and so to a very good
approximation we may simply take
$\Omega_{ab}(\Omega_{\rmm},\Omega_{\Lambda})\rightarrow
\Omega_{ab}(\Omega_{\rmm}=1,\Omega_{\Lambda}=0)$, and keep
\eqn{eq:gensol}. We shall use this approximation throughout the
remainder of this paper.


\subsection{Initial conditions}\label{ssec:ics1}

We must now specify the initial conditions for the $2N$ variables. For
cosmological structure formation an interesting case is when, for each
fluid component, $\tilde{\del}_i(\kv,\eta=0)$ and
$\tilde{\tht}_i(\kv,\eta=0)$ $(i=1,2,\ldots,N)$ are proportional to
the same random field 
\footnote{This is not the only physically interesting case: one may
also consider the case where the density fluctuations are proportional
to their relevant transfer functions and a primordial perturbation,
but that the velocity fields are proportional to the {\em total
matter} fluctuation. In this study, which primarily builds the
formalism, we shall restrict attention to the case where $\delta_i$
and $\theta_i$ are proportional to the same random field.}.
However, the initial random field can differ from component to
component. In this case we can write
\beq
\left[\phi^{(0)}_a(\kv)\right]^{T} = 
\left[
  u_1 \del_{1}^{(0)}(\kv),\,
  u_2 \del_{1}^{(0)}(\kv),\, 
  \dots,\,
  u_{2N-1}\del_{N}^{(0)}(\kv),\,
  u_{2N} \del_{N}^{(0)}(\kv)
  \right]\, ,
\label{eq:InitCondsN}
\eeq
where the $u_a$, $(a=1,2,\ldots,2N)$ are simply numerical coefficients.

For the case of the $N=1$ component fluid, ``pure growing'' mode
initial conditions can be obtained by setting
$u_a=u^{(1)}_a\equiv(1,1)$, and ``pure decaying'' mode ones by setting
$u_a=u^{(2)}_a\equiv(2/3,-1)$. These are the eigenvectors of the
matrix $g_{ab}$. In going to the more complex case of the $N=2$
multi-fluid, there are, unfortunately, no such simple choices for the
$u_a$ components to obtain pure ``growing mode'' or pure ``decaying
modes'' at all $k$-modes, unless all of the $\del_{i}^{(0)}(\kv)$ are
equivalent.  To see why this is so, consider the eigenvectors of the
linear propagator $g_{ab}$, these are:
\beq 
u^{(1)}_a=
\left(
\begin{array}{c}
  1\\
  1\\
  1\\
  1
\end{array}
\right)
\ ;\
u^{(2)}_a=
\left(
\begin{array}{c}
  2/3\\
  -1\\
  2/3\\
  -1
\end{array}
\right)
\ ;\ 
u^{(3,1)}_a=
\left(
\begin{array}{c}
  w_2\\
  0\\
  -w_1\\
  0
\end{array}
\right)
\ ;\
u^{(4,1)}_a=
\left(
\begin{array}{c}
  2w_2\\
  -w_2\\
  -2w_1\\
  w_1
\end{array}
\right)
\ .
\eeq
On dotting $g_{ab}$ with $u^{(1)}$ we find $g_{ab}u_b^{(1)}=e^{\eta}
u_a^{(1)}$, and we have a pure growing mode solution. However, to
propagate the initial modes we require $g_{ab}$ dotted with
$\phi^{(0)}_b(\bk)$. If we do this and set $u_a=u^{(1)}_a$ then we
find:
\beq
g_{ab}(\eta)\phi^{(0)}_b(\bk)=\sum_b g_{ab}(\eta)u^{(1)}_b\delta_b^{(0)}(\bk)
= e^{\eta}u^{(1)}_{a}\delta^{(0)}(\bk)\ = e^{\eta}\phi^{(0)}_a(\bk) \ ,\eeq
if and only if $\delta^{(0)}_b(\bk)=\delta^{(0)}(\bk)$ for all
$b$. However, this situation is not physically interesting since, in
the absence of any process such as pressure support, there would be no
discernible difference between treating the $N$-component fluid as an
effective 1-component fluid.

The more interesting physical case will be the situation where each
fluid component evolves from a distinct set of initial conditions,
i.e.~$\del_{i}^{(0)}\ne\del_{j}^{(0)}$, with $(i\ne j)$. What then
are the appropriate choices for the $u_a$? The distinct initial
conditions that are of interest are: the dark matter, baryon, neutrino
and photon, perturbations present in the post-recombination
Universe. As described earlier the dynamics of these species are
generally followed from the post-inflationary universe and through the
recombination era using the coupled, linearized Einstein--Boltzmann
equations \citep[see for example][]{MaBertschinger1995}. In this case
the fluctuations in $k$-space are all proportional to the same random
field, which is related to the primordial curvature perturbation, but
with some $k$-dependent transfer function that takes into account the
physical processes that affects each species. We shall use {\tt
CMBFAST} \citep{SeljakZaldarriaga1996} for following the time
evolution of the perturbations up to some large redshift say $z=100$ as
the initial condition for our non-linear computation; and the
information on the different components is given as a set of transfer
functions, which we denote $T_i(k)$. Thus for the $N$-component fluid,
we shall express our initial conditions, more specifically as:
\beq
\left[\phi^{(0)}_a(\kv)\right]^{T} = 
\left[
  u_1 T_1(k),\,
  u_2 T_1(k),\, 
  \dots,\,
  u_{2N-1} T_N(k),\,
  u_{2N} T_N(k)
  \right]\delta_0(\bk)\, .
\label{eq:InitCondsN_V2}
\eeq
Importantly for the standard $\Lambda$CDM framework of adiabatic near
scale-invariant fluctuations, all of the $T_i(k)\rightarrow 1$ as
$k\rightarrow0$. Hence, on large scales if we choose $u_a=u_a^{(1)}$,
then $\phi^{(0)}_a(\kv)$ {\em is} an eigenvector of the linear
propagator and so we have pure growing mode initial conditions. On
smaller scales there will, however, be some mixing of growing and
decaying modes in the initial conditions.


\subsection{Application to a CDM and baryon fluid}\label{ssec:application}

As a demonstration of the formalism, we now consider the specific case
of the linear evolution under gravity of a two-component fluid of CDM
and baryons. In this calculation we set the initial cosmological model
to be that identified from the WMAP5 analysis \cite{Komatsuetal2009}:
$\{\Omega_{\rc}=0.2275,\,\Omega_{\rb}=0.0456,\,
\Omega_{\Lambda}=0.7268,\,h=0.705,\,n_s=0.96,\,\sigma_8=0.812\}$; and
assume negligible contribution from neutrinos. Thus, we have
$\{w_1=\Omega_{\rc}/{\Omega_\rmm}=(1-\fb),w_2=\Omega_{\rb}/\Omega_{\rmm}=\fb\}$. We
use {\tt CMBFAST} \citep{SeljakZaldarriaga1996} to generate CDM and
baryon transfer functions at the initial redshift $z_i=100$: hence
$T_1(k)=T^{\rc}(k,z=100)\equiv T^{\rc}(k)$ and
$T_2(k)=T^{\rb}(k,z=100)\equiv T^{\rb}(k)$.


\begin{figure}
\centering{
  \includegraphics[scale=1.0,clip=]{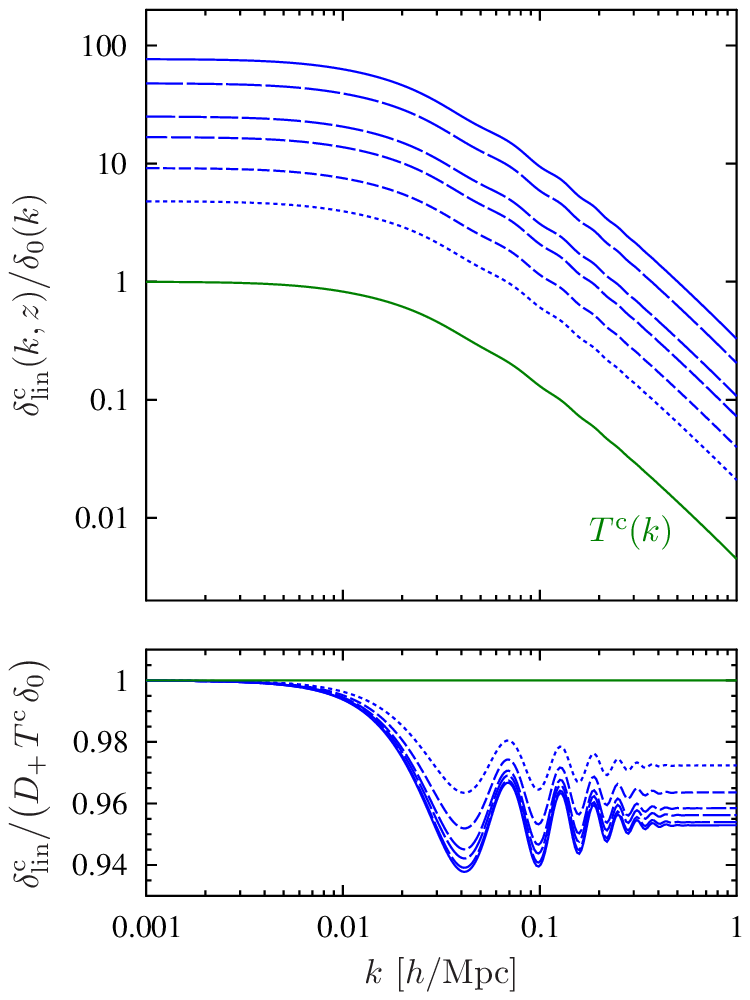}\hspace{0.4cm}
  \includegraphics[scale=1.0,clip=]{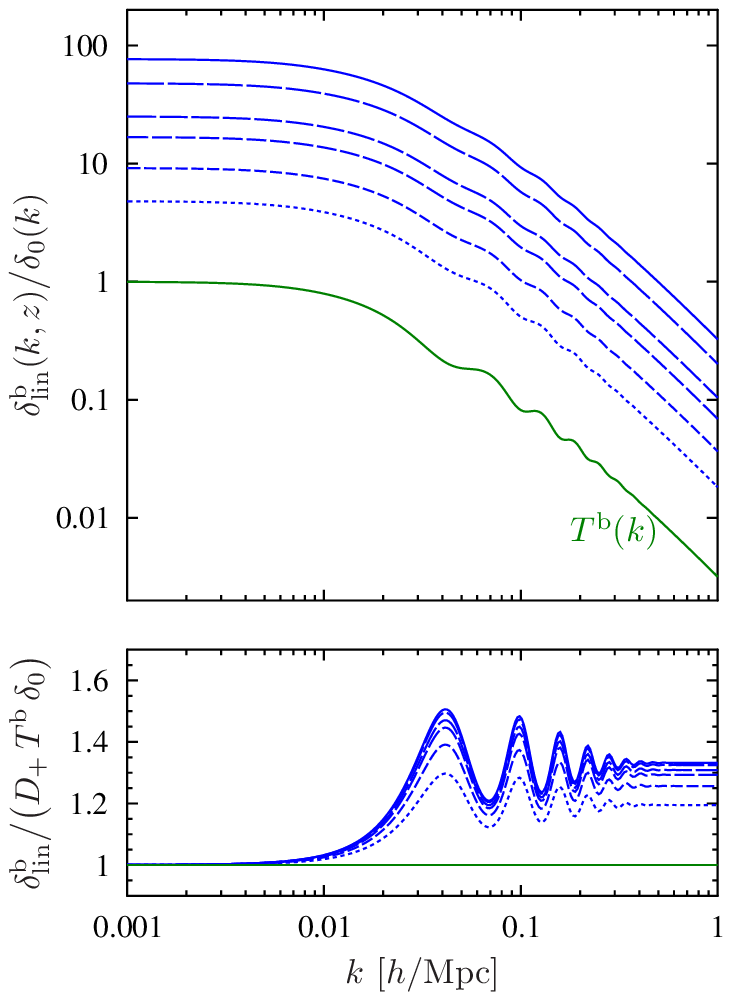}}
\caption{\small{Evolution of the initial density perturbations for the
    CDM and baryons under the linear propagator, relative to the
    primordial density perturbation, as a function of spatial
    wavenumber and redshift. {\em Top left and right panels}:
    evolution of the CDM perturbations $\delta^{\rc}$, and baryons,
    $\delta^{\rb}$, respectively.  Going from top to bottom, the blue
    solid through to dashed lines show results for redshifts
    $z=\{0,1.0,3.0,5.0,10.0,20.0\}$}. The solid green line at the
    bottom shows the initial mode. {\em Bottom left and right panels}:
    ratio of the linearly evolved CDM and baryon perturbations to the
    initial perturbations evolved under the standard linear growth
    factor for matter perturbations. Line styles are as for
    above.}\label{fig:onemode}
\end{figure}


The implicit solution for $\Psi_{a}(\bk,\eta)$, \eqn{eq:gensol},
can be linearized, and the zeroth order term can be written:
$\Psi_{a}^{(0)}(\bk,\eta)=g_{ab}(\eta)\phi^{(0)}_b(\bk)$
(c.f.~\eqn{eq:PT0}). Following our discussions in
\Sect{sec:intro} and \Sect{ssec:ics1}, we are interested in the
case where the initial conditions are those of the post-recombination
era Universe. Prior to this time, electrons and photons are tightly
coupled through Thompson scattering, and photons drag electrons and
baryons out of the potential wells. Thus at the end of recombination,
baryons and dark matter are spatially displaced from one another
(c.f.~\fig{fig:baryonbias}). In a somewhat simplified view,
once the scattering processes are switched off, the evolution of the
fluids can be captured by the two perturbations relaxing under gravity
and so the evolving fields are simply some linear combination of the
initial fields, with appropriate weight factors.

\Fig{fig:onemode} graphically demonstrates this. Here we show
the ratio of the initial CDM (left panel) and baryon (right panel)
perturbations evolving under the action of the linear propagator
$g_{ab}(\eta)$, divided through by the amplitude of the initial
primordial density perturbation that sourced them. For large-scale
growing mode initial conditions $(u_a=u_a^{(1)})$, then we have:
\ba 
\hspace{-0.4cm}
\delta^{\rc}_{\lin}(\bk,\eta)/\delta_0(k) =
\Psi^{(0)}_{1}(\bk,\eta)/\delta_0(k) 
&  =  &
\left[g_{11}(\eta)+g_{12}(\eta)\right]T^{\rc}(k)+
\left[g_{13}(\eta)+g_{14}(\eta)\right]T^{\rb}(k)\ \nn \\
&  =  &
\left[(1-\fb)\re^{\eta}+3\fb(1-2\re^{-\eta/2})\right]T^{\rc}(k)+
\fb\left[\re^{\eta}-3+2\re^{-\eta/2}\right]T^{\rb}(k)\ ;\label{eq:Lin_c} \\
\hspace{-0.4cm}
\delta^{\rb}_{\lin}(\bk,\eta)/\delta_0(k) = 
\Psi^{(0)}_{3}(\bk,\eta)/\delta_0(k) 
& =  &
\left[g_{31}(\eta)+g_{32}(\eta)\right]T^{\rc}(k)+
\left[g_{33}(\eta)+g_{34}(\eta)\right]T^{\rb}(k)\ \nn \\
&  =  &
(1-\fb)\left[\re^{\eta}-3+2\re^{-\eta/2}\right]T^{\rc}(k)+
\left[\fb\re^{\eta}+(1-\fb)(3-2\re^{-\eta/2})\right]T^{\rb}(k) \label{eq:Lin_b}\ .
\ea
In the limit of small and large $\eta$ these relations simplify to:
\ba 
\delta^{\rc}_{\lin}(\bk,\eta)/\delta_0(k) & \approx & \left\{
\begin{array}{ll}
T^{\rc}(k) & \quad (\eta\ll1)  \\
\re^{\eta}\left[(1-\fb) T^{\rc}(k)+\fb T^{\rb}(k)\right] & \quad (\eta\gg1)  
\end{array}
\right. \ ;\\
\delta^{\rb}_{\lin}(\bk,\eta)/\delta_0(k) &  \approx  & \left\{
\begin{array}{ll}
T^{\rb}(k) & \quad (\eta\ll1)   \\
\re^{\eta}\left[(1-\fb) T^{\rc}(k)+\fb T^{\rb}(k)\right] & \quad (\eta\gg1) 
\end{array}
\right. \ ,
\ea
where the $\eta\gg1$ relations, show that the fields evolve to an
equilibrium state.

On inspecting \fig{fig:onemode}, we note that on large scales at
the initial time (green solid lines) the two perturbations are
perfectly correlated with the primordial fluctuation. On smaller
scales we see that both the CDM and baryon modes are damped, and this
is due to the M\'esz\'aros effect
\citep[see][]{Peebles1980,Dodelson2003}. Also the waves have an
oscillating structure as a function of $k$, these are the Baryon
Acoustic Oscillations (BAO) \citep[for a discussion
  see][]{EisensteinHu1998,MeiksinWhitePeacock1999}.  We also note that
the BAO are almost absent in the dark matter distribution at
$z_i=100$, whereas for the baryons they are strongly present.  

As the system evolves linearly under gravity, we now see that on large
scales ($k<0.01\kMpc$) both solutions scale with time in accordance
with the standard linear growth factor for the total matter
perturbation on large scales: i.e., $\delta^{\rmm}(\kv,\eta)=
\lim_{k\rightarrow0}[w_1 \delta^{\rc}(\kv,\eta)+w_2
  \delta^{\rb}(\kv,\eta)] =D_{+}(\eta)\delta^{\rmm}(\kv,\eta_0)$,
which we obtain using the model of \cite{Carrolletal1992} with
$\Omega_{\rmm}=\Omega_{\rc}+\Omega_{\rb}$.  On smaller scales
($k>0.01\kMpc$) the fluctuations grow with time, but now we note that
the BAO features present in the baryon modes become damped, whilst
those in the CDM grow. This can be seen more clearly by dividing each
mode by $D_{+}(\eta)\delta^{(0)}(k)=D_{+}(\eta)T(k)\delta_0(k)$, for
each component and this is what is plotted in the bottom panels of
\fig{fig:onemode}. Again, in the large scale limit, all
$T_i(k)\rightarrow 1$, and so $g_{1b}(\eta)u^{(1)}_b\rightarrow
D_{+}(\eta)$. Hence
$\delta^{\rc}_{\lin}(k,\eta)/[D_+(\eta)T^{\rc}(k)\delta_0(k)]\rightarrow
1$, and likewise for the baryons. Thus we see that the amplitude of
the BAO in the CDM are indeed increasing, and those in the baryons are
decaying.

Note that for the CDM distribution, by $z=5$ the BAO are almost fully
in place with $<1\%$ changes to the oscillation amplitude as the
distribution is evolved to $z=0$. However between $z=10$ and $5$ there
is a significant difference and one must be careful to account for
these changes when modeling BAO at high redshift.  We now draw
attention to the important fact that the baryon distributions display
more significant changes, there being an almost 50\% difference
between the initial and final distributions on large-scales. As noted
this will lead to significant baryon bias for statistical probes that
are sensitive primarily to the distribution of baryons in the Universe
and not the dark matter. In the following sections, we explore how
non-linear evolution changes these results.


\section{Perturbation theory}\label{sec:PT}


\subsection{Solutions}\label{ssec:solns}

\begin{figure}
\centering{
  \includegraphics[scale=0.8,clip=]{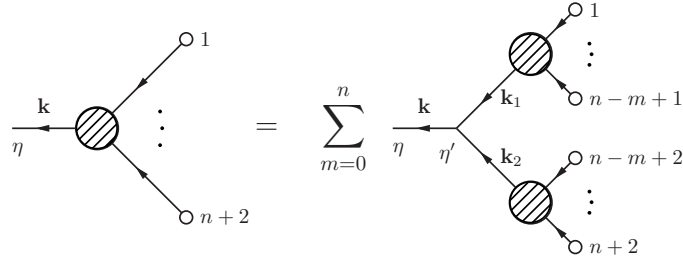}}
\caption{\small{Graphical representation of \eqn{eq:PertSoln}. The
    blob on the left hand side corresponds to $\Psij{n+1}$ (which
    clearly involves $n+2$ initial conditions), while the two blobs on
    the right hand side correspond to $\Psij{n-m}$ and $\Psij{m}$
    (involving $n-m+1$ and $m+1$ initial conditions) respectively.
    The trivalent vertex denotes the vertex matrix
    $\gas$.}\label{fig:rec_sol}}
\end{figure}

Following \citep{CrocceScoccimarro2006a} we look for power series
solutions to the equations of motion, \eqn{eq:eom}
\beq
\Psi_a(\kv,\eta) = \sum_{j=0}^{\infty} \Psi^{(j)}_a(\kv,\eta)
\label{eq:PTsol}\ 
\eeq
(in our notation the linear solution is $\Psij{0}$ as opposed to the
more usual $\Psij{1}$). On inserting the above expression into the
general solution of \eqn{eq:gensol}, we find that
\bal
\Psij{0}_a(\kv,\eta) 
	&=  g_{ab}(\eta)\phi^{(0)}_b(\kv) \, ; 
\label{eq:PT0} 
\\
\Psij{1}_a(\kv,\eta)  
	&=  \int_0^\eta \rd\eta' g_{ab}(\eta-\eta')
	\gas_{bcd}(\kv,\kv_1,\kv_2)\Psi_c^{(0)}(\kv_1,\eta')\Psi_d^{(0)}(\kv_2,\eta')\, ; 
\label{eq:PT1}
\\
\Psij{2}_a(\kv,\eta)  
	&=  2 \int_0^\eta \rd\eta' g_{ab}(\eta-\eta')
	\gas_{bcd}(\kv,\kv_1,\kv_2)\Psi_c^{(0)}(\kv_1,\eta')\Psi_d^{(1)}(\kv_2,\eta')\, ; 
\label{eq:PT2}
\\
\vdots \hspace{1cm} &\hspace{0.15cm} \vdots \hspace{2cm} \vdots  \hspace{2cm} \vdots
\hspace{2cm} \vdots \nn \\ 
\Psij{n+1}_a(\kv,\eta)  
	&=  \int_0^\eta \rd\eta' g_{ab}(\eta-\eta')
	\gas_{bcd}(\kv,\kv_1,\kv_2)
	\sum_{m=0}^n\Psij{n-m}_c(\kv_1,\eta')\Psij{m}_d(\kv_2,\eta')\, .
\label{eq:PertSoln}
\eal
\Fig{fig:rec_sol} shows a graphical representation of
\eqn{eq:PertSoln}.


We may develop these solutions further by inserting into the above
expressions the linear propagator $g_{ab}(\eta)$ given by
\eqn{eq:gab}. Thus, for the linear evolution of the initial
waves, we have:
\beq
\Psij{0}_a(\kv,\eta) = \sum_l\re^{l\eta}g_{ab,l}\,\phi^{(0)}_b(\kv)\, .
\label{eq:PertSol0}
\eeq 
Similarly, for the first order solution we have
\ba
\Psij{1}_a(\kv,\eta) & = &  \int_0^\eta \rd\eta' g_{ab}(\eta-\eta')
	\gas_{bcd}(\kv,\kv_1,\kv_2)g_{cc'}(\eta')\phi_{c'}(\kv_1)g_{dd'}(\eta')\phi_{d'}(\kv_2)\ ;	
\nn \\ & = & 
	\int_0^\eta \rd\eta' \sum_l\re^{l(\eta-\eta')}g_{ab,l}
	\gas_{bcd}(\kv,\kv_1,\kv_2)
	\sum_m\re^{m\eta'}g_{cc',m}\,\phi^{(0)}_{c'}(\kv_1)
	\sum_n\re^{n\eta'}g_{dd',n}\,\phi^{(0)}_{d'}(\kv_2)
\nn \\ & = &
	\int_0^\eta \rd\eta' \sum_{l,m,n}\re^{l\eta+(m+n-l)\eta'}g_{ab,l}
	\gas_{bcd}(\kv,\kv_1,\kv_2)
	g_{cc',m}\,g_{dd',n}\,\phi^{(0)}_{c'}(\kv_1)\phi^{(0)}_{d'}(\kv_2)
\nn \\ & = &
	\sum_{l,m,n}\re^{l\eta}I_1(m+n-l;\eta)
	g_{ab,l}\,\bgas_{bcd}(\kv_1,\kv-\kv_1)g_{cc',m}\,g_{dd',n}\,
	\phi^{(0)}_{c'}(\kv_1)\phi^{(0)}_{d'}(\kv-\kv_1)\ ,
\label{eq:PertSol1}
\ea
where in the last line we have introduced the function:
\beq
I_1(l;\eta) = \int_0^\eta\rd\eta'\,\re^{l\eta'} = 
	\left\{\begin{array}{ll}
	\displaystyle{(\re^{l\eta}-1)/l} & l\ne 0
	\\
	\eta & l=0
	\end{array}\right. \ ,
\label{eq:I1}
\eeq
and performed the $\kv_2$ integral using the Dirac delta function
implicit in $\gas_{bcd}(\kv,\kv_1,\kv_2)$.

In what follows we shall also need to make use of the 2nd order
solution for the fields and after a similar procedure as above we
find:
\ba
\Psij{2}_a(\kv,\eta) & = & 
	2\sum_{l,m}\sum_{p,q,r}
	\re^{l\eta}I_2(m+p-l,q+r-p;\eta)\,
        g_{ab,l}\,\bgas_{bcd}(\kv_1,\kv-\kv_1)\,g_{cc',m}\,	
\nn \\ & \times &
        g_{df,p}\,\bgas_{fgh}(\kv_2,\kv-\kv_1-\kv_2)\,g_{gg',q}\,g_{hh',r}\,
        \phi^{(0)}_{c'}(\kv_1)\phi^{(0)}_{g'}(\kv_2)\phi^{(0)}_{h'}(\kv-\kv_1-\kv_2) \ . 
\label{eq:PertSol2}
\ea
In the above we have introduced a further auxiliary function,
\beq
I_2(l_2,l_1;\eta) = \int_0^\eta\rd\eta\, \re^{l_2\eta'}I_1(l_1;\eta')\ 
= \left\{
\begin{array}{ll}
\left[l_2e^{(l_1+l_2)\eta}-(l_1+l_2)e^{l_2\eta}+l_1\right]/\left[l_1l_2(l_1+l_2)\right] 
& (l_1\ne0,l_2\ne0) \\[.5ex]
\left[e^{l_1\eta}-1-l_1\eta\right]/l_1^2 & (l_1\ne0,l_2=0)\\[.5ex]
\left[e^{l_2\eta}(l_2\eta-1)+1\right]/l_2^2 & (l_1=0,l_2\ne0) \\[.5ex]
\eta^2/2 & (l_1=0,l_2=0)
\end{array}\right.\ .
\label{eq:I2}
\eeq
%


\subsection{Feynman rules for the PT solutions}\label{ssec:feynman}

\begin{figure}[t]
\centering{
  \includegraphics[scale=0.8,clip=]{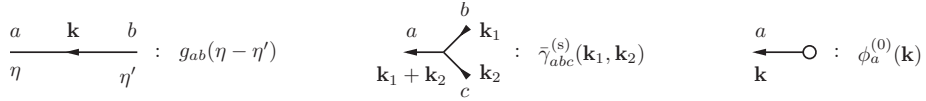}}
\caption{ \small{Diagrammatic representation of the basic mathematical
  building blocks of the solutions. {\em Left panel}: the linear propagator
  $g_{ab}(\eta-\eta')$. {\em Middle panel}: the interaction vertex 
  $\bgas_{abc}(\kv_1,\kv_2)$. The sticks are not part of the vertex and are
  present only to show the correct number of incoming and outgoing lines.
  {\em Right panel}: the initial field $\phi^{(0)}_{a}(\kv)$, denoted by the empty
  dot. The stick with the arrow serves only to show that an initial condition
  must always be connected to a diagram with a line going into the diagram,
  i.e.~oriented away from the initial condition itself.
  \label{fig:feynrule}}
}
\end{figure}

As was shown by \citep{CrocceScoccimarro2006a} there is a simple
interpretation for the solutions $\Psi^{(n)}_a(\bk)$ given by
\eqnss{eq:PT0}{eq:PertSoln}. The first equation in the hierarchy
denotes an initial condition $\phi_a^{(0)}$, evolved under the linear
propagator $g_{ab}$. The second denotes the non-linear coupling
between two initial waves of incoming momentum $\bk_1$ and $\bk_2$
through the interaction vertex $\gamma^{(s)}_{abc}(\bk,\bk_1,\bk_2)$,
to produce an outgoing wave of momentum $\bk=\bk_1+\bk_2$, which is
one order higher in the PT series and evolving under the action of
$g_{ab}$. For the next equation in the series we see that the solution
structure is now repeated, and this recurs for all higher order
terms. Thus a convenient graphical representation for the solutions is
obtained by ``iterating'' the graph in \fig{fig:rec_sol}, until on the
right-hand-side we are left with diagrams involving only trivalent
vertices and no ``blobs''.
%
%
Associated to this graphical representation is a simple set of {\em
  Feynman rules} for obtaining the $n$-th order solution,
$\Psij{n}(\kv,\eta)$; these may be stated as follows:
\begin{enumerate}
\item \label{step:draw} Draw all {\em topologically distinct,
  connected tree diagrams} containing $n$ vertices and a directed line
  coming into the diagram from the right for each of $n+1$ initial
  conditions, and a directed line going to the left out of the diagram
  for the final wave. Draw any number of directed internal lines
  running from one vertex to another, as required to give each vertex
  exactly two incoming and one outgoing attached line. It is most
  straightforward to label the direction of the lines by drawing
  arrows on them.
\item \label{step:mark} Assign each vertex its own time variable,
  $s_j$, and each line its own momentum, $\kv_j$, such that momentum
  is conserved at each vertex, with the momenta considered to flow in
  the direction of the arrows. For the single outgoing line, assign
  the momentum $\kv$ and final time $\eta$.
\item \label{step:math} Identify the various pieces of each diagram
  with the mathematical expressions as shown in \fig{fig:feynrule}.
\item \label{step:int} Integrate over all intermediate times $s_j$,
  each between $[0,\eta]$. Integrate over all the independent
  wavevectors with the measure $\rd^3 \kv_j$. (Note that the momenta
  of all internal lines and a single incoming line are fixed by
  momentum conservation, thus there will be $n$ independent momenta to
  integrate over.) Sum over all internal field indices $b$, $c$, etc.
\item \label{step:symm} Each diagram carries a factor of $2^r$ if
  there are precisely $r$ vertices that are not symmetric with respect
  to interchanging their two incoming waves.
\item \label{step:sum} The value of $\Psi_a^{(n)}(\kv,\eta)$ is given
  by the sum over the values of these diagrams.
\end{enumerate}


\begin{figure}[h]
\centering{
  \includegraphics[angle=-90,scale=0.8,clip=]{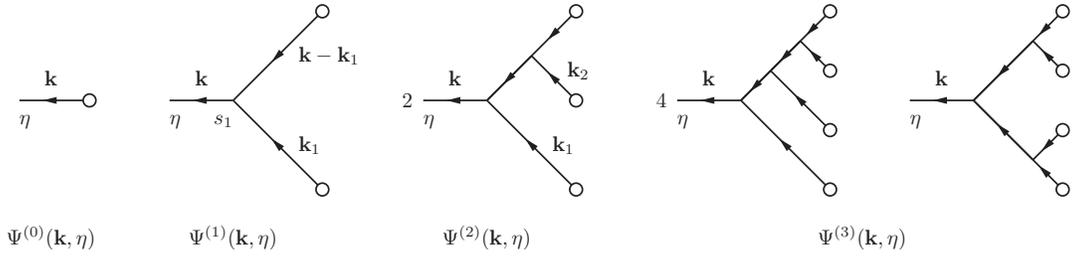}}
\caption{\small{Graphical representation of the topologically distinct
    Feynman diagrams for the perturbation theory up to third
    order. From left to right we show, $\Psi_a^{(n)}(\bk)$, for
    $n\in\{0,1,2,3\}$, with the last two diagrams contributing to
    $n=3$.}\label{fig:sols}}
\end{figure}

\Fig{fig:sols} shows all diagrams up to $n=3$, together with their
symmetry factors calculated as in Step~\ref{step:symm} above.  As an
example of the use of the Feynman rules, the expressions corresponding
to the first two diagrams can be written,
\bal
\Psij{0}_a(\kv,\eta) &= g_{ab}(\eta) \phi^{(0)}_b(\kv)\,,
\label{eq:Psi0fromFrules}
\\
\Psij{1}_a(\kv,\eta) &=
	\int_{0}^{\eta} \rd s_1\, g_{ab}(\eta-s_1)
	\int \rd^3 \kv_1\, \bgas_{bcd}(\kv_1,\kv-\kv_1)
	g_{cc'}(s_1)\phi^{(0)}_{c'}(\kv_1) g_{dd'}(s_1)\phi^{(0)}_{d'}(\kv-\kv_1)\,,
\label{eq:Psi1fromFrules}
\eal
which are nothing but \eqns{eq:PertSol0}{eq:PertSol1} (after
performing the $s_1$ integral in \eqn{eq:Psi1fromFrules}).  Similarly,
evaluating the third diagram, we obtain \eqn{eq:PertSol2}, and so on.


\section{Statistics}\label{sec:stats}

\subsection{Two-point covariance matrix of density and velocity modes}

In cosmology rather than modeling $\Psi_a(\bk)$ at a particular point,
we are more interested in computing statistical averages of various
products of the fields at different points in space. We shall
represent these statistical averages as: $\left<\dots\right>$. For
mean zero fields, we must have $\left<\Psi_a(\bk)\right>=0$. Thus the
lowest order statistic of interest to us is the two-point correlation
function, or power spectrum, of $\Psi_a(\bk)$, which can be defined:
\beq
\left<\Psi_a(\kv,\eta)\Psi_b(\kv',\eta)\right\rangle
= P_{ab}(\kv,\eta)\Ddel(\kv+\kv')\, .
\label{eq:Pdef}
\eeq


\subsection{Two-point covariance matrix of initial fields}\label{ssec:ics2}

Throughout we will assume that the initial conditions are
Gaussian. Consequently the statistical properties of the fields
$\phi^{(0)}_a(\kv)$ are then completely characterized by the two-point
covariance matrix:
\beq
\left<\phi^{(0)}_a(\kv)\phi^{(0)}_b(\kv')\right> 
\equiv    \cP^{(0)}_{ab}(k)\Ddel(\kv+\kv') \label{eq:Phi0def} \ .
\eeq
For the case of a CDM and baryon fluid this matrix takes the explicit
form:
\beq
\cP^{(0)}_{ab}(k) = 
\left[\begin{array}{llll}
u_1^2 [T^{\rc}(k)]^2   & u_1 u_2 [T^{\rc}(k)]^2 & u_1 u_3 T^{\rc}(k)T^{\rb}(k) & u_1 u_4 T^{\rc}(k)T^{\rb}(k) \\
u_1 u_2 [T^{\rc}(k)]^2  & u_2^2 [T^{\rc}(k)]^2  & u_2 u_3 T^{\rc}(k)T^{\rb}(k) & u_2 u_4 T^{\rc}(k)T^{\rb}(k) \\
u_1 u_3 T^{\rc}(k)T^{\rb}(k) & u_2 u_3 T^{\rc}(k)T^{\rb}(k) & u_3^2   [T^{\rb}(k)]^2 & u_3 u_4 [T^{\rb}(k)]^2 \\
u_1 u_4 T^{\rc}(k)T^{\rb}(k) & u_2 u_4 T^{\rc}(k)T^{\rb}(k) & u_3 u_4 [T^{\rb}(k)]^2 & u_4^2   [T^{\rb}(k)]^2
\end{array}\right]{\mathcal P}_{0}(k) \ ,
\label{eq:Phi0def2}
\eeq
where we defined ${\mathcal P}_{0}(k)\Ddel(\kv+\kv') =
\left<\del_{0}(\kv)\del_{0}(\kv')\right\rangle$, to be the power
spectrum of primordial density fluctuations: i.e.~${\mathcal
  P}_0\propto k^{n}$, with $n=1$ for the usual Harrison--Zel'dovich
spectrum. The two-point correlation matrix of the initial fields may
be expressed more compactly if we choose large-scale growing mode
initial conditions, ($u_a=u_a^{(1)}=(1,1,1,1)$), whereupon
\beq \cP^{(0)}_{ab}(k) = T_a(k) T_b(k) {\mathcal P}_{0}(k) \ ,\eeq
with
$\left[T_a(k)\right]^T \equiv \left[T^{\rc}(k),T^{\rc}(k),T^{\rb}(k),T^{\rb}(k)\right]$.

For Gaussian initial conditions all of the multi-point correlators of
the initial $\phi^{(0)}_a(\kv)$ can be computed using Wick's theorem:
all higher order correlators involving an odd number of fields vanish,
while for an even number, there are $(2n-1)!!$ contributions
corresponding to the number of different pairings of the $2n$ fields:
\beq
\left<\phi^{(0)}_{a_1}(\kv_1)\ldots\phi^{(0)}_{a_{2n}}(\kv_{2n})\right\rangle =
\sum_{\mbox{\scriptsize all pair associations}} \;\;
\prod_{\mbox{\scriptsize p pairs of } (i,j)}
\left<\phi^{(0)}_{a_i}(\kv_i)\phi^{(0)}_{a_j}(\kv_j)\right\rangle\,.
\eeq

\begin{figure}[h!]
\centering{
  \includegraphics[scale=0.8,clip=]{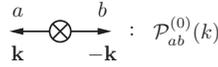}}
\caption{\small{Diagrammatic representation of the initial power
    spectrum matrix. The averaged pair of initial fields is denoted by
    the crossed dot, with the sticks and arrows serving only to show
    that each initial power spectrum will have two outgoing lines
    attached in any diagram.}\label{fig:P0}}
\end{figure}

For later use, we now also introduce a diagrammatic representation of
the initial power spectrum matrix, see \fig{fig:P0}.


\subsection{The non-linear propagator}\label{ssec:multipoint}

As the non-linear interactions are turned on, the evolution of the
initial field breaks away from the simple scaling, $\Psi_a(\bk,\eta)=
g_{ab}(\eta)\phi^{(0)}_b(\bk)$. The deviations can be thought of as,
{\em erasing the memory of the mode to its initial conditions}. This
concept can be made more concrete, through introducing the non-linear
propagator \citep{CrocceScoccimarro2006a,CrocceScoccimarro2006b}:
\beq
G_{ab}(\kv,\eta)\Ddel(\kv-\kv') \equiv 
	\left<\frac{\del\Psi_a(\kv,\eta)[\phi^{(0)}]}{\del\phi_b^{(0)}(\kv')}\right>\,.
\label{eq:Gabdef}
\eeq
This means that we take the functional derivative of the full solution
$\Psi_a(\kv,\eta)$ with respect to the initial condition
$\phi^{(0)}_b(\kv')$, and then perform the ensemble average. Note that
the non-linear propagator $G_{ab}$ is defined with the momentum
conserving delta function factored out. The expectation is taken
since, again, it is the average evolution of the mode away from its
initial conditions that we are interested in.

In perturbation theory, the full solution is given as in
\eqn{eq:PTsol}, and thus 
\beq
G_{ab}(\kv,\eta) = \sum_{l=0}^{\infty}\dGj{l}_{ab}(\kv,\eta)\,,
\label{eq:Gabpert}
\eeq
where
\beq
\dGj{l}_{ab}(\kv,\eta)\Ddel(\kv-\kv') = 
	\left<\frac{\del\Psij{2l}_a(\kv,\eta)[\phi^{(0)}]}{\del\phi_b^{(0)}(\kv')}\right>\,.
\label{eq:Glabdef}
\eeq
Notice that in the above equation we only consider contributions to
the propagator that come from terms in the expansion of
$\Psi_a(\bk,\eta)$ with an odd number of initial conditions, i.e.~only
terms $\Psi^{(j)}_a(\bk,\eta)$ with $j$ an even number contribute.
This owes to the fact that, for non-linear fields involving an even
number of initial conditions, after performing the functional
differentiation we are left with products that involve an isolated
initial field, which vanishes on taking the expectation:
$\langle\phi^{(0)}_a(\bk)\rangle=0$. Recall that we take $j=0$ as the
first term in the perturbative series for $\Psi_a$.  Thus the total
non-linear propagator up to one-loop level can be written:
\beq 
G_{ab}^{(1)}(\kv,\eta)  =  g_{ab}(\eta)+\dGj{1}_{ab}(\kv,\eta)\  ,
\eeq 
where we used the fact that $\del G_{ab}^{(0)}(\kv,\eta) =
g_{ab}(\eta)$, and where \citep{CrocceScoccimarro2006b},
\ba
\delta G_{ab}^{(1)}(\bk,\eta)  & = & 
4\int_0^{\eta} \rd\eta'\int_0^{\eta} \rd\eta'' g_{ab'}(\eta-\eta') 
\int \rd^3 \kv_1\,
\bgas_{b'cd}(\bk_1,\bk-\bk_1)  g_{de}(\eta'-\eta'') 
\nn \\
& \times &
\bgas_{efg}(-\bk_1,\bk)g_{cc'}(\eta') g_{ff'}(\eta'')
g_{gb}(\eta'')(\bk_1)\cP^{(0)}_{c'f'}(k_1)\ .
\label{eq:dGab1compact}
\ea
In \app{app:propagator} we present full details of the calculation for
\eqn{eq:Glabdef} explicitly for $j=0,1,2$.

As for the case of the perturbative solutions to $\Psi_a$, the series
expansion for the non-linear propagator with arbitrary numbers of loops
can also be constructed in a diagrammatic fashion.  Assuming Gaussian
initial conditions, the Feynman rules for computing the $n$-loop
correction to the propagator, $\dGj{n}_{ab}(\kv,\eta)$ are:
\begin{enumerate}
\item Draw all tree diagrams corresponding to $\Psij{2n}(\kv,\eta)$ as
  described in \Sect{ssec:feynman}. Recall that these diagrams
  involve $2n+1$ initial fields each.
\item For any specific tree, drop any one initial condition, labeling
  the incoming line thus created with the momentum $\kv$ (this
  corresponds to the functional derivative and dropping the delta
  function). Pair the remaining $2n$ initial fields in all possible
  ways into $n$ initial power spectra, as dictated by Wick's theorem
  (corresponding to the ensemble averaging). Any specific tree then
  leads to $(2n+1)(2n-1)!!=(2n+1)!!$ $n$-loop diagrams, not all of
  which need to be topologically distinct. Diagrams containing a
  ``tadpole'' as a subdiagram (i.e.~a subdiagram with a single external line) 
  may be dropped, since they are zero, as explained below.
\item \label{step:Gmark} Follow Steps~\ref{step:mark}--\ref{step:int}
  of \Sect{ssec:feynman}. When assigning the lines their momenta,
  ensure that the two lines emanating from an initial power spectrum
  carry equal and opposite momenta. Identify the various pieces of
  each diagram with the mathematical expressions as shown in
  \figs{fig:feynrule}{fig:P0}.  (The number of independent momenta to
  integrate over in Step~\ref{step:int} will be just the number of
  independent loops, $n$.)
\item \label{step:Gsymm} Each diagram carries the appropriate factor
  of $2^r$ inherited form the original tree diagram. As noted above,
  it is possible that a specific loop diagram arises more than
  once. Of course such a diagram needs to be computed only once, but
  one must take account of its multiplicity explicitly.
\item The value of $\dGj{n}_{ab}(\kv,\eta)$ is given by the sum over
  the values of all diagrams coming from all trees.
\end{enumerate}

\begin{figure}[t]
\centering{
  \includegraphics[scale=0.8,clip=]{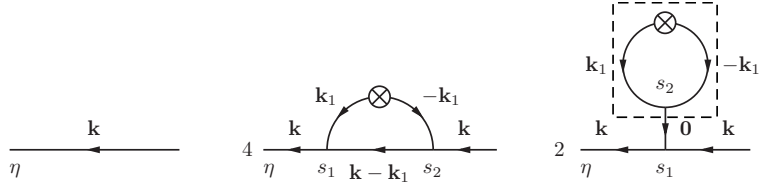}}
\caption{\small{The diagrams for the one-loop non-linear propagator
    $G^{(1)}(\kv,\eta)$. The first diagram represents \Eqn{eq:gab};
    the second \Eqn{eq:dGab1compact}; and the last diagram contains a
    tadpole (marked with the dashed box) and thus evaluates to
    zero.}\label{fig:G1graphs}}
\end{figure}

\Fig{fig:G1graphs} shows all diagrams that arise in the computation of
the one-loop propagator, including the symmetry factor computed in
Step~\ref{step:Gsymm} above. Only the first two diagrams give
non-vanishing contributions. The last diagram contains a tadpole
(marked with the dashed box) and this gives zero contribution, since
for each individual tadpole momentum conservation implies:
$\alpha(\kv,-\kv)=\beta(\kv,-\kv)=0$. Hence, diagrams containing
tadpoles vanish.

The power of the above graphical approach is that if one inspects the
non-vanishing one-loop diagrams and uses the Feynman rules, then one
can immediately write down the perturbative correction, as in
\eqn{eq:dGab1compact}.


\subsection{Perturbative representation of the power spectrum}\label{ssec:perturbpow}

A perturbative expansion for the non-linear power spectrum may be
obtained through insertion of the non-linear fields $\Psi_a^{(n)}$
(c.f.~\eqn{eq:PertSoln}) into \eqn{eq:Pdef}, whereupon
\beq
P_{ab}(\kv,\eta) = \sum_{l=0}^{\infty} \del P_{ab}^{(l)}(\kv,\eta)
\label{eq:Ppertser}\ .
\eeq
On assuming Gaussian initial conditions, then from Wick's theorem we
find that only those products of fields that involve an even number of
initial conditions contribute to the non-linear power
spectrum. Consequently we may write,
\beq
\del P_{ab}^{(l)}(\kv,\eta)\Ddel(\kv+\kv') =
\sum_{j=0}^{2l}\left<\Psij{j}_a(\kv,\eta)\Psij{2l-j}_b(\kv',\eta)\right\rangle\,.
\label{eq:Ppertcont}
\eeq
On taking all contributions to the power spectrum matrix up to
one-loop, $l=1$, we have: 
\beq P_{ab}=\del P_{ab}^{(0)}+\del P_{ab}^{(1)}\ ;\eeq
where the linear theory contribution (assuming large-scale growing
mode initial conditions) is given by:
\ba
\del P^{(0)}_{ab}(\kv,\eta)\Ddel(\kv+\kv') & = &  
	\left< \Psi_a^{(0)}(\kv,\eta) \Psi_b^{(0)}(\kv',\eta) \right> \ \nn \\ 
	& = &
	g_{aa'}(\eta)T_{a'}(k)g_{bb'}(\eta)T_{b'}(k){\mathcal P}_{0}(k)\Ddel(\kv+\kv') .
\label{eq:LinPowEvol}
\ea
We may also write the above expression using conventional matrix
notation: 
\beq
{\bf P}^{(0)}(\kv,\eta) = 
{\bf g}(\eta){\bf T}(k){\bf T}^{T}(k){\bf g}^{T}(\eta){\mathcal P}_{0}(k)\,.
\label{eq:P0mat}
\eeq
The one-loop corrections to the power spectrum are written:
\ba 
\del P^{(1)}_{ab}(\kv,\eta)\Ddel(\kv+\kv') & = & 
    \left< \Psi_a^{(0)}(\kv,\eta) \Psi_b^{(2)}(\kv',\eta) \right>
   +\left< \Psi_a^{(1)}(\kv,\eta) \Psi_b^{(1)}(\kv',\eta) \right>
   +\left< \Psi_a^{(2)}(\kv,\eta) \Psi_b^{(0)}(\kv',\eta) \right>
\ \,. 
\label{eq:P1loop}
\ea
Consider the sum of the first and last terms, we shall call these the
``reducible'' contributions and denote the sum of these terms: $\delta
P^{(1)}_{ab,\mathrm{R}}$.  It was shown by
\citep{CrocceScoccimarro2006b} that the non-linear propagator plays
the role of a Green's function in the two-point sense,
$\langle\Psi_a\phi^{(0)}_b\rangle=G_{ac}\langle\phi^{(0)}_c\phi^{(0)}_b\rangle$,
and furthermore that this property holds order-by-order:
$\langle\Psij{2n}_a\phi^{(0)}_b\rangle=\dGj{n}_{ac}\langle\phi^{(0)}_c\phi^{(0)}_b\rangle$.
On applying the above relations, it is straightforward to see that we
must have
\ba
\del P^{(1)}_{ab,\mathrm{R}}(\kv,\eta)\Ddel(\kv+\kv') & = &
	\left[\left< \Psi_a^{(0)}(\kv,\eta) \Psi_b^{(2)}(\kv',\eta) \right>
	+\left< \Psi_a^{(2)}(\kv,\eta) \Psi_b^{(0)}(\kv',\eta) \right>\right]
\nn \\ & = &
\left[g_{ac}(\eta) \dGj{1}_{bd}(\kv',\eta) \cP^{(0)}_{cd}(\kv')
	+\dGj{1}_{ac}(\kv,\eta) g_{bd}(\eta) \cP^{(0)}_{cd}(\kv)\right]
	\Ddel(\kv+\kv')\,.
\label{eq:dP1abR}
\ea
Specializing to growing mode initial conditions, and using momentum
conservation, we find
\beq
\del P^{(1)}_{ab,\mathrm{R}}(\kv,\eta) = 
	\left[g_{ac}(\eta) \dGj{1}_{bd}(\kv,\eta)
	+\dGj{1}_{ac}(\kv,\eta) g_{bd}(\eta)\right] T_c(k)T_d(k){\mathcal P}_0(k)\,.
\label{eq:dP1Rgm}
\eeq
In particular, we can write this in conventional matrix notation as
\beq 
\del{\bf P}^{(1)}_{\mathrm{R}}(\kv,\eta) = 
\left[{\bf g}(\eta){\bf T}(k){\bf T}^{T}(k)\big[\del {\bf G}^{(1)}\big]^{T}(\kv,\eta) +  
\del {\bf G}^{(1)}(\kv,\eta){\bf T}(k){\bf T}^{T}(k){\bf g}^T(\eta)\right]
{\mathcal P}_{0}(k)\,.
\label{eq:dP1mat}
\eeq
Adding \eqns{eq:P0mat}{eq:dP1mat}, we see that the sum $\big[{\bf
    P}^{(0)}(\kv,\eta) + \del{\bf
    P}^{(1)}_{\mathrm{R}}(\kv,\eta)\big]$ is nothing but the one-loop
truncation of the all-order expression 
\beq {\bf P}_{\mathrm{R}}(\kv,\eta) = {\bf G}(\kv,\eta) {\bf T}(k){\bf
  T}^{T}(k){\bf G}^{T}(\kv,\eta)\cP_0(k)\,.  \eeq
The all-order reducible contribution collects all corrections to ${\bf
  P}(\kv,\eta)$ that are proportional to the initial power spectrum at
the same scale $k$.

Turning now to the second term in \eqn{eq:P1loop}, conventionally
called the ``mode-coupling'' contribution, we find that it can be
written \citep{CrocceScoccimarro2006a}:
\beq
\bsp
\del P^{(1)}_{ab,\mathrm{MC}}(\kv,\eta) &=
	2\int_0^\eta \rd \eta' \int_0^\eta \rd \eta'' \int \rd^3\kv_1\,
	g_{ac}(\eta-\eta') \bgas_{cde}(\kv_1,\kv-\kv_1) g_{dd'}(\eta') g_{ee'}(\eta')
\\
&\quad\times
	g_{bf}(\eta-\eta'') \bgas_{fgh}(-\kv_1,\kv_1-\kv) g_{gg'}(\eta'') g_{hh'}(\eta'')
	\cP^{(0)}_{d'g'}(k_1) \cP^{(0)}_{e'h'}(|\kv-\kv_1|)\,. 
\esp
\label{eq:dP1mc}
\eeq 
The mode-coupling piece includes corrections coming from modes other
than $k$ in the initial power spectrum. In particular, we see from
\eqn{eq:dP1mc} that the one-loop mode-coupling correction involves the
convolution of two initial power spectra with a specific kernel. At
higher loop orders, the mode-coupling pieces involve convolutions of
more than two initial powers.

The upshot is that for growing mode initial conditions, we can write
the full power spectrum as
\beq 
{\bf P}(\kv,\eta) = 
{\bf G}(\kv,\eta) {\bf T}(k){\bf T}^{T}(k){\bf G}^{T}(\kv,\eta) {\mathcal P}_{0}(k)
+{\bf P}_{\mathrm{MC}}(\kv,\eta)\,.\label{eq:Pupshot}
\eeq

At this point we may extend our diagrammatic description: assuming
Gaussian initial conditions, the Feynman rules for computing the
$n$-loop correction to the power spectrum, $\del
P^{(n)}_{ab}(\kv,\eta)$ are:
\begin{enumerate}
\item Draw the sum of all tree diagrams corresponding to
  $\Psij{j}(\kv,\eta)$, and the sum of diagrams corresponding to
  $\Psij{2n-j}(\kv',\eta)$ as described in \Sect{ssec:feynman},
  including the arrows on the lines.
\item Formally ``multiply'' these two sums using distributivity. It is
  convenient to draw one set of diagrams flipped around the vertical
  axis, so that after the pairing of diagrams for $\Psij{j}(\kv,\eta)$
  with those for $\Psij{2n-j}(\kv,\eta)$, their initial conditions
  face each other.  Next, pair the $2n+2$ initial fields into $n+1$
  initial power spectra (some pairings may lead to the same loop
  diagram). Disregard diagrams that contain a tadpole as a
  subdiagram. (This includes disconnected diagrams in particular.) 
\item Follow Step~\ref{step:Gmark} of \Sect{ssec:multipoint}.
\item \label{step:Psymm} Each diagram carries the appropriate factor
  of $2^{r+s}$ inherited form the two original tree diagrams. As noted
  above, it is possible that a specific loop diagram arises more than
  once. Such a diagram needs to be computed only once, but its
  multiplicity must be taken into account.
\item The value of $\del P^{(n)}_{ab}(\kv,\eta)$ is given by the sum
  over the the values of these diagrams for all $j=0,\ldots,2n$.
\end{enumerate}

\begin{figure}[t]
\centering{
  \includegraphics[angle=-90,scale=0.8,clip=]{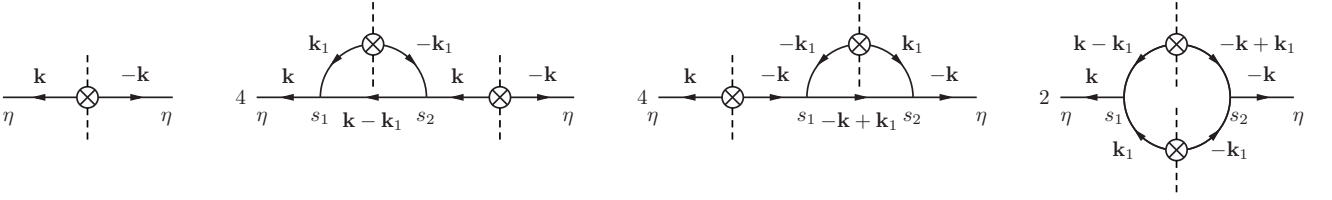}}
\caption{\small{The non-zero diagrams for the one-loop power spectrum
    $P^{(1)}(\kv,\eta)$. The dashed lines indicate the points at which
    the two trees were glued together. The first diagram represents
    \Eqn{eq:LinPowEvol}; the second and third graphs are the two
    reducible contributions in \Eqn{eq:dP1mat}; and the fourth graph
    represents the mode-coupling term of
    \Eqn{eq:dP1mc}.}\label{fig:P1graphs}}
\end{figure}

\Fig{fig:P1graphs} shows all non-zero diagrams for the power spectrum
up to one-loop order, including the symmetry factor computed in
Step~\ref{step:Psymm} above. In particular, notice how \eqn{eq:dP1abR}
is very simple to write down by inspecting the second and third
graphs. The last diagram corresponds to the mode-coupling
piece. Evidently \eqn{eq:dP1mc} may be written down immediately by
inspecting this graph.


\section{One-loop corrections to the power spectrum}\label{sec:OneLoopPower}


\subsection{The non-linear propagator}\label{ssec:1Lprop}


In \Sect{sec:stats}, we defined the non-linear propagator
$G_{ab}(\bk,\eta)$ as the ensemble average of the functional
derivative of the full solution $\Psi_a(\kv,\eta)$ by the initial
condition $\phi^{(0)}_b(\kv')$, without the delta function for
momentum conservation (c.f.~\eqn{eq:Gabdef}).  It was shown to have a
perturbative expansion (c.f.~\eqns{eq:Gabpert}{eq:Glabdef}) and terms
up to one-loop level were computed in \app{app:propagator}. Here we
further develop the one-loop expression for the case of $N=2$ fluids.

To begin, let us reconsider \eqn{eq:dGab1compact}, we may expand all
of the linear propagators, $g_{ab}(\eta)$, using \eqn{eq:gab} to
obtain,
\ba 
\delta G^{(1)}_{aa'}(\bk,\eta)
& = & 
4\int_0^{\eta} \rd\eta' \sum_{l_1} g_{ab,l_1}\re^{l_1(\eta-\eta')}
\int \rd^3 \bk_1 \bar{\gamma}^{(s)}_{bcd}(\bk_1,\bk-\bk_1)
\sum_{l_2} g_{cc',l_2}\re^{l_2\eta'}
\int_0^{\eta'} \rd\eta'' 
\sum_{l_3} g_{de,l_3}\re^{l_3(\eta'-\eta'')}\nn \\
& \times &  
 \bar{\gamma}^{(s)}_{efg}(-\bk_1,\bk') 
\sum_{l_4} g_{ff',l_4}\re^{l_4\eta''}
\sum_{l_5} g_{ga',l_5}\re^{l_5\eta''}
{\mathcal P}_{c'f'}^{(0)}(k_1) \ .
\ea
On rearranging the above expression we find,
\ba 
\delta G^{(1)}_{aa'}(\bk,\eta)
& = & 
4 \sum_{l_1,\dots,l_5} \re^{l_1\eta} 
\int_0^{\eta}  \rd\eta'   \re^{(l_2+l_3-l_1)\eta'}
\int_0^{\eta'} \rd\eta''  \re^{(l_4+l_5-l_3)\eta''}\nn \\
& \times & g_{ab,l_1}\, \int \rd^3 \bk_1 \bar{\gamma}^{(s)}_{bcd}(\bk_1,\bk-\bk_1)
g_{cc',l_2}\, g_{de,l_3}\,\bar{\gamma}^{(s)}_{efg}(-\bk_1,\bk') g_{ff',l_4}\,g_{ga',l_5}
{\mathcal P}_{c'f'}^{(0)}(k_1) \ .
\ea
Considering the integrals over $\eta'$ and $\eta''$, these may be
conveniently evaluated using the algebraic function
$I_2(l_2,l_1,\eta)$ (c.f.~\eqn{eq:I2}). Thus our final expression for
the one-loop propagator takes the form:
\ba
\dGj{1}_{aa'}(\kv,\eta) &=&
	4\sum_{l_1,\dots,l_5}\re^{l\eta}I_2(l_2+l_3-l_1,l_4+l_5-l_3;\eta) \nn \\ 
	& \times & g_{ab,l_1} \int \rd^3 \bk_1\,\bgas_{bcd}(\kv_1,\kv-\kv_1)g_{cc',l_2}\,g_{de,l_3}
	\bgas_{efg}(-\kv_1,\kv)g_{ff',l_4}\,g_{ga',l_5}\, {\mathcal P}_{c'f'}^{(0)}(k_1)\ .
\ea
This expression may be reorganized so that the time dependent
algebraic functions, the constant coefficient matrices $g_{ab,l}$ and
the $k$-dependent matrices are grouped together. Further, if we take
large-scale growing mode initial conditions,
$u_a=u_a^{(1)}=(1,1,1,1)$, then we find that the propagator may be
written,
\ba
\dGj{1}_{aa'}(\kv,\eta) &=&
	4\sum_{l_1,\dots,l_5}\re^{l\eta}I_2(l_2+l_3-l_1,l_4+l_5-l_3;\eta) 
	\,g_{ab,l_1}\, g_{cc',l_2}\,g_{de,l_3}\, g_{ff',l_4}\,g_{ga',l_5}\, \nn \\ 
	& \times & 
	\int \rd^3 \bk_1\,\bgas_{bcd}(\kv_1,\kv-\kv_1) \bgas_{efg}(-\kv_1,\kv) T_{c'}(k_1)T_{f'}(k_1) 
	     {\mathcal P}_{0}(k_1)\ .
\ea
In \app{app:Gkints} we show that the integral over the product of
vertex matrices and the power spectrum matrix leads to five linearly
independent functions of $k$ for each independent product
$T_i(k_1)T_j(k_1)$. Hence, it is possible to write the propagator,
\def\dc{{\delta^{\rc}}}
\def\db{{\delta^{\rb}}}
\beq
\dGj{1}_{ab}(\kv,\eta) = 
	\sum_{n=1}^{5} [f^{\dc\dc}_n(k) A^{\dc\dc}_{ab,n}(\eta) + f^{\dc\db}_{n}(k) A^{\dc\db}_{ab,n}(\eta) 
	+ f^{\db\db}_{n}(k) A^{\db\db}_{ab,n}(\eta)]\,,
\label{eq:dG1final}
\eeq
where the summation in $n$ runs over the set of independent integrals
$f^{\del^i\del^j}_{n}(k)$ that are defined in \app{app:Gkints}.
The complete results for the $A^{\dc\dc}_{ab,n}(\eta)$,
$A^{\dc\db}_{ab,n}(\eta)$ and $A^{\db\db}_{ab,n}(\eta)$ functions are
too lengthy to reproduce here, but their general structure is the
following:
\beq
A^{\del^i \del^j}_{ab,n}(\eta) =
	\sum_l P^{\del^i\del^j}_{ab,n,l}(w_2) \re^{l\eta}
	+Q^{\del^i\del^j}_{ab,n}(w_2) \eta\,,
\label{eq:Adidjabn}
\eeq 
where the summation in $l$ (in this equation only) runs over the set
$\{3,2,3/2,1,0,-1/2,-1,-3/2,-2,-5/2\}$, and the
$P^{\del^i\del^j}_{ab,n,l}(w_2)$ and $Q^{\del^i\del^j}_{ab,n}(w_2)$
functions multiplying each $\eta$-structure are at most third order
polynomials in $w_2$ (we used $w_1=1-w_2$).


\subsection{Mode-coupling power spectrum}

Consider next the mode-coupling contribution to the power spectrum. At
the one-loop level, it is given by \eqn{eq:dP1mc}. We may develop this
expression further by again expanding all of the linear propagators
using \eqn{eq:gab} to obtain,
\ba
\del P^{(1)}_{ab,\mathrm{MC}}(\kv,\eta) & = &
	2\int_0^\eta \rd \eta' \int_0^\eta \rd \eta'' \int \rd^3\kv_1\,
	\sum_{l_1}g_{ac,l_1}\re^{l_1(\eta-\eta')} \bgas_{cde}(\kv_1,\kv-\kv_1) 
	\sum_{l_2}g_{dd',l_2}\re^{l_2\eta'} \sum_{l_3}g_{ee',l_3}\re^{l_3\eta'}
\nn \\
& \times & 
	\sum_{l_4}g_{bf,l_4}\re^{l_4(\eta-\eta'')} \bgas_{fgh}(-\kv_1,\kv_1-\kv) 
	\sum_{l_5}g_{gg',l_5}\re^{l_5\eta''} \sum_{l_6}g_{hh',l_6}\re^{l_6\eta''}
	\cP^{(0)}_{d'g'}(k_1) \cP^{(0)}_{e'h'}(|\kv-\kv_1|)\,. 
\label{eq:dP1mcE1}
\ea
On reorganizing this equation and collecting together all of the time
dependent terms, we find
\ba
\del P^{(1)}_{ab,\mathrm{MC}}(\kv,\eta) & = & 
	2\int_0^\eta \rd \eta' \sum_{l_1,l_2,l_3} \re^{l_1\eta}\re^{(l_2+l_3-l_1)\eta'}
	\int_0^\eta \rd \eta'' \sum_{l_4,l_5,l_6} \re^{l_4\eta}\re^{(l_5+l_6-l_4)\eta''}
\nn \\
& & \hspace{-2cm} \times 
	\int \rd^3\kv_1\,	
	g_{ac,l_1} \bgas_{cde}(\kv_1,\kv-\kv_1) g_{dd',l_2} g_{ee',l_3}
	g_{bf,l_4} \bgas_{fgh}(-\kv_1,\kv_1-\kv) g_{gg',l_5} g_{hh',l_6}
	\cP^{(0)}_{d'g'}(k_1) \cP^{(0)}_{e'h'}(|\kv-\kv_1|)\,. 
\ea
The integrals over $\eta'$ and $\eta''$ are easily performed, and the
result can be conveniently written in terms of the $I_1(l,\eta)$
function introduced in \eqn{eq:I1}. Hence,
\ba
\del P^{(1)}_{ab,\mathrm{MC}}(\kv,\eta) & = &
	2 \sum_{l_1,l_2,l_3} \re^{l_1\eta}I_1(l_2+l_3-l_1;\eta)
	\sum_{l_4,l_5,l_6} \re^{l_4\eta}I_1(l_5+l_6-l_4;\eta) \nn \\
& & \hspace{-2cm}\times
	\int \rd^3\kv_1\,	
	g_{ac,l_1} \bgas_{cde}(\kv_1,\kv-\kv_1) g_{dd',l_2} g_{ee',l_3}
	g_{bf,l_4} \bgas_{fgh}(-\kv_1,\kv_1-\kv) g_{gg',l_5} g_{hh',l_6}
	\cP^{(0)}_{d'g'}(k_1) \cP^{(0)}_{e'h'}(|\kv-\kv_1|)\,,
\ea
which is our final expression for the mode-coupling contribution to
the one-loop power spectrum. Upon setting large-scale growing mode
initial conditions, $u_a = u_a^{(1)}=(1,1,1,1)$, the above expression
may be written as,
\ba
\del P^{(1)}_{ab,\mathrm{MC}}(\kv,\eta) & = & 
	2\int \rd^3\kv_1\,\cP_{0}(k_1) \cP_{0}(|\kv-\kv_1|)
\nn \\
& \times &
	\sum_{l_1,l_2,l_3} \re^{l_1\eta}I_1(l_2+l_3-l_1;\eta)
	g_{ac,l_1} \bgas_{cde}(\kv_1,\kv-\kv_1) g_{dd',l_2} g_{ee',l_3}
	T_{d'}(k_1) T_{e'}(|\kv-\kv_1|)
\nn \\
& \times &
	\sum_{l_4,l_5,l_6} \re^{l_4\eta}I_1(l_5+l_6-l_4;\eta)
	g_{bf,l_4} \bgas_{fgh}(-\kv_1,\kv_1-\kv) g_{gg',l_5} g_{hh',l_6}
	T_{g'}(k_1) T_{h'}(|\kv-\kv_1|)\,.
\ea
In \app{app:Pkints} we show that the integration over the vertex
matrices and power spectrum matrices leads to four linearly
independent functions of $k$ for each independent product
$T_i(k_1)T_j(k_1)T_k(|\kv-\kv_1|)T_l(|\kv-\kv_1|)$ if $(ij)=(kl)$, and
six independent functions otherwise. Furthermore, if $(ij)\ne (kl)$,
then merely exchanging $(ij)\leftrightarrow (kl)$ does not lead to new
independent integrals. Thus, the mode-coupling correction to the
one-loop power spectrum can be written:
\ba
\del P^{(1)}_{ab,\mathrm{MC}}(\kv,\eta) & = &
	\sum_{n=1}^{4}[f^{\dc\dc\dc\dc}_n(k) A^{\dc\dc\dc\dc}_{ab,n}(\eta)
		+f^{\dc\db\dc\db}_n(k) A^{\dc\db\dc\db}_{ab,n}(\eta)
		+f^{\db\db\db\db}_n(k) A^{\db\db\db\db}_{ab,n}(\eta)]
\nn \\
& + &
	\sum_{n=1}^{6}[f^{\dc\dc\dc\db}_n(k) A^{\dc\dc\dc\db}_{ab,n}(\eta)
		+f^{\dc\dc\db\db}_n(k) A^{\dc\dc\db\db}_{ab,n}(\eta)
		+f^{\dc\db\db\db}_n(k) A^{\dc\db\db\db}_{ab,n}(\eta)]\,,
\label{eq:dP1MCfinal}
\ea
where the independent integrals $f^{\del^i\del^j\del^k\del^l}_n(k)$
are defined in \app{app:Pkints}. The full results for the
$A^{\del^i\del^j\del^k\del^l}_{ab,n}(\eta)$ coefficients are too
lengthy to reproduce here, but their general structure is the
following:
\beq
A^{\del^i\del^j\del^k\del^l}_{ab,n}(\eta) =
	\sum_l P^{\del^i\del^j\del^k\del^l}_{ab,n,l}(w_2) \re^{l\eta}\,,
\eeq	
with $l \in \{4, 3, 5/2, 2, 3/1, 1, 1/2, 0, -1/2, -1, -3/2, -2, -5/2,
-3\}$ (in this equation only), and
$P^{\del^i\del^j\del^k\del^l}_{ab,n,l}(w_2)$ an (at most) fourth order
polynomial in $w_2$ (we used $w_1=1-w_2$).


\section{Results: non-linear evolution of CDM and baryon fluid}\label{sec:results}

\subsection{Propagator}

\begin{figure}
\centering{
  \includegraphics[width=8cm,clip=]{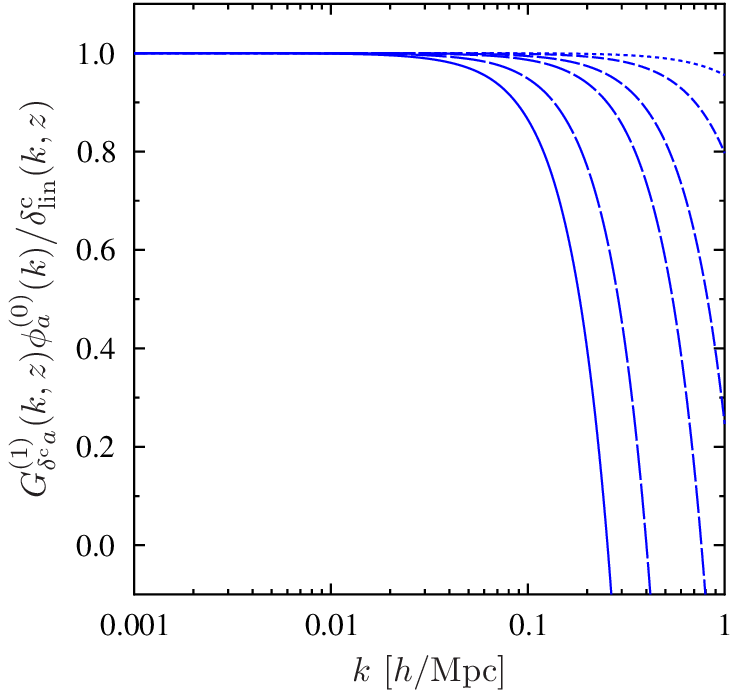}\hspace{0.4cm}
  \includegraphics[width=8cm,clip=]{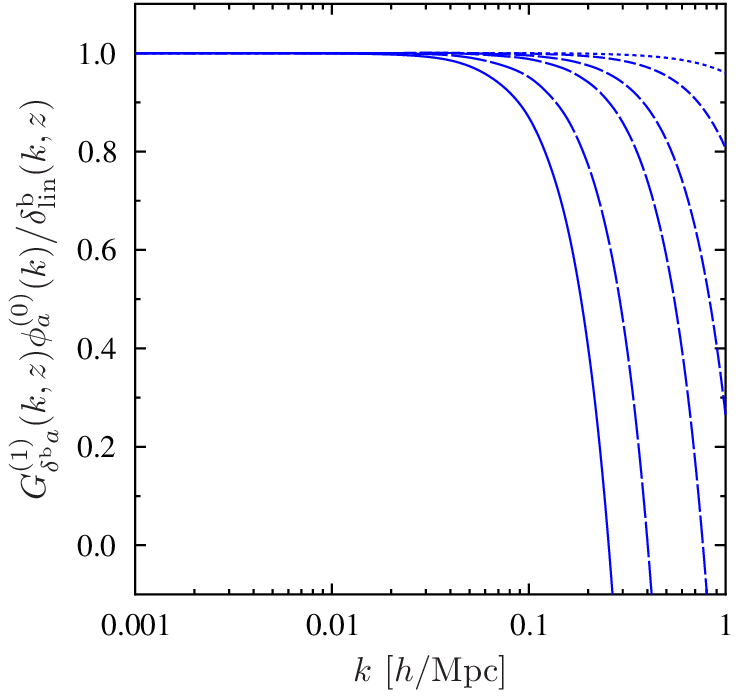}}
\caption{\small{Evolution of the initial density perturbations in the
    CDM plus baryon fluid under the action of the non-linear
    propagator up to one-loop level, as a function of spatial
    wavenumber $k$. Left and right panels present results for the CDM
    and baryon fluctuations, respectively. The linear evolution of the
    initial perturbation in each species has been scaled out, hence
    all lines approach unity on large scales. Solid through to dashed
    lines show results for redshifts
    $z=\{0,1.0,3.0,5.0,10.0,20.0\}$.}\label{fig:OneLoopProp}}
\end{figure}

In \fig{fig:OneLoopProp} we present the evolution of the initial CDM
(left panel) and baryon (right panel) density fluctuations under the
action of the non-linear propagator, up to the one-loop level in the
perturbation series. To show the effects of the non-linearity on the
dynamics, we have divided through by the linearly evolved initial
fields.  We see that, for both CDM and baryons, on very large scales
($k<0.01\kMpc$), the modes retain memory of their initial
configurations. However, on smaller scales the propagator drops away
from unity as memory is rapidly lost.  The non-linear decay scale of
the propagator can be represented as the point where
$G^{(1)}_{ia}\phi_{a}^{(0)}/g_{ib}\phi_{b}^{(0)}=\re^{-1/2}$ with
$i\in\{1,3\}$ for CDM and baryon density perturbations. We notice that
the decay scale becomes smaller as redshift increases, and that for
both fluid species is essentially the same for all epochs. We also
notice that the propagator eventually crosses zero, this is unphysical
behaviour and signifies the breakdown of the validity of the one-loop
expression.


\begin{figure}
\centering{
  \includegraphics[width=8cm,clip=]{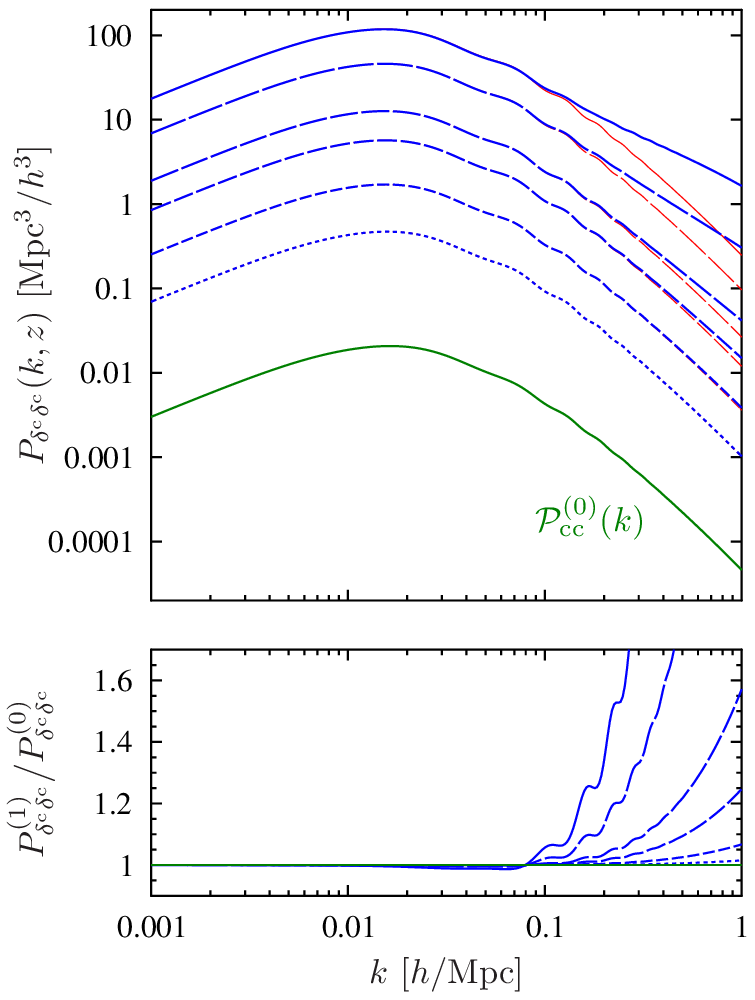}\hspace{0.4cm}
  \includegraphics[width=8cm,clip=]{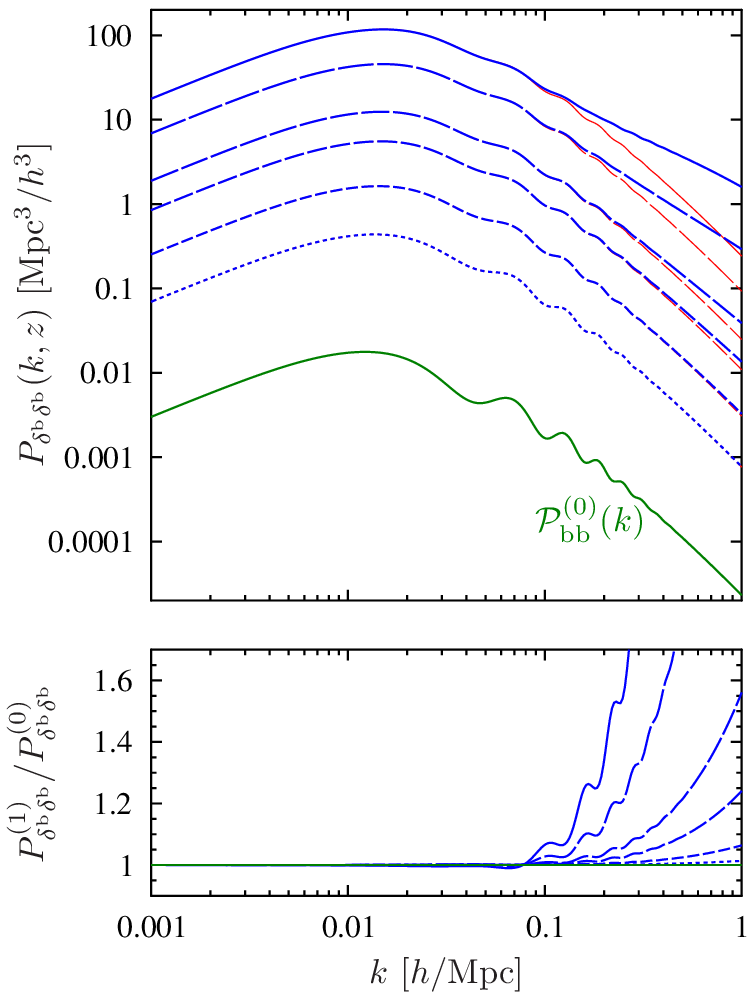}}
\caption{\small{Comparison of the linear and one-loop non-linear power
spectrum for the CDM and baryons. {\em Top panels:} the absolute power
as function of wavenumber.  Left panels, show results for CDM, and
right for the baryons. Thin (red) lines represent linear theory, and
thick (blue) ones non-linear theory.  The solid through to dashed
lines show results for redshifts
$z=\{0,1.0,3.0,5.0,10.0,20.0\}$. Lowest solid green line shows the
initial power spectrum. {\em Bottom panels:} show the ratio of the
one-loop non-linear to linear theory power spectrum.}\label{fig:MM1}}
\end{figure}
 

\subsection{Power spectra of CDM and baryons}
\label{ssec:results_PS}

\Fig{fig:MM1} presents the results for the linear and non-linear
evolution of the power spectra in the 2-component fluid model. The
left and right panels show results for the CDM and the baryons,
respectively.  As expected from our discussion in
\Sect{ssec:application} and \fig{fig:onemode}, we see that the initial
CDM power spectrum ($z_i=100$) is very smooth, whereas that for the
baryon distribution is highly oscillatory (green solid lines labeled
${\mathcal P}^{(0)}_{\rc\rc}$ and ${\mathcal P}^{(0)}_{\rb\rb}$ in the
plot). As the two fluids evolve under gravity the power grows, and in
linear theory (thin red lines) larger amplitude BAO are induced in the
CDM spectrum, whilst those in the baryon distribution are slowly
damped.

The figure also shows how these results change when the non-linear
evolution of the density fields is turned on. The non-linear power
spectra are represented as thick blue lines, and we have included all
terms up to one-loop level (c.f.~\eqn{eq:Pupshot}).  At $z=0$ and on
very large scales ($k<0.02\kMpc$), the linear and non-linear
calculations appear to agree well (solid lines). However, on smaller
scales the agreement begins to breakdown. To investigate this in more
detail, we take the ratio of the two spectra, and this is plotted in
the lower panels of \fig{fig:onemode}. From this we clearly see that
there is a small suppression of power on scales of the order
$k\sim0.07\kMpc$, this is known as the pre-virialization feature
\citep{Smithetal2007}. This is followed by a strong non-linear
amplification at smaller scales. Interestingly, the relative
non-linear boost at $z=0$ appears to possess almost identical
structure for both CDM and baryons. If we consider earlier times $z>0$
(in the plots denoted by increasingly dashed line styles), then,
besides the non-linear boosts being reduced, we note that the curves
still maintain the same relative structure. As we will show in
\Sect{sec:approx}, this can be attributed to the fact that the
structure of the leading in $\eta$, one-loop corrections, take the
same form for both CDM and baryons.


\subsection{Comparison of the exact 2-component fluid with the
approximate 1-component fluid}
\label{ssec:results_2to1}

We now explore how our exact 2-component fluid results differ from
those that would have been obtained had we used the 1-component fluid
approach, along with the approximate recipe as was described in
\Sect{sec:intro}. In order to concentrate solely on the approximations
introduced by using the 1-component fluid approach, we slightly modify
the procedure for obtaining the initial power spectrum, from that set
out in \Sect{sec:intro}:
\begin{enumerate}
\item[1'\!.] Fix the cosmological model, specifying $\Omega_{\rc}$ and
  $\Omega_{\rb}$, and hence $\fb$. Solve for the evolution of all
  perturbed species using the linearized coupled Einstein--Boltzmann
  equations. Obtain transfer functions for the CDM and baryons at some
  large redshift say $z=100$.
\item[2'\!.] Use these transfer functions to generate the linear power
  spectrum matrix in the 2-component fluid theory, at the present day: i.e.
$P_{ab}(k,z=0)\approx g_{aa'}(z=0) T_{a'}(k) g_{bb'}(z=0) T_{b'}(k) A
  k^n\ .$
  Using this power spectrum matrix, define the present day matter
  power spectrum as
\beq
P_{\bar{\delta}\bar{\delta}}(\bk,z=0) = 
	(1-\fb)^2 P_{\dc\dc}(\bk,z=0) 
	+2 (1-\fb) \fb P_{\dc\db}(\bk,z=0)
	+(\fb)^2 P_{\db\db}(\bk,z=0)\,.
\label{eq:N1initconds}
\eeq
One could further make the approximation that at the present day
$\bar{\delta} = (1-f_{\rb})\dc + f_{\rb}\db \approx \dc$, 
hence approximating the matter power spectrum as
\beq
P_{\bar{\delta}_{\mathrm{A}}\bar{\delta}_{\mathrm{A}}}(\bk,z=0) \approx 
	P_{\dc\dc}(\bk,z=0)\,.
\label{eq:N1initcondsA}
\eeq
\end{enumerate}
There rest of the steps are unchanged from those discussed in
\Sect{sec:intro}. We shall refer to the initial conditions obtained
using \eqn{eq:N1initconds} as the ``exact'', and those produced using
\eqn{eq:N1initcondsA} as the ``approximate'' 1-component fluid initial
conditions, respectively.

\begin{figure}
\centering{
  \includegraphics[width=8cm,clip=]{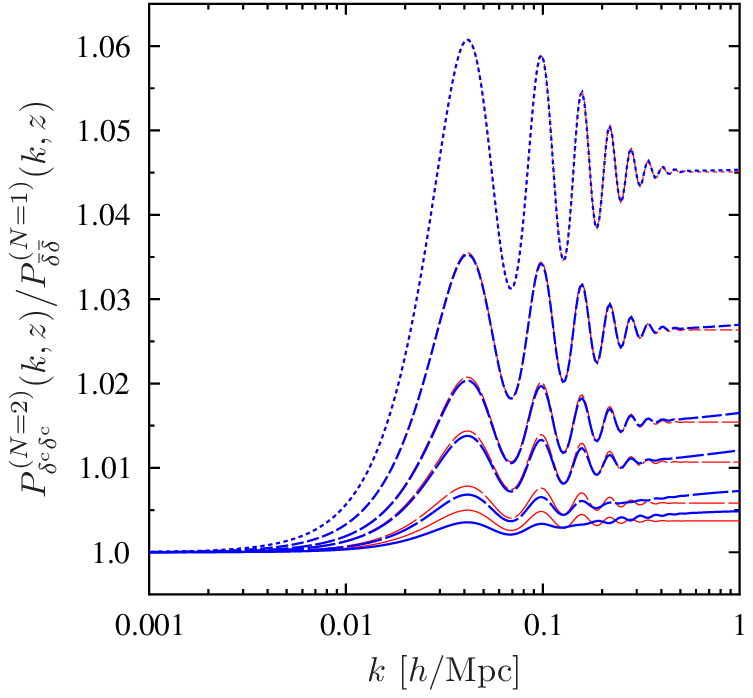}\hspace{0.4cm}
  \includegraphics[width=8cm,clip=]{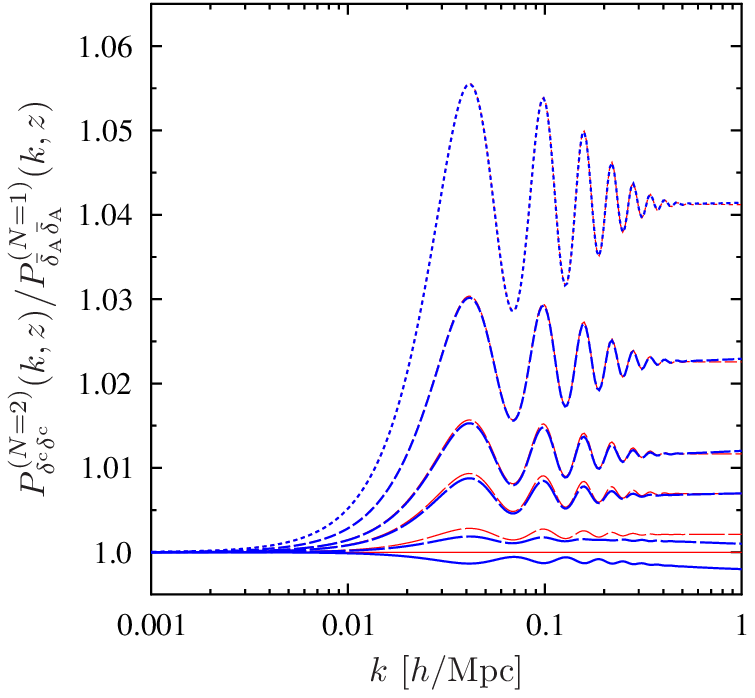}}
\caption{\small{Ratio of the CDM power spectrum obtained from the CDM
    and baryon 2-component fluid model, to the total matter spectrum
    obtained from the 1-component fluid approximation, as a function
    of wavenumber. Thin (red) lines show the ratio of power spectra in
    linear theory, while thick (blue) lines show the ratio of one-loop
    non-linear power spectra. {\em Left and right panels}: show
    results for the exact and approximate 1-component fluid initial
    conditions, respectively. Again, the solid through to dashed lines
    show results for redshifts
    $z=\{0,1.0,3.0,5.0,10.0,20.0\}$. \label{fig:MM2c}}}
\vspace{0.3cm}
\centering{
  \includegraphics[width=8cm,clip=]{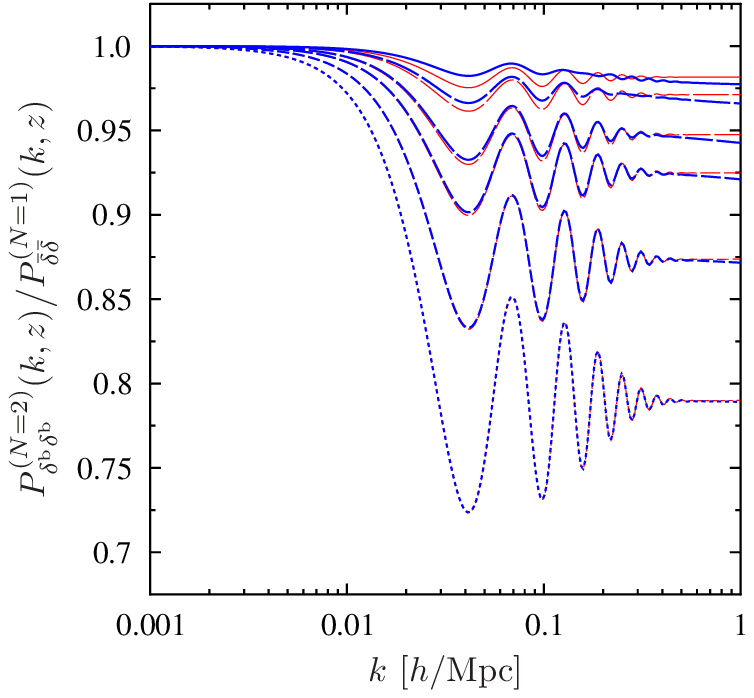}\hspace{0.4cm}
  \includegraphics[width=8cm,clip=]{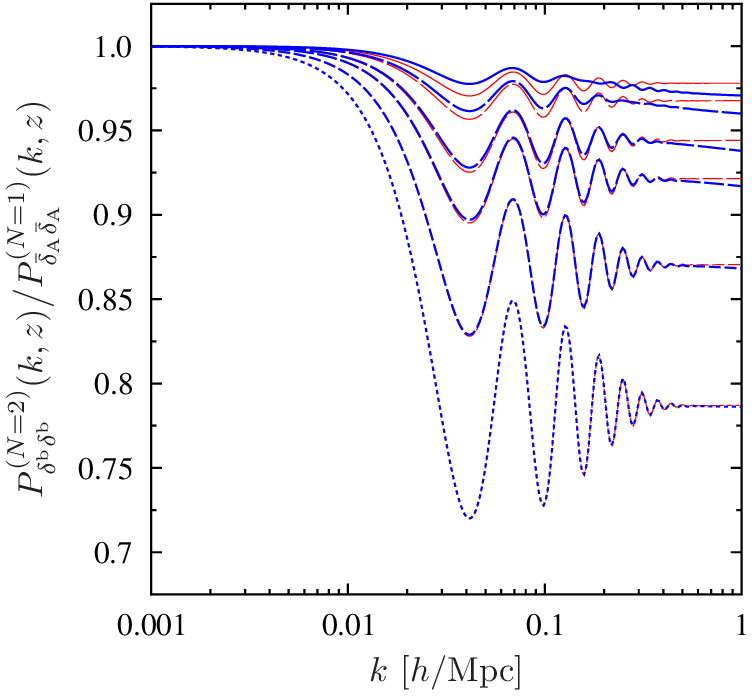}}
\caption{\small{Ratio of the baryon power spectrum obtained from the
    CDM and baryon 2-component fluid model, to the total matter
    spectrum obtained from the 1-component fluid approximation, as a
    function of wavenumber. Thin (red) lines show the ratio of power
    spectra in linear theory, while thick (blue) lines show the ratio
    of one-loop non-linear power spectra. {\em Left and right panels}:
    show results for the exact and approximate 1-component fluid
    initial conditions, respectively. Again, the solid through to
    dashed lines show results for redshifts
    $z=\{0,1.0,3.0,5.0,10.0,20.0\}$. \label{fig:MM2b}}}
\end{figure}

\Figs{fig:MM2c}{fig:MM2b}, present the ratio of the exact linear and
non-linear CDM (\fig{fig:MM2c}) and baryon (\fig{fig:MM2b}) power
spectra, with the power spectrum obtained from modeling CDM and
baryons as an effective single matter fluid. The left and right panels
show results obtained using the exact and approximate initial
conditions for the 1-component fluid calculation.  In all cases, the
non-linear evolution is followed using the the RPT framework,
including all terms up to the one-loop level. For the case of CDM, we
clearly see that at high redshift the approximation is rather poor,
there being significantly more amplitude in the exact $N=2$ model than
the effective $N=1$ model. We also note that there are strong BAO
features in the ratio, and these increase with increasing
redshift. These features are due to the fact that the $z=0$ transfer
function is used to model the initial conditions of the $N=1$ matter
fluid, and therefore can suppress power and artificially enhance the
BAO. At early stages in the evolution, the systematic errors are
between $\sim4-5\%$ at $z=20$, and between $\sim2-3\%$ at
$z=10$. However, at later times the approximation is much better, and
by $z=3$ the errors are $<1\%$, and by $z=0$ are $<0.5\%$.

On the other hand, the case for the baryons is worse. We see that the
exact $N=2$ results show less power than the approximate $N=1$
counterpart and that this suppression increases with increasing
redshift. Again we notice that there are BAO features in the residual,
and the oscillation amplitude is much more significant. At early
stages in the evolution the systematic errors are large: at $z=20$
they are $\sim25\%$, and by $z=10$ are $\sim15\%$. At later times the
approximation is still quite poor: at $z=3$ the errors are of the
order $\sim 5\%$, and by $z=0$ are still between $\sim2-3\%$.


\clearpage


\subsection{Impact on total matter power spectrum}
\label{ssec:results_2to1_total_mass}
  
Since some cosmological probes are only sensitive to the total mass
distribution, such as weak gravitational lensing, it is interesting to
ask, what is the error between the exact $N=2$ and the approximate
$N=1$ fluid dynamics, incurred when modeling the total matter power
spectrum. Recall that the total matter power spectrum is defined as,
\beq
P_{\bar{\delta}\bar{\delta}}(\bk,z)
=(1-\fb)^2P_{\delta^{\rc}\delta^{\rc}}(\bk,z)+2(1-\fb)\fb
P_{\delta^{\rc}\delta^{\rb}}(\bk,z)+(\fb)^2
P_{\delta^{\rb}\delta^{\rb}}(\bk,z) \ .\eeq

\begin{figure}
\centering{
  \includegraphics[width=8cm,clip=]{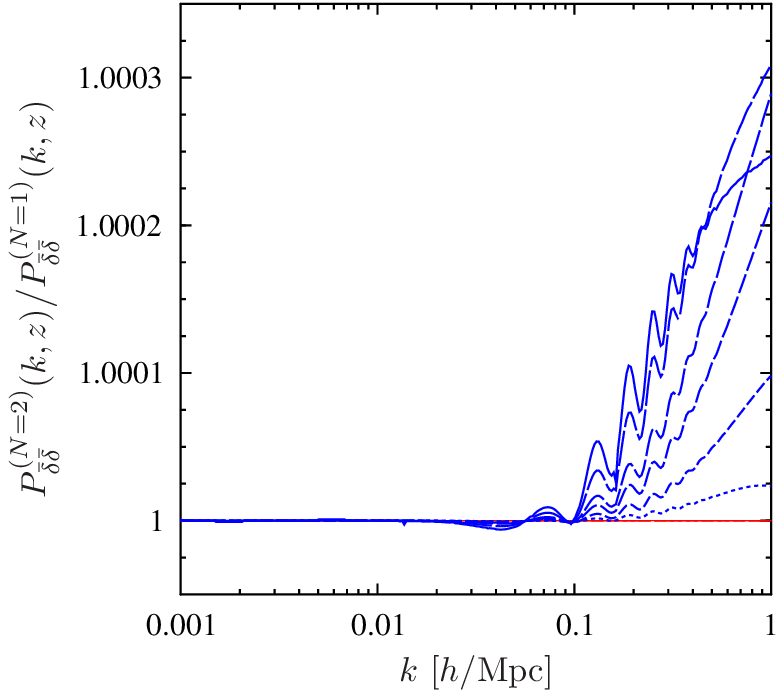}\hspace{0.4cm}
  \includegraphics[width=8cm,clip=]{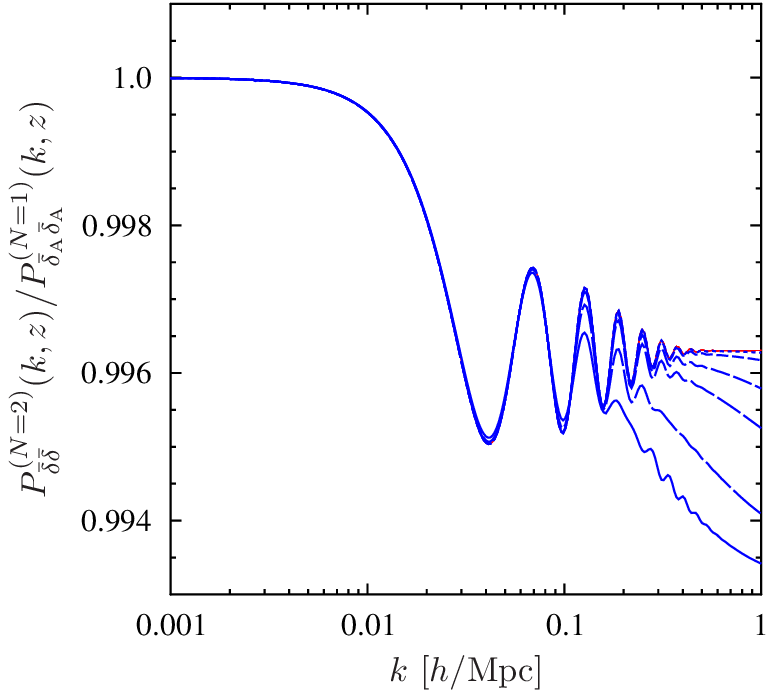}}
\caption{\small{Ratio of the total matter power spectrum obtained from
    the 2-component fluid model for CDM plus baryons, to that obtained
    from the 1-component fluid approximation, as a function of
    wavenumber. Again, thin (red) lines show the ratio of power
    spectra in linear theory, and thick (blue) lines the ratio of
    one-loop non-linear power spectra. The solid through to dashed
    lines show results for redshifts
    $z=\{0,1.0,3.0,5.0,10.0,20.0\}$. See end of
    \Sect{ssec:results_2to1_total_mass} for an explanation as to why
    the linear theory ratios are independent of time \label{fig:MM3}}}
\end{figure}

The results obtained by using the exact and approximate initial
conditions for the 1-component fluid model are plotted in the left and
right panels of \fig{fig:MM3} respectively.  We see that the
single-fluid approximation based on exact initial conditions is
excellent, being accurate to $<0.03\%$ for all times and scales
considered. As we will show in \Sect{sec:approx}, this spectacular
success can be attributed to the fact that the leading in $\eta$,
one-loop corrections, depend only on the total mass distribution.

Using the approximate initial conditions for the single-fluid
calculation still leads to very good agreement with the full
2-component fluid result on large scales $k\sim0.2\kMpc$, with
$<0.5\%$ accuracy for all the times considered. On smaller scales the
approximation becomes worse, being of order $\sim0.7\%$ at
$k\sim1.0\kMpc$. However, this departure is unlikely to be accurate,
since we can not trust the PT results on these small scales. This owes
to the fact that the {\em missing} higher order loop corrections
become increasingly more important. Nevertheless, this point is rather
academic, since using the exact initial conditions will give better
than 1\% precision for the non-linear matter power spectrum down to
scales of the order $k\sim1.0\kMpc$, as required by future weak
lensing surveys, such as Euclid.

Finally, we note that one can easily show that the {\em linear} mass
power spectrum simply scales with the square of the linear growth
factor, even in the 2-component fluid theory. Hence, the ratio of the
2-component fluid linear mass power spectrum to the 1-component fluid
model is constant in time. Furthermore, this constant is simply unity,
if the 1-component fluid model is set up with exact initial
conditions.


\section{Approximate description of the multi-fluid dynamics}
\label{sec:approx}


Given that the full expressions for the one-loop corrections to the
power spectrum are quite formidable and time consuming to compute, it
is of use to attempt to develop an approximate description of these
expressions. We present this below. Our approximation scheme is based
on insights concerning the late-time behaviour of the one-loop power
spectrum, and it is valid for growing mode initial conditions and,
although here we present the case of a 2-component fluid, can be
generalized for $N$-component matter fluids.
  

\subsection{Approximate form of the reducible correction}
\label{ssec:approxR}

To begin, we recall that the one-loop reducible correction to the
power spectrum reads (c.f.~\eqn{eq:dP1abR})
\beq \del P^{(1)}_{ab,\mathrm{R}}(\kv,\eta) =
\left[g_{ac}(\eta) \dGj{1}_{bd}(\kv,\eta) + \dGj{1}_{ac}(\kv,\eta)
g_{bd}(\eta)\right]\cP^{(0)}_{cd}(k)\,.
\label{eq:dP1Ragain}
\eeq
In the limit of late times, $\eta\to\infty$, we can discard all but
the fastest growing modes of the linear propagator, $g_{ab}$, and
one-loop correction to the propagator, $\dGj{1}_{ab}$. For the linear
propagator we have simply $g_{ab}(\eta) \to g_{ab,1}\re^\eta$, while
for the one-loop correction we find
\beq
\def\fdd{{f(k)}}
\def\gdd{{g(k)}}
\dGj{1}_{ab}(\kv,\eta) \to \frac{1}{5}
	\left[\begin{array}{cccc}
		3w_1 \fdd & 2w_1 \fdd & 3w_2 \fdd & 2w_2 \fdd \\
		3w_1 \gdd & 2w_1 \gdd & 3w_2 \gdd & 2w_2 \gdd \\
		3w_1 \fdd & 2w_1 \fdd & 3w_2 \fdd & 2w_2 \fdd \\
		3w_1 \gdd & 2w_1 \gdd & 3w_2 \gdd & 2w_2 \gdd 
	\end{array}\right] \re^{3\eta}\,,
\label{eq:dG1higheta}
\eeq
where we have 
\ba
f(k) & = &
	\int \frac{\rd^3\kv_1}{504 k^3 k_1^5}
	\left[6 k^7 k_1 -79  k^5 k_1^3 +50
   k^3 k_1^5 -21 k k_1^7 +\frac{3}{4}
   (k^2-k_1^2)^3 (2 k^2+7 k_1^2) \ln
   \frac{|k-k_1|^2}{|k+k_1|^2}\right]
   \cP^{(0)}_{\bar{\del} \bar{\del}}(k_1)\,,
\\
g(k) & = &
	\int \frac{\rd^3\kv_1}{168 k^3 k_1^5}
	\left[6 k^7 k_1 - 41  k^5 k_1^3 + 2
   k^3 k_1^5 - 3 k k_1^7 +\frac{3}{4}
   (k^2-k_1^2)^3 (2 k^2+k_1^2) \ln
   \frac{|k-k_1|^2}{|k+k_1|^2}\right]
   \cP^{(0)}_{\bar{\del} \bar{\del}}(k_1)\,\ .
\ea
These are just the $f$ and $g$ functions of
\citep{CrocceScoccimarro2006b}, and $\cP^{(0)}_{\bar{\del}
  \bar{\del}}(k)$ is the initial power spectrum of the total matter
fluctuation, $\bar{\del}(k)=w_1\del_1(k)+w_2\del_2(k)$:
\beq
\cP^{(0)}_{\bar{\del} \bar{\del}}(k) = 
	w_1^2\cP^{(0)}_{11}(k) + 2w_1 w_2\cP^{(0)}_{12}(k) + w_2^2\cP^{(0)}_{22}(k) =
	[w_1 T_1(k) + w_2 T_2(k)]^2 \cP_0(k)\,,
\eeq
where we have assumed growing mode initial conditions. Upon evaluating
\eqn{eq:dP1Ragain} with the approximate form of the propagators as
given above, we obtain
\beq
\def\fdd{{f(k)}}
\def\gdd{{g(k)}}
\del P^{(1)}_{ab,\mathrm{R}}(\kv,\eta) \to
	\left[\begin{array}{cccc}
		2\fdd & \fdd+\gdd & 2\fdd & \fdd+\gdd \\
		\fdd+\gdd & 2\gdd & \fdd+\gdd & 2\gdd \\
		2\fdd & \fdd+\gdd & 2\fdd & \fdd+\gdd \\
		\fdd+\gdd & 2\gdd & \fdd+\gdd & 2\gdd
	\end{array}\right] \cP^{(0)}_{\bar{\del}\bar{\del}}(k) \re^{4\eta}\,.
\label{eq:dP1Rhigheta}
\eeq

For early times on the other hand, as $\eta\to 0$ we have simply that
$\del P^{(1)}_{ab,\mathrm{R}}(\kv,\eta) \to 0$. In fact, explicit
computation shows that in this limit, the reducible correction
vanishes as $\del P^{(1)}_{ab,\mathrm{R}}(\kv,\eta) \to\propto
(\re^\eta-1)^2$. Since the vanishing of $\del
P^{(1)}_{ab,\mathrm{R}}(\kv,\eta)$ as $\eta\to 0$ is required on
physical grounds, we would like to implement it even in the
approximate description. To do so, we will simply change the time
dependence of \eqn{eq:dP1Rhigheta} as $\re^{4\eta}\to
\re^{2\eta}(\re^\eta-1)^2$. Thus we define the approximate one-loop
reducible correction to the power spectrum as
\beq
\def\fdd{{f(k)}}
\def\gdd{{g(k)}}
\del P^{(1)}_{ab,\mathrm{R,A}}(\kv,\eta) \equiv
	\left[\begin{array}{cccc}
		2\fdd & \fdd+\gdd & 2\fdd & \fdd+\gdd \\
		\fdd+\gdd & 2\gdd & \fdd+\gdd & 2\gdd \\
		2\fdd & \fdd+\gdd & 2\fdd & \fdd+\gdd \\
		\fdd+\gdd & 2\gdd & \fdd+\gdd & 2\gdd
	\end{array}\right] \cP^{(0)}_{\bar{\del}\bar{\del}}(k)  \re^{2\eta}(\re^\eta-1)^2\,.
\label{eq:dP1R-A}
\eeq
Needless to say, the fastest growing mode is not affected by this
substitution.


\subsection{Approximate form of the mode-coupling correction}
\label{ssec:approxMC}

Keeping only the fastest growing mode of the one-loop mode-coupling
correction to the power spectrum, we find
\beq
\def\fdddd{{F(k)}}
\def\gdddd{{G(k)}}
\def\hdddd{{H(k)}}
\del P^{(1)}_{ab,\mathrm{MC}}(\kv,\eta) \to
	\left[\begin{array}{cccc}
		2\fdddd & \hdddd & 2\fdddd & \hdddd \\
		\hdddd & 2\gdddd & \hdddd & 2\gdddd \\
		2\fdddd & \hdddd & 2\fdddd & \hdddd \\
		\hdddd & 2\gdddd & \hdddd & 2\gdddd 
	\end{array}\right] \re^{4\eta}\,,
\label{eq:dP1Ihigheta}
\eeq
where we have defined
\bal
F(k) &=
	\int \rd^3\kv_1\, 
	\frac{k^4 \left[7 k x+k_1 \left(3-10 x^2\right)\right]^2}{196 k_1^2
   \left(k^2-2 k k_1 x +k_1^2\right)^2}
	\cP^{(0)}_{\bar{\delta}\bar{\delta}}(k_1)
	\cP^{(0)}_{\bar{\delta}\bar{\delta}}\left(\sqrt{k^2-2 k k_1 x+k_1^2}\right)\,,
\label{eq:FMC}
\\
G(k) &=
	\int \rd^3\kv_1\, 
	\frac{k^4 \left[7 k x - k_1(1+6x^2)\right]^2}{196 k_1^2 \left(k^2-2 k k_1 x
   +k_1^2\right)^2}
	\cP^{(0)}_{\bar{\delta}\bar{\delta}}(k_1)
	\cP^{(0)}_{\bar{\delta}\bar{\delta}}\left(\sqrt{k^2-2 k k_1 x+k_1^2}\right)\,,
\label{eq:GMC}
\\
H(k) &=
	\int \rd^3\kv_1\, 
	\frac{k^4 \left[\left(60 x^4-8 x^2-3\right) k_1^2-14 k k_1 x \left(8
   x^2-1\right) +49 k^2 x^2\right]}{98 k_1^2 \left(k^2-2 k k_1 x
   +k_1^2\right)^2}
	\cP^{(0)}_{\bar{\delta}\bar{\delta}}(k_1)
	\cP^{(0)}_{\bar{\delta}\bar{\delta}}\left(\sqrt{k^2-2 k k_1 x+k_1^2}\right)\,,
\label{eq:HMC}
\eal
where $x$ is the cosine of the angle between $\kv$ and $\kv_1$, i.e.~$\kv\cdot\kv_1=k k_1 x$.

For early times, the situation is completely analogous to the case of
the reducible correction: as $\eta\to 0$ we have that $\del
P^{(1)}_{ab,\mathrm{MC}}(\kv,\eta) \to 0$ and in fact, $\del
P^{(1)}_{ab,\mathrm{MC}}(\kv,\eta) \to\propto (\re^\eta-1)^2$ in this
limit. So again we implement the vanishing of the mode-coupling
correction as $\eta\to 0$ by simply changing the time dependence of
\eqn{eq:dP1Ihigheta} as $\re^{4\eta}\to
\re^{2\eta}(\re^\eta-1)^2$. Then the approximate one-loop
mode-coupling correction to the power spectrum is defined as
\beq
\def\fdddd{{F(k)}}
\def\gdddd{{G(k)}}
\def\hdddd{{H(k)}}
\del P^{(1)}_{ab,\mathrm{MC,A}}(\kv,\eta) \equiv
	\left[\begin{array}{cccc}
		2\fdddd & \hdddd & 2\fdddd & \hdddd \\
		\hdddd & 2\gdddd & \hdddd & 2\gdddd \\
		2\fdddd & \hdddd & 2\fdddd & \hdddd \\
		\hdddd & 2\gdddd & \hdddd & 2\gdddd 
	\end{array}\right]\re^{2\eta}(\re^\eta-1)^2\,.
\label{eq:dP1I-A}
\eeq
Clearly the fastest growing mode is not affected by this substitution.


\begin{figure}
\centering{
  \includegraphics[width=8cm,clip=]{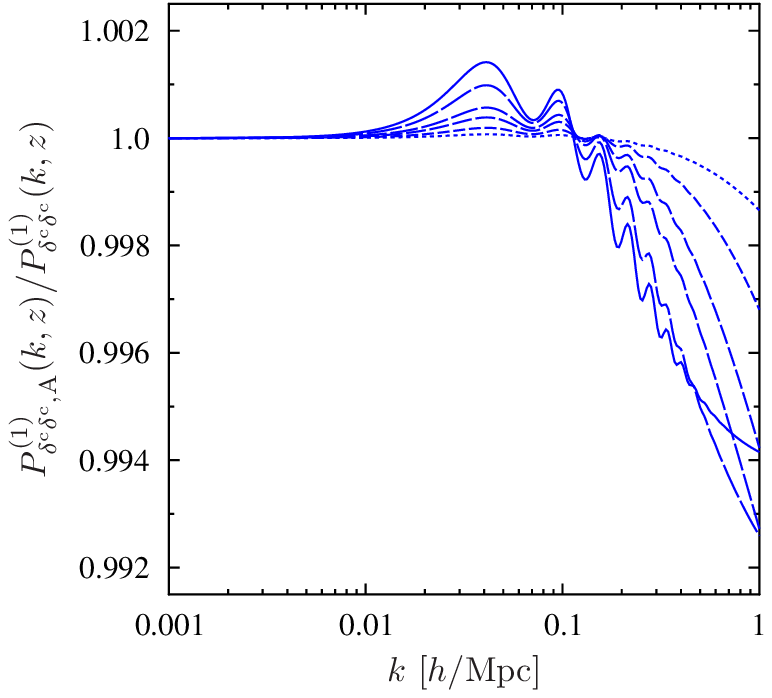}\hspace{0.4cm}
  \includegraphics[width=8cm,clip=]{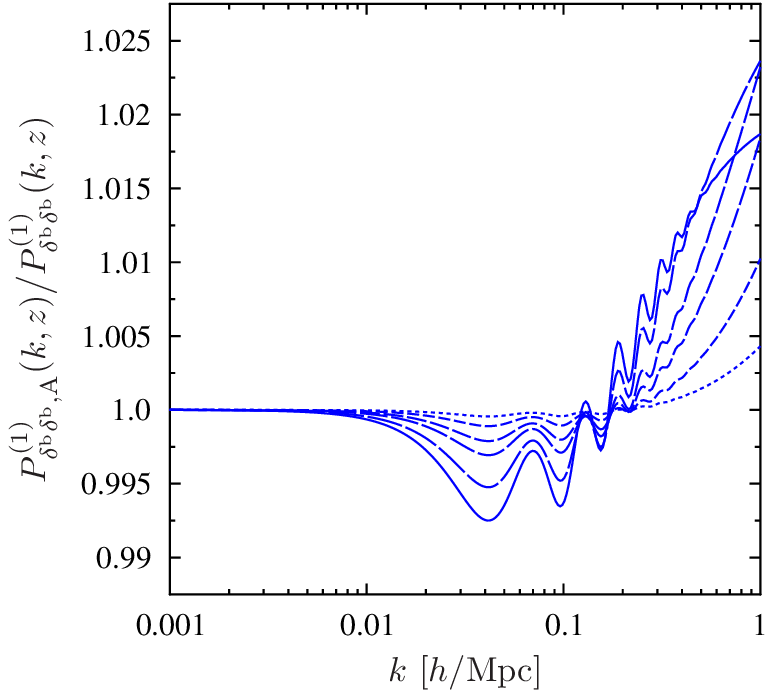}}
\caption{\small{Ratio of the approximate to exact one-loop non-linear
    matter power spectra for CDM ({\em left panel}) and baryons ({\em
      right panel}) in the 2-component fluid model for CDM plus
    baryons. As before, the solid through to dashed lines show results
    for redshifts
    $z=\{0,\,1.0,\,3.0,\,5.0,\,10.0,\,20.0\}$. Approximations are good
    to better than 1\% up to
    $k\approx0.2\kMpc$.\label{fig:ApproxToExact}}}
\end{figure}


\subsection{Comparison of the approximate and exact 2-component fluid models}

\Fig{fig:ApproxToExact} shows the ratio of the approximate non-linear
CDM (left panel) and baryon (right panel) power spectra with the exact
2-component fluid non-linear power spectra:
\beq
\frac{P^{(1)}_{ab,\mathrm{A}}(k,z)}{P^{(1)}_{ab}(k,z)} \equiv
	\frac{P^{(0)}_{ab}(k,z) + \del P^{(1)}_{ab,\mathrm{R,A}} + \del P^{(1)}_{ab,\mathrm{MC,A}}(k,z)}{P^{(0)}_{ab}(k,z) + \del P^{(1)}_{ab,\mathrm{R}} + \del P^{(1)}_{ab,\mathrm{MC}}(k,z)}\,,
\label{eq:P2approxToP2full}
\eeq
where the first term in the numerator is given by \Eqn{eq:P0mat} with
$g_{ab}$ being the full linear propagator. For the case of CDM, the
agreement is very good, being $<0.2\%$ for $k<0.2 \kMpc$ (i.e.~on
scales large enough for one-loop PT to be trusted), and for all times
considered.  The approximation performs worse for baryons, however,
the agreement is still quite good, being $<0.7\%$, on scales of up to
$k\sim 0.2 \kMpc$, again for all considered times.

We note in passing that the success of the approximation -- which was
based on the late time behaviour of the one-loop correction -- for
relatively early times (say $z=10$--$20$) is simply a consequence of
the fact that at early times the one-loop correction to the power
spectrum is small. Hence, even if the approximation of the one-loop
correction itself is not very accurate at such times, this does not
influence the fact that the full non-linear power spectrum is still
well approximated. Clearly, if all of the $\del P^{(1)}$s are small
compared to $P^{(0)}$ in \eqn{eq:P2approxToP2full}, then the ratio is
close to one, irrespective of whether the individual approximations
$\del P^{(1)} \approx \del P^{(1)}_{\mathrm{A}}$ are particularly good
or not.

We thus conclude that the one-loop corrections to the power spectrum
for both CDM and baryons are very well-modeled by the leading late
time terms, for all scales and times considered.  Incidentally, this
observation explains two phenomena we encountered earlier.  Firstly,
as noted in \Sect{ssec:results_PS}, the structure of the non-linear to
linear power spectra for the CDM and baryons (see bottom panels of
\fig{fig:MM1}) are almost identical. This is because the leading late
time corrections in \eqns{eq:dP1R-A}{eq:dP1I-A} for both CDM and
baryons are seen to take {\em precisely the same form}.  Secondly, the
spectacular success of the single-fluid approximation in modeling the
total mass power spectrum, noted in
\Sect{ssec:results_2to1_total_mass} (see \fig{fig:MM3}), is due to the
fact the leading late time terms depend only on the average initial
field, and not on the two fields separately. Thus, treating the {\em
  total} 2-component fluid as an effective 1-component fluid is {\em
  exact at the level of the leading non-linear corrections}.

Finally, this approximate theory is of great practical advantage,
since it reduces the number of numerical integrals to be evaluated by
roughly a factor of 10 for the case of the 2-component fluid, so
giving an order of magnitude speed up in computational time. Further,
the approximate theory can be represented by a compact set of
equations.


\section{Summary and Conclusions}\label{sec:conclusions}

In this paper we have generalized the matrix based RPT formalism of
\citet{CrocceScoccimarro2006a} to the problem of dealing with
$N$-component fluids, each of which evolves from a distinct set of
initial conditions and contributes to the matter-density and velocity
fields.

In \Sect{sec:EoM} we showed that, for $N$-component fluids, the
essential structure of the RPT framework remained the same as for the
$N=1$ case -- that is, the equations of motion could be solved in
exactly the same way. However, the details of the building blocks of
the theory did change: i.e.~the vertex, propagator and initial
conditions. For $N=2$ the vertex matrix offered no surprises. The
linear propagator was more interesting, and we showed that besides the
usual growing and decaying modes $\{D_{+}\propto \re^{\eta},
D^{(1)}_{-}\propto \re^{-3\eta/2}\}$, there were two additional
eigenvalues: a static and decaying mode $\{D_{\rm stat}\propto {\rm
  const}, D^{(2)}_{-}\propto \re^{-\eta/2}\}$. Interestingly, in going
to $N>2$ no further time dependence was gained. Unlike the 1-component
fluid case, it was no longer possible to choose pure growing modes,
unless all of the fluids evolved from identical initial
conditions. However, we showed that one can still set up the initial
conditions to possess pure large-scale growing modes. All our results
were obtained for this specific choice of initial conditions.

In \Sect{sec:PT} it was shown that the equations of motion for the
$N$-component fluid case could be solved using a power series solution
and that the solutions could be recovered order by order. A graphical
representation of the solutions was also described.

In \Sect{sec:stats} we discussed the statistics of the fields: in
particular the covariance matrix of the $N$-component fluid
perturbations, and the fully non-linear propagator.  This latter
quantity can be understood to be the statistic that provides
information on the memory of the evolved field to the initial
conditions. It also possesses a perturbative expansion and we gave
expressions up to one-loop order. A diagrammatic representation was
also discussed.

As for the case of the 1-component fluid RPT, we then showed that for
the $N$-component fluid RPT the full non-linear power spectrum also
possessed a series expansion, and that the series could be grouped
into two sub-series, ``reducible'' and ``mode-coupling''. It was shown
that the reducible terms were proportional to the initial power
spectrum and non-linear propagators. Again we gave complete details
for the series up to the one-loop level in the expansions. A
diagrammatic description was also discussed.

In \Sect{sec:OneLoopPower} we developed further the one-loop
expressions for the propagator and the mode-coupling contribution to
the power spectrum. These expressions were then given for the explicit
case of a 2-component fluid.

In \Sect{sec:results} we applied the formalism to the problem of
modeling the non-linear evolution of a CDM and baryon fluid in the
$\Lambda$CDM paradigm. This enabled us to test the validity of the
approximation of treating CDM and baryons as an effective 1-component
fluid evolving from a single set of initial conditions, as is
currently standard practice.

For the case of CDM, it was shown that the approximation was very good
for $z<3$, with the exact and approximate CDM power spectra differing
by $<1\%$. However, for higher redshifts the approximation became
progressively worse, it being of the order $\sim3\%$ at $z=10$, with
the power in the exact theory being amplified and having weaker BAO
features than the approximate theory.

For the case of baryons the situation was worse: at $z=20$ the exact
and approximate spectra differed by $\sim25\%$, and at $z=10$ by
$\sim15\%$. At later times the approximation was still quite poor, and
at $z=3$ the errors were of the order $\sim 5\%$, and by $z=0$ were
still between $\sim2-3\%$. The power in the exact theory was
suppressed in amplitude but possessed stronger BAO than the
corresponding power for the approximate theory.

These conclusions remained essentially unaffected, when the so-called
approximate 1-component fluid initial conditions were employed for the
effective 1-component fluid computation.

Lastly, we computed the total non-linear matter power spectrum and
found that the exact and approximate theory agreed to within
$<0.03\%$, when the so-called exact 1-component fluid initial
conditions were used. Employing the approximate 1-component fluid
initial conditions on the other hand leads to an agreement to within
$<0.5\%$ on scales where the perturbation theory was valid
$k<0.2\kMpc$ and for all times of interest. Deviations on the order of
$\sim0.7\%$ were found on smaller scales with this approximation.

\vspace{0.2cm}

Our main conclusions therefore are:

\begin{itemize}
\item For theoretical modeling of the low-redshift CDM distribution,
  the approximate 1-component fluid treatment provides a good
  approximation. For higher redshift studies, such as those that probe
  the epoch of reionization, or attempt to model the first objects
  that form (stars/haloes), then one must be careful to use the
  appropriate CDM transfer function for the redshift of
  interest. Otherwise, systematic errors will be present at the level
  of several percent.
\item For theoretical modeling of baryons in the Universe, a
  1-component fluid treatment leads to significant $>1\%$ errors at
  all times. This implies that cosmological probes that are primarily
  sensitive to the distributions of baryons in the universe, such as:
  21 cm radiation from neutral hydrogen, or the Lyman alpha forest
  absorption lines in high redshift quasars, can not be modeled using
  dark matter only simulations at high precision. Instead the baryons
  must be modeled using a 2-component fluid system of CDM and baryons,
  with the initial distributions being specified by the linear
  Einstein--Boltzmann solutions. The correct treatment of CDM and
  baryon initial conditions may also affect how galaxies form.
\item Observational probes for cosmology that are sensitive to the
  total mass distribution, such as weak gravitational lensing, can be
  accurately interpreted using an effective single CDM plus baryon
  fluid. Thus current modeling technology is good enough for high
  precision work, but it is highly desirable to use the exact
  prescription when specifying the initial conditions for the
  effective single fluid.
\end{itemize}


\section*{Acknowledgments}

We thank Ravi Sheth and Patrick McDonald for comments on the text.  We
acknowledge Martin Crocce, Zhaoming Ma, Roman Scoccimarro and Romain
Teyssier for useful discussions. GS acknowledges support from the
Hungarian Scientific Research Fund grant OKTA K-60432 and the Swiss
National Science Foundation under contract 200020-117602.  RES
acknowledges support from a Marie Curie Reintegration Grant and
partial support from the Swiss National Foundation under contract
200021-116696/1 and WCU grant R32-2008-000-10130-0.




\begin{thebibliography}{48}
\expandafter\ifx\csname natexlab\endcsname\relax\def\natexlab#1{#1}\fi
\expandafter\ifx\csname bibnamefont\endcsname\relax
  \def\bibnamefont#1{#1}\fi
\expandafter\ifx\csname bibfnamefont\endcsname\relax
  \def\bibfnamefont#1{#1}\fi
\expandafter\ifx\csname citenamefont\endcsname\relax
  \def\citenamefont#1{#1}\fi
\expandafter\ifx\csname url\endcsname\relax
  \def\url#1{\texttt{#1}}\fi
\expandafter\ifx\csname urlprefix\endcsname\relax\def\urlprefix{URL }\fi
\providecommand{\bibinfo}[2]{#2}
\providecommand{\eprint}[2][]{\url{#2}}

\bibitem[{\citenamefont{{Crocce} and
  {Scoccimarro}}(2006{\natexlab{a}})}]{CrocceScoccimarro2006a}
\bibinfo{author}{\bibfnamefont{M.}~\bibnamefont{{Crocce}}} \bibnamefont{and}
  \bibinfo{author}{\bibfnamefont{R.}~\bibnamefont{{Scoccimarro}}},
  \bibinfo{journal}{\prd} \textbf{\bibinfo{volume}{73}},
  \bibinfo{pages}{063519} (\bibinfo{year}{2006}{\natexlab{a}}),
  \eprint{arXiv:astro-ph/0509418}.

\bibitem[{\citenamefont{{Peebles} and {Ratra}}(2003)}]{PeeblesRatra2003}
\bibinfo{author}{\bibfnamefont{P.~J.} \bibnamefont{{Peebles}}}
  \bibnamefont{and} \bibinfo{author}{\bibfnamefont{B.}~\bibnamefont{{Ratra}}},
  \bibinfo{journal}{Reviews of Modern Physics} \textbf{\bibinfo{volume}{75}},
  \bibinfo{pages}{559} (\bibinfo{year}{2003}), \eprint{arXiv:astro-ph/0207347}.

\bibitem[{\citenamefont{{Spergel} et~al.}(2007)\citenamefont{{Spergel}, {Bean},
  {Dor{\'e}}, {Nolta}, {Bennett}, {Dunkley}, {Hinshaw}, {Jarosik}, {Komatsu},
  {Page} et~al.}}]{Spergeletal2007}
\bibinfo{author}{\bibfnamefont{D.~N.} \bibnamefont{{Spergel}}},
  \bibinfo{author}{\bibfnamefont{R.}~\bibnamefont{{Bean}}},
  \bibinfo{author}{\bibfnamefont{O.}~\bibnamefont{{Dor{\'e}}}},
  \bibinfo{author}{\bibfnamefont{M.~R.} \bibnamefont{{Nolta}}},
  \bibinfo{author}{\bibfnamefont{C.~L.} \bibnamefont{{Bennett}}},
  \bibinfo{author}{\bibfnamefont{J.}~\bibnamefont{{Dunkley}}},
  \bibinfo{author}{\bibfnamefont{G.}~\bibnamefont{{Hinshaw}}},
  \bibinfo{author}{\bibfnamefont{N.}~\bibnamefont{{Jarosik}}},
  \bibinfo{author}{\bibfnamefont{E.}~\bibnamefont{{Komatsu}}},
  \bibinfo{author}{\bibfnamefont{L.}~\bibnamefont{{Page}}},
  \bibnamefont{et~al.}, \bibinfo{journal}{\apjs}
  \textbf{\bibinfo{volume}{170}}, \bibinfo{pages}{377} (\bibinfo{year}{2007}),
  \eprint{arXiv:astro-ph/0603449}.

\bibitem[{\citenamefont{{Komatsu} et~al.}(2009)\citenamefont{{Komatsu},
  {Dunkley}, {Nolta}, {Bennett}, {Gold}, {Hinshaw}, {Jarosik}, {Larson},
  {Limon}, {Page} et~al.}}]{Komatsuetal2009}
\bibinfo{author}{\bibfnamefont{E.}~\bibnamefont{{Komatsu}}},
  \bibinfo{author}{\bibfnamefont{J.}~\bibnamefont{{Dunkley}}},
  \bibinfo{author}{\bibfnamefont{M.~R.} \bibnamefont{{Nolta}}},
  \bibinfo{author}{\bibfnamefont{C.~L.} \bibnamefont{{Bennett}}},
  \bibinfo{author}{\bibfnamefont{B.}~\bibnamefont{{Gold}}},
  \bibinfo{author}{\bibfnamefont{G.}~\bibnamefont{{Hinshaw}}},
  \bibinfo{author}{\bibfnamefont{N.}~\bibnamefont{{Jarosik}}},
  \bibinfo{author}{\bibfnamefont{D.}~\bibnamefont{{Larson}}},
  \bibinfo{author}{\bibfnamefont{M.}~\bibnamefont{{Limon}}},
  \bibinfo{author}{\bibfnamefont{L.}~\bibnamefont{{Page}}},
  \bibnamefont{et~al.}, \bibinfo{journal}{\apjs}
  \textbf{\bibinfo{volume}{180}}, \bibinfo{pages}{330} (\bibinfo{year}{2009}),
  \eprint{0803.0547}.

\bibitem[{\citenamefont{{Ma} and {Bertschinger}}(1995)}]{MaBertschinger1995}
\bibinfo{author}{\bibfnamefont{C.-P.} \bibnamefont{{Ma}}} \bibnamefont{and}
  \bibinfo{author}{\bibfnamefont{E.}~\bibnamefont{{Bertschinger}}},
  \bibinfo{journal}{\apj} \textbf{\bibinfo{volume}{455}}, \bibinfo{pages}{7}
  (\bibinfo{year}{1995}), \eprint{arXiv:astro-ph/9506072}.

\bibitem[{\citenamefont{{Seljak} and
  {Zaldarriaga}}(1996)}]{SeljakZaldarriaga1996}
\bibinfo{author}{\bibfnamefont{U.}~\bibnamefont{{Seljak}}} \bibnamefont{and}
  \bibinfo{author}{\bibfnamefont{M.}~\bibnamefont{{Zaldarriaga}}},
  \bibinfo{journal}{\apj} \textbf{\bibinfo{volume}{469}}, \bibinfo{pages}{437}
  (\bibinfo{year}{1996}), \eprint{arXiv:astro-ph/9603033}.

\bibitem[{\citenamefont{{Bernardeau} et~al.}(2002)\citenamefont{{Bernardeau},
  {Colombi}, {Gazta{\~n}aga}, and {Scoccimarro}}}]{Bernardeauetal2002}
\bibinfo{author}{\bibfnamefont{F.}~\bibnamefont{{Bernardeau}}},
  \bibinfo{author}{\bibfnamefont{S.}~\bibnamefont{{Colombi}}},
  \bibinfo{author}{\bibfnamefont{E.}~\bibnamefont{{Gazta{\~n}aga}}},
  \bibnamefont{and}
  \bibinfo{author}{\bibfnamefont{R.}~\bibnamefont{{Scoccimarro}}},
  \bibinfo{journal}{\physrep} \textbf{\bibinfo{volume}{367}},
  \bibinfo{pages}{1} (\bibinfo{year}{2002}), \eprint{arXiv:astro-ph/0112551}.

\bibitem[{\citenamefont{{Bertschinger}}(2001)}]{Bertschinger2001}
\bibinfo{author}{\bibfnamefont{E.}~\bibnamefont{{Bertschinger}}},
  \bibinfo{journal}{\apjs} \textbf{\bibinfo{volume}{137}}, \bibinfo{pages}{1}
  (\bibinfo{year}{2001}), \eprint{arXiv:astro-ph/0103301}.

\bibitem[{\citenamefont{{Efstathiou} et~al.}(1985)\citenamefont{{Efstathiou},
  {Davis}, {White}, and {Frenk}}}]{Efstathiouetal1985}
\bibinfo{author}{\bibfnamefont{G.}~\bibnamefont{{Efstathiou}}},
  \bibinfo{author}{\bibfnamefont{M.}~\bibnamefont{{Davis}}},
  \bibinfo{author}{\bibfnamefont{S.~D.~M.} \bibnamefont{{White}}},
  \bibnamefont{and} \bibinfo{author}{\bibfnamefont{C.~S.}
  \bibnamefont{{Frenk}}}, \bibinfo{journal}{\apjs}
  \textbf{\bibinfo{volume}{57}}, \bibinfo{pages}{241} (\bibinfo{year}{1985}).

\bibitem[{\citenamefont{{Thomas} and {Couchman}}(1992)}]{ThomasCouchman1992}
\bibinfo{author}{\bibfnamefont{P.~A.} \bibnamefont{{Thomas}}} \bibnamefont{and}
  \bibinfo{author}{\bibfnamefont{H.~M.~P.} \bibnamefont{{Couchman}}},
  \bibinfo{journal}{\mnras} \textbf{\bibinfo{volume}{257}}, \bibinfo{pages}{11}
  (\bibinfo{year}{1992}).

\bibitem[{\citenamefont{{Couchman} et~al.}(1995)\citenamefont{{Couchman},
  {Thomas}, and {Pearce}}}]{Couchmanetal1995}
\bibinfo{author}{\bibfnamefont{H.~M.~P.} \bibnamefont{{Couchman}}},
  \bibinfo{author}{\bibfnamefont{P.~A.} \bibnamefont{{Thomas}}},
  \bibnamefont{and} \bibinfo{author}{\bibfnamefont{F.~R.}
  \bibnamefont{{Pearce}}}, \bibinfo{journal}{\apj}
  \textbf{\bibinfo{volume}{452}}, \bibinfo{pages}{797} (\bibinfo{year}{1995}),
  \eprint{arXiv:astro-ph/9409058}.

\bibitem[{\citenamefont{{Sugiyama}}(1995)}]{Sugiyama1995}
\bibinfo{author}{\bibfnamefont{N.}~\bibnamefont{{Sugiyama}}},
  \bibinfo{journal}{\apjs} \textbf{\bibinfo{volume}{100}}, \bibinfo{pages}{281}
  (\bibinfo{year}{1995}), \eprint{arXiv:astro-ph/9412025}.

\bibitem[{\citenamefont{{Eisenstein} and {Hu}}(1998)}]{EisensteinHu1998}
\bibinfo{author}{\bibfnamefont{D.~J.} \bibnamefont{{Eisenstein}}}
  \bibnamefont{and} \bibinfo{author}{\bibfnamefont{W.}~\bibnamefont{{Hu}}},
  \bibinfo{journal}{\apj} \textbf{\bibinfo{volume}{496}}, \bibinfo{pages}{605}
  (\bibinfo{year}{1998}), \eprint{arXiv:astro-ph/9709112}.

\bibitem[{\citenamefont{{Meiksin} et~al.}(1999)\citenamefont{{Meiksin},
  {White}, and {Peacock}}}]{MeiksinWhitePeacock1999}
\bibinfo{author}{\bibfnamefont{A.}~\bibnamefont{{Meiksin}}},
  \bibinfo{author}{\bibfnamefont{M.}~\bibnamefont{{White}}}, \bibnamefont{and}
  \bibinfo{author}{\bibfnamefont{J.~A.} \bibnamefont{{Peacock}}},
  \bibinfo{journal}{\mnras} \textbf{\bibinfo{volume}{304}},
  \bibinfo{pages}{851} (\bibinfo{year}{1999}), \eprint{arXiv:astro-ph/9812214}.

\bibitem[{\citenamefont{{Pearce} et~al.}(2001)\citenamefont{{Pearce},
  {Jenkins}, {Frenk}, {White}, {Thomas}, {Couchman}, {Peacock}, and
  {Efstathiou}}}]{Pearceetal2001}
\bibinfo{author}{\bibfnamefont{F.~R.} \bibnamefont{{Pearce}}},
  \bibinfo{author}{\bibfnamefont{A.}~\bibnamefont{{Jenkins}}},
  \bibinfo{author}{\bibfnamefont{C.~S.} \bibnamefont{{Frenk}}},
  \bibinfo{author}{\bibfnamefont{S.~D.~M.} \bibnamefont{{White}}},
  \bibinfo{author}{\bibfnamefont{P.~A.} \bibnamefont{{Thomas}}},
  \bibinfo{author}{\bibfnamefont{H.~M.~P.} \bibnamefont{{Couchman}}},
  \bibinfo{author}{\bibfnamefont{J.~A.} \bibnamefont{{Peacock}}},
  \bibnamefont{and}
  \bibinfo{author}{\bibfnamefont{G.}~\bibnamefont{{Efstathiou}}},
  \bibinfo{journal}{\mnras} \textbf{\bibinfo{volume}{326}},
  \bibinfo{pages}{649} (\bibinfo{year}{2001}), \eprint{arXiv:astro-ph/0010587}.

\bibitem[{\citenamefont{{Teyssier}}(2002)}]{Teyssier2002}
\bibinfo{author}{\bibfnamefont{R.}~\bibnamefont{{Teyssier}}},
  \bibinfo{journal}{\aap} \textbf{\bibinfo{volume}{385}}, \bibinfo{pages}{337}
  (\bibinfo{year}{2002}), \eprint{arXiv:astro-ph/0111367}.

\bibitem[{\citenamefont{{Springel}}(2005)}]{Springel2005}
\bibinfo{author}{\bibfnamefont{V.}~\bibnamefont{{Springel}}},
  \bibinfo{journal}{\mnras} \textbf{\bibinfo{volume}{364}},
  \bibinfo{pages}{1105} (\bibinfo{year}{2005}),
  \eprint{arXiv:astro-ph/0505010}.

\bibitem[{\citenamefont{{Springel}}(2009)}]{Springel2009}
\bibinfo{author}{\bibfnamefont{V.}~\bibnamefont{{Springel}}},
  \bibinfo{journal}{ArXiv e-prints}  (\bibinfo{year}{2009}),
  \eprint{0901.4107}.

\bibitem[{\citenamefont{{Albrecht} et~al.}(2006)\citenamefont{{Albrecht},
  {Bernstein}, {Cahn}, {Freedman}, {Hewitt}, {Hu}, {Huth}, {Kamionkowski},
  {Kolb}, {Knox} et~al.}}]{DETF2006}
\bibinfo{author}{\bibfnamefont{A.}~\bibnamefont{{Albrecht}}},
  \bibinfo{author}{\bibfnamefont{G.}~\bibnamefont{{Bernstein}}},
  \bibinfo{author}{\bibfnamefont{R.}~\bibnamefont{{Cahn}}},
  \bibinfo{author}{\bibfnamefont{W.~L.} \bibnamefont{{Freedman}}},
  \bibinfo{author}{\bibfnamefont{J.}~\bibnamefont{{Hewitt}}},
  \bibinfo{author}{\bibfnamefont{W.}~\bibnamefont{{Hu}}},
  \bibinfo{author}{\bibfnamefont{J.}~\bibnamefont{{Huth}}},
  \bibinfo{author}{\bibfnamefont{M.}~\bibnamefont{{Kamionkowski}}},
  \bibinfo{author}{\bibfnamefont{E.~W.} \bibnamefont{{Kolb}}},
  \bibinfo{author}{\bibfnamefont{L.}~\bibnamefont{{Knox}}},
  \bibnamefont{et~al.}, \bibinfo{journal}{ArXiv Astrophysics e-prints}
  (\bibinfo{year}{2006}), \eprint{arXiv:astro-ph/0609591}.

\bibitem[{\citenamefont{{Peacock} et~al.}(2006)\citenamefont{{Peacock},
  {Schneider}, {Efstathiou}, {Ellis}, {Leibundgut}, {Lilly}, and
  {Mellier}}}]{ESO2006}
\bibinfo{author}{\bibfnamefont{J.~A.} \bibnamefont{{Peacock}}},
  \bibinfo{author}{\bibfnamefont{P.}~\bibnamefont{{Schneider}}},
  \bibinfo{author}{\bibfnamefont{G.}~\bibnamefont{{Efstathiou}}},
  \bibinfo{author}{\bibfnamefont{J.~R.} \bibnamefont{{Ellis}}},
  \bibinfo{author}{\bibfnamefont{B.}~\bibnamefont{{Leibundgut}}},
  \bibinfo{author}{\bibfnamefont{S.~J.} \bibnamefont{{Lilly}}},
  \bibnamefont{and}
  \bibinfo{author}{\bibfnamefont{Y.}~\bibnamefont{{Mellier}}},
  \bibinfo{type}{Tech. Rep.} (\bibinfo{year}{2006}).

\bibitem[{\citenamefont{{Zhang} et~al.}(2004)\citenamefont{{Zhang}, {Pen}, and
  {Trac}}}]{Zhangetal2004}
\bibinfo{author}{\bibfnamefont{P.}~\bibnamefont{{Zhang}}},
  \bibinfo{author}{\bibfnamefont{U.-L.} \bibnamefont{{Pen}}}, \bibnamefont{and}
  \bibinfo{author}{\bibfnamefont{H.}~\bibnamefont{{Trac}}},
  \bibinfo{journal}{\mnras} \textbf{\bibinfo{volume}{355}},
  \bibinfo{pages}{451} (\bibinfo{year}{2004}), \eprint{arXiv:astro-ph/0402115}.

\bibitem[{\citenamefont{{Iliev} et~al.}(2006)\citenamefont{{Iliev}, {Mellema},
  {Pen}, {Merz}, {Shapiro}, and {Alvarez}}}]{Ilievetal2006}
\bibinfo{author}{\bibfnamefont{I.~T.} \bibnamefont{{Iliev}}},
  \bibinfo{author}{\bibfnamefont{G.}~\bibnamefont{{Mellema}}},
  \bibinfo{author}{\bibfnamefont{U.-L.} \bibnamefont{{Pen}}},
  \bibinfo{author}{\bibfnamefont{H.}~\bibnamefont{{Merz}}},
  \bibinfo{author}{\bibfnamefont{P.~R.} \bibnamefont{{Shapiro}}},
  \bibnamefont{and} \bibinfo{author}{\bibfnamefont{M.~A.}
  \bibnamefont{{Alvarez}}}, \bibinfo{journal}{\mnras}
  \textbf{\bibinfo{volume}{369}}, \bibinfo{pages}{1625} (\bibinfo{year}{2006}),
  \eprint{arXiv:astro-ph/0512187}.

\bibitem[{\citenamefont{{Furlanetto} et~al.}(2006)\citenamefont{{Furlanetto},
  {Oh}, and {Briggs}}}]{Furlanettoetal2006}
\bibinfo{author}{\bibfnamefont{S.~R.} \bibnamefont{{Furlanetto}}},
  \bibinfo{author}{\bibfnamefont{S.~P.} \bibnamefont{{Oh}}}, \bibnamefont{and}
  \bibinfo{author}{\bibfnamefont{F.~H.} \bibnamefont{{Briggs}}},
  \bibinfo{journal}{\physrep} \textbf{\bibinfo{volume}{433}},
  \bibinfo{pages}{181} (\bibinfo{year}{2006}), \eprint{arXiv:astro-ph/0608032}.

\bibitem[{\citenamefont{{Pillepich} et~al.}(2007)\citenamefont{{Pillepich},
  {Porciani}, and {Matarrese}}}]{PillepichPorciani2007}
\bibinfo{author}{\bibfnamefont{A.}~\bibnamefont{{Pillepich}}},
  \bibinfo{author}{\bibfnamefont{C.}~\bibnamefont{{Porciani}}},
  \bibnamefont{and}
  \bibinfo{author}{\bibfnamefont{S.}~\bibnamefont{{Matarrese}}},
  \bibinfo{journal}{\apj} \textbf{\bibinfo{volume}{662}}, \bibinfo{pages}{1}
  (\bibinfo{year}{2007}), \eprint{arXiv:astro-ph/0611126}.

\bibitem[{\citenamefont{{Croft} et~al.}(1998)\citenamefont{{Croft}, {Weinberg},
  {Katz}, and {Hernquist}}}]{Croftetal1998}
\bibinfo{author}{\bibfnamefont{R.~A.~C.} \bibnamefont{{Croft}}},
  \bibinfo{author}{\bibfnamefont{D.~H.} \bibnamefont{{Weinberg}}},
  \bibinfo{author}{\bibfnamefont{N.}~\bibnamefont{{Katz}}}, \bibnamefont{and}
  \bibinfo{author}{\bibfnamefont{L.}~\bibnamefont{{Hernquist}}},
  \bibinfo{journal}{\apj} \textbf{\bibinfo{volume}{495}}, \bibinfo{pages}{44}
  (\bibinfo{year}{1998}), \eprint{arXiv:astro-ph/9708018}.

\bibitem[{\citenamefont{{McDonald} et~al.}(2006)\citenamefont{{McDonald},
  {Seljak}, {Burles}, {Schlegel}, {Weinberg}, {Cen}, {Shih}, {Schaye},
  {Schneider}, {Bahcall} et~al.}}]{McDonaldetal2006}
\bibinfo{author}{\bibfnamefont{P.}~\bibnamefont{{McDonald}}},
  \bibinfo{author}{\bibfnamefont{U.}~\bibnamefont{{Seljak}}},
  \bibinfo{author}{\bibfnamefont{S.}~\bibnamefont{{Burles}}},
  \bibinfo{author}{\bibfnamefont{D.~J.} \bibnamefont{{Schlegel}}},
  \bibinfo{author}{\bibfnamefont{D.~H.} \bibnamefont{{Weinberg}}},
  \bibinfo{author}{\bibfnamefont{R.}~\bibnamefont{{Cen}}},
  \bibinfo{author}{\bibfnamefont{D.}~\bibnamefont{{Shih}}},
  \bibinfo{author}{\bibfnamefont{J.}~\bibnamefont{{Schaye}}},
  \bibinfo{author}{\bibfnamefont{D.~P.} \bibnamefont{{Schneider}}},
  \bibinfo{author}{\bibfnamefont{N.~A.} \bibnamefont{{Bahcall}}},
  \bibnamefont{et~al.}, \bibinfo{journal}{\apjs}
  \textbf{\bibinfo{volume}{163}}, \bibinfo{pages}{80} (\bibinfo{year}{2006}),
  \eprint{arXiv:astro-ph/0405013}.

\bibitem[{\citenamefont{{Crocce} and
  {Scoccimarro}}(2006{\natexlab{b}})}]{CrocceScoccimarro2006b}
\bibinfo{author}{\bibfnamefont{M.}~\bibnamefont{{Crocce}}} \bibnamefont{and}
  \bibinfo{author}{\bibfnamefont{R.}~\bibnamefont{{Scoccimarro}}},
  \bibinfo{journal}{\prd} \textbf{\bibinfo{volume}{73}},
  \bibinfo{pages}{063520} (\bibinfo{year}{2006}{\natexlab{b}}),
  \eprint{arXiv:astro-ph/0509419}.

\bibitem[{\citenamefont{{Crocce} and
  {Scoccimarro}}(2008)}]{CrocceScoccimarro2008}
\bibinfo{author}{\bibfnamefont{M.}~\bibnamefont{{Crocce}}} \bibnamefont{and}
  \bibinfo{author}{\bibfnamefont{R.}~\bibnamefont{{Scoccimarro}}},
  \bibinfo{journal}{\prd} \textbf{\bibinfo{volume}{77}},
  \bibinfo{pages}{023533} (\bibinfo{year}{2008}), \eprint{arXiv:0704.2783}.

\bibitem[{\citenamefont{{Nusser}}(2000)}]{Nusser2000}
\bibinfo{author}{\bibfnamefont{A.}~\bibnamefont{{Nusser}}},
  \bibinfo{journal}{\mnras} \textbf{\bibinfo{volume}{317}},
  \bibinfo{pages}{902} (\bibinfo{year}{2000}), \eprint{arXiv:astro-ph/0002317}.

\bibitem[{\citenamefont{{Matarrese} and
  {Mohayaee}}(2002)}]{MatarreseMohayaee2002}
\bibinfo{author}{\bibfnamefont{S.}~\bibnamefont{{Matarrese}}} \bibnamefont{and}
  \bibinfo{author}{\bibfnamefont{R.}~\bibnamefont{{Mohayaee}}},
  \bibinfo{journal}{\mnras} \textbf{\bibinfo{volume}{329}}, \bibinfo{pages}{37}
  (\bibinfo{year}{2002}), \eprint{arXiv:astro-ph/0102220}.

\bibitem[{\citenamefont{{Shoji} and {Komatsu}}(2009)}]{ShojiKomatsu2009}
\bibinfo{author}{\bibfnamefont{M.}~\bibnamefont{{Shoji}}} \bibnamefont{and}
  \bibinfo{author}{\bibfnamefont{E.}~\bibnamefont{{Komatsu}}},
  \bibinfo{journal}{\apj} \textbf{\bibinfo{volume}{700}}, \bibinfo{pages}{705}
  (\bibinfo{year}{2009}), \eprint{0903.2669}.

\bibitem[{\citenamefont{{Matarrese} and
  {Pietroni}}(2007)}]{MatarresePietroni2007}
\bibinfo{author}{\bibfnamefont{S.}~\bibnamefont{{Matarrese}}} \bibnamefont{and}
  \bibinfo{author}{\bibfnamefont{M.}~\bibnamefont{{Pietroni}}},
  \bibinfo{journal}{Journal of Cosmology and Astro-Particle Physics}
  \textbf{\bibinfo{volume}{6}}, \bibinfo{pages}{26} (\bibinfo{year}{2007}),
  \eprint{arXiv:astro-ph/0703563}.

\bibitem[{\citenamefont{{Matarrese} and
  {Pietroni}}(2008)}]{MatarresePietroni2008}
\bibinfo{author}{\bibfnamefont{S.}~\bibnamefont{{Matarrese}}} \bibnamefont{and}
  \bibinfo{author}{\bibfnamefont{M.}~\bibnamefont{{Pietroni}}},
  \bibinfo{journal}{Modern Physics Letters A} \textbf{\bibinfo{volume}{23}},
  \bibinfo{pages}{25} (\bibinfo{year}{2008}), \eprint{arXiv:astro-ph/0702653}.

\bibitem[{\citenamefont{{Bernardeau} et~al.}(2008)\citenamefont{{Bernardeau},
  {Crocce}, and {Scoccimarro}}}]{Bernardeauetal2008}
\bibinfo{author}{\bibfnamefont{F.}~\bibnamefont{{Bernardeau}}},
  \bibinfo{author}{\bibfnamefont{M.}~\bibnamefont{{Crocce}}}, \bibnamefont{and}
  \bibinfo{author}{\bibfnamefont{R.}~\bibnamefont{{Scoccimarro}}},
  \bibinfo{journal}{\prd} \textbf{\bibinfo{volume}{78}},
  \bibinfo{pages}{103521} (\bibinfo{year}{2008}), \eprint{0806.2334}.

\bibitem[{\citenamefont{{Pietroni}}(2008)}]{Pietroni2008}
\bibinfo{author}{\bibfnamefont{M.}~\bibnamefont{{Pietroni}}},
  \bibinfo{journal}{Journal of Cosmology and Astro-Particle Physics}
  \textbf{\bibinfo{volume}{10}}, \bibinfo{pages}{36} (\bibinfo{year}{2008}),
  \eprint{0806.0971}.

\bibitem[{\citenamefont{{Matsubara}}(2008)}]{Matsubara2008}
\bibinfo{author}{\bibfnamefont{T.}~\bibnamefont{{Matsubara}}},
  \bibinfo{journal}{\prd} \textbf{\bibinfo{volume}{78}},
  \bibinfo{pages}{083519} (\bibinfo{year}{2008}), \eprint{0807.1733}.

\bibitem[{\citenamefont{{Wong}}(2008)}]{Wong2008}
\bibinfo{author}{\bibfnamefont{Y.~Y.~Y.} \bibnamefont{{Wong}}},
  \bibinfo{journal}{Journal of Cosmology and Astro-Particle Physics}
  \textbf{\bibinfo{volume}{10}}, \bibinfo{pages}{35} (\bibinfo{year}{2008}),
  \eprint{0809.0693}.

\bibitem[{\citenamefont{{Saito} et~al.}(2008)\citenamefont{{Saito}, {Takada},
  and {Taruya}}}]{Saitoetal2008}
\bibinfo{author}{\bibfnamefont{S.}~\bibnamefont{{Saito}}},
  \bibinfo{author}{\bibfnamefont{M.}~\bibnamefont{{Takada}}}, \bibnamefont{and}
  \bibinfo{author}{\bibfnamefont{A.}~\bibnamefont{{Taruya}}},
  \bibinfo{journal}{Physical Review Letters} \textbf{\bibinfo{volume}{100}},
  \bibinfo{pages}{191301} (\bibinfo{year}{2008}), \eprint{0801.0607}.

\bibitem[{\citenamefont{{Ichiki} et~al.}(2009)\citenamefont{{Ichiki}, {Takada},
  and {Takahashi}}}]{Ichikietal2009}
\bibinfo{author}{\bibfnamefont{K.}~\bibnamefont{{Ichiki}}},
  \bibinfo{author}{\bibfnamefont{M.}~\bibnamefont{{Takada}}}, \bibnamefont{and}
  \bibinfo{author}{\bibfnamefont{T.}~\bibnamefont{{Takahashi}}},
  \bibinfo{journal}{\prd} \textbf{\bibinfo{volume}{79}},
  \bibinfo{pages}{023520} (\bibinfo{year}{2009}), \eprint{0810.4921}.

\bibitem[{\citenamefont{{Brandbyge} et~al.}(2008)\citenamefont{{Brandbyge},
  {Hannestad}, {Haugb{\o}lle}, and {Thomsen}}}]{Brandbygeetal2008}
\bibinfo{author}{\bibfnamefont{J.}~\bibnamefont{{Brandbyge}}},
  \bibinfo{author}{\bibfnamefont{S.}~\bibnamefont{{Hannestad}}},
  \bibinfo{author}{\bibfnamefont{T.}~\bibnamefont{{Haugb{\o}lle}}},
  \bibnamefont{and}
  \bibinfo{author}{\bibfnamefont{B.}~\bibnamefont{{Thomsen}}},
  \bibinfo{journal}{Journal of Cosmology and Astro-Particle Physics}
  \textbf{\bibinfo{volume}{8}}, \bibinfo{pages}{20} (\bibinfo{year}{2008}),
  \eprint{0802.3700}.

\bibitem[{\citenamefont{{Brandbyge} and
  {Hannestad}}(2009{\natexlab{a}})}]{BrandbygeHannestad2009a}
\bibinfo{author}{\bibfnamefont{J.}~\bibnamefont{{Brandbyge}}} \bibnamefont{and}
  \bibinfo{author}{\bibfnamefont{S.}~\bibnamefont{{Hannestad}}},
  \bibinfo{journal}{Journal of Cosmology and Astro-Particle Physics}
  \textbf{\bibinfo{volume}{5}}, \bibinfo{pages}{2}
  (\bibinfo{year}{2009}{\natexlab{a}}), \eprint{0812.3149}.

\bibitem[{\citenamefont{{Brandbyge} and
  {Hannestad}}(2009{\natexlab{b}})}]{BrandbygeHannestad2009b}
\bibinfo{author}{\bibfnamefont{J.}~\bibnamefont{{Brandbyge}}} \bibnamefont{and}
  \bibinfo{author}{\bibfnamefont{S.}~\bibnamefont{{Hannestad}}},
  \bibinfo{journal}{ArXiv e-prints}  (\bibinfo{year}{2009}{\natexlab{b}}),
  \eprint{0908.1969}.

\bibitem[{\citenamefont{{Lesgourgues} et~al.}(2009)\citenamefont{{Lesgourgues},
  {Matarrese}, {Pietroni}, and {Riotto}}}]{Lesgourguesetal2009}
\bibinfo{author}{\bibfnamefont{J.}~\bibnamefont{{Lesgourgues}}},
  \bibinfo{author}{\bibfnamefont{S.}~\bibnamefont{{Matarrese}}},
  \bibinfo{author}{\bibfnamefont{M.}~\bibnamefont{{Pietroni}}},
  \bibnamefont{and} \bibinfo{author}{\bibfnamefont{A.}~\bibnamefont{{Riotto}}},
  \bibinfo{journal}{Journal of Cosmology and Astro-Particle Physics}
  \textbf{\bibinfo{volume}{6}}, \bibinfo{pages}{17} (\bibinfo{year}{2009}),
  \eprint{0901.4550}.

\bibitem[{\citenamefont{{Scoccimarro} et~al.}(1998)\citenamefont{{Scoccimarro},
  {Colombi}, {Fry}, {Frieman}, {Hivon}, and {Melott}}}]{Scoccimarroetal1998}
\bibinfo{author}{\bibfnamefont{R.}~\bibnamefont{{Scoccimarro}}},
  \bibinfo{author}{\bibfnamefont{S.}~\bibnamefont{{Colombi}}},
  \bibinfo{author}{\bibfnamefont{J.~N.} \bibnamefont{{Fry}}},
  \bibinfo{author}{\bibfnamefont{J.~A.} \bibnamefont{{Frieman}}},
  \bibinfo{author}{\bibfnamefont{E.}~\bibnamefont{{Hivon}}}, \bibnamefont{and}
  \bibinfo{author}{\bibfnamefont{A.}~\bibnamefont{{Melott}}},
  \bibinfo{journal}{\apj} \textbf{\bibinfo{volume}{496}}, \bibinfo{pages}{586}
  (\bibinfo{year}{1998}), \eprint{arXiv:astro-ph/9704075}.

\bibitem[{\citenamefont{{Peebles}}(1980)}]{Peebles1980}
\bibinfo{author}{\bibfnamefont{P.~J.~E.} \bibnamefont{{Peebles}}},
  \emph{\bibinfo{title}{{The large-scale structure of the universe}}}
  (\bibinfo{publisher}{Research supported by the National Science
  Foundation.~Princeton, N.J., Princeton University Press, 1980.~435 p.},
  \bibinfo{year}{1980}).

\bibitem[{\citenamefont{{Dodelson}}(2003)}]{Dodelson2003}
\bibinfo{author}{\bibfnamefont{S.}~\bibnamefont{{Dodelson}}},
  \emph{\bibinfo{title}{{Modern cosmology}}} (\bibinfo{publisher}{Modern
  cosmology / Scott Dodelson.~Amsterdam (Netherlands): Academic Press.~ISBN
  0-12-219141-2, 2003, XIII + 440 p.}, \bibinfo{year}{2003}).

\bibitem[{\citenamefont{{Carroll} et~al.}(1992)\citenamefont{{Carroll},
  {Press}, and {Turner}}}]{Carrolletal1992}
\bibinfo{author}{\bibfnamefont{S.~M.} \bibnamefont{{Carroll}}},
  \bibinfo{author}{\bibfnamefont{W.~H.} \bibnamefont{{Press}}},
  \bibnamefont{and} \bibinfo{author}{\bibfnamefont{E.~L.}
  \bibnamefont{{Turner}}}, \bibinfo{journal}{\araa}
  \textbf{\bibinfo{volume}{30}}, \bibinfo{pages}{499} (\bibinfo{year}{1992}).

\bibitem[{\citenamefont{{Smith} et~al.}(2007)\citenamefont{{Smith},
  {Scoccimarro}, and {Sheth}}}]{Smithetal2007}
\bibinfo{author}{\bibfnamefont{R.~E.} \bibnamefont{{Smith}}},
  \bibinfo{author}{\bibfnamefont{R.}~\bibnamefont{{Scoccimarro}}},
  \bibnamefont{and} \bibinfo{author}{\bibfnamefont{R.~K.}
  \bibnamefont{{Sheth}}}, \bibinfo{journal}{\prd}
  \textbf{\bibinfo{volume}{75}}, \bibinfo{pages}{063512}
  (\bibinfo{year}{2007}), \eprint{arXiv:astro-ph/0609547}.

\end{thebibliography}


\appendix


\section{A relation involving the conformal expansion rate}
\label{app:Hubble}

We prove that $\rd \cH/\rd\ts =
-\cH^2\left[\Omega_{\rmm}(a)/2-\Omega_{\Lambda}(a)\right]$. To begin, we
recall that for a homogeneous and isotropic universe, the Einstein
field equations reduce to:
\beq 
\frac{3\ddot{a}}{a} = -4\pi G(3p+\rho)\,; \quad\quad
\dot{a}^2+K = \frac{8\pi G\rho a^2}{3}\,,
\eeq
where dots denote differentiation with respect to cosmological time
$t$. Then for the $\Lambda$CDM model with matter density
$\rho=\rho_m+\rho_{\Lambda}$ and pressure
$p=p_{\Lambda}=-\rho_{\Lambda}$, we find
\beq \frac{\ddot{a}}{a} = -\frac{4\pi G\rho_m}{3}+\frac{8\pi
  G\rho_{\Lambda}}{3} =
-H^2(a)\left[\frac{\Omega_{\rmm}(a)}{2}-\Omega_{\Lambda}(a)\right]\,
.\eeq
Consider now,
\beq
\frac{\rd \cH}{\rd\ts} = \frac{\rd}{\rd\ts}(aH) = 
\frac{\rd t}{\rd\ts}\frac{\rd}{\rd t}(\dot{a}) = a \ddot{a}\,.
\eeq
The last two equations together give the desired result
\beq
\frac{\rd \cH}{\rd\ts} = -\cH^2\left[\frac{\Omega_{\rmm}(a)}{2}-\Omega_{\Lambda}(a)\right]\,.
\eeq
%


\section{Calculating the propagator}\label{app:propagator}


\subsection{The non-linear propagator}

In this appendix we provide explicit details of the calculation of the
propagator for the PT solutions up to third order in the expansion for
$\Psi_a^{(n)}$. To begin, recall that the propagator is defined
through \eqn{eq:Gabdef}, as
\beq G_{ab}(\kv,\eta)\Ddel(\kv-\kv') \equiv
\left<\frac{\del\Psi_a(\kv,\eta)[\phi^{(0)}]}{\del\phi_b^{(0)}(\kv')}\right>\,.
\label{eq:Gabdef2} 
\eeq
If we insert the series expansion for $\Psi^{(n)}_{a}$ as given by
\eqn{eq:PTsol}, then we obtain a series expansion for the
propagator. 
\beq G_{ab}(\kv,\eta)\Ddel(\kv-\kv') = 
\sum_{n=0}\left<\frac{\del\Psi^{(n)}_a(\kv,\eta)[\phi^{(0)}]}
{\del\phi_b^{(0)}(\kv')}\right>\,.
\label{eq:Gabdef3} 
\eeq
The next step is to calculate the functional derivatives of the
perturbative solutions.  Functionals are essentially functions that
depend on the full shape of another function over a certain domain. We
shall denote functionals using the notation, $F[y(x)]$, and this has
the specific meaning,
\beq F[y(x)]=\int_{a}^{b} \rd x\, g(y(x)) \ .\label{eq:functional}\eeq
where $g(y)$ is some arbitrary continuous function.  Functional
derivatives may be taken using the relation,
\beq 
\rd F \equiv \int_{a}^{b} \rd x \left.\frac{\delta F}{\delta
y(x)}\right|_{y^0(x)}\delta y(x) \ , \label{eq:funderiv}\eeq
where in the above we identify $\left.\delta F/\delta
y(x)\right|_{y^{0}(x)}$ as the functional derivative about the
function, $y^{0}(x)$.


\subsection{Propagator for solutions: $n=\{0,1,2\}$}

To see how we may use \eqn{eq:funderiv}, let us explicitly apply
it to the solutions $\Psi^{(n)}$, for $n=\{0,1,2\}$.

\begin{itemize}
\item {Solution for $n=0$:}
\end{itemize}
For the linear theory solution we have, $\Psi_a^{(0)}(\bk,\eta) =
g_{ab}(\eta)\phi_b^{(0)}(\bk).$
This is not in the form of \eqn{eq:functional}, but we can easily
write it in such a form using the Dirac delta function,
\beq \Psi_a^{(0)}(\bk,\eta)[\phi_b^{(0)}(\bk')] = \int \rd^3 \bk' 
\Ddel(\bk-\bk')g_{ab}(\eta)\phi_b^{(0)}(\bk') \ .\eeq
Now consider a small change $\delta \phi^{(0)}_{b}(\bk')$ to the
initial conditions $\phi^{(0)}_{b}(\bk')$, whereupon we have
\ba 
\Psi_a^{(0)}(\bk,\eta)[\phi^{(0)}_{b}(\bk')+\delta\phi^{(0)}_{b}(\bk')] & = & 
\int \rd^3 \bk' \Ddel(\bk-\bk')g_{ab}(\eta)
\left[ \phi_b^{(0)}(\bk')+\delta\phi^{(0)}_b(\bk') \right]\  \nn\\
& = & \Psi_a^{(0)}(\bk,\eta)[\phi^{(0)}_{b}(\bk')] + \int \rd^3 \bk'
\Ddel(\bk-\bk')g_{ab}(\eta)\delta\phi^{(0)}_{b}(\bk')\ .
\ea
Taking the first term on the right-hand-side over to the left, we find
\ba 
\rd\Psi_a^{(0)}(\bk,\eta)[\phi^{(0)}_{b}(\bk')] & = & \int \rd^3 \bk'
\Ddel(\bk-\bk')g_{ab}(\eta)\delta\phi^{(0)}_{b}(\bk') \ .
\ea
and on comparing the above with our definition from
\eqn{eq:funderiv}, we identify the functional derivative as,
\beq \frac{\delta\Psi_a^{(0)}(\bk,\eta)}{\delta \phi_{b}^{(0)}(\bk')}
=\Ddel(\bk-\bk')g_{ab}(\eta)\ .\eeq
Finally, we are interested in the ensemble average evolution of the
mode, and since there are no initial fields remaining there is no
stochasticity left in the relation and we simply have,
\beq \left<\frac{\delta\Psi_a^{(0)}(\bk,\eta)}{\delta \phi_{b}^{(0)}(\bk')}\right>
=\Ddel(\bk-\bk')g_{ab}(\eta)\ .\eeq


\begin{itemize}
\item {Solution for $n=1$:}
\end{itemize}
Let us now turn to the more complicated situation of computing the
functional derivatives of the first order solution. The first order
solution is a functional of the form,
\beq \Psi_a^{(1)}(\bk,\eta)[\phi^{(0)}_{c'}(\bk_1),\phi^{(0)}_{d'}(\bk_2)]
=\int_0^{\eta} \rd\eta' g_{ab}(\eta-\eta')\int \rd^3 \bk_1 \rd^3 \bk_2\, 
\gas_{bcd}(\bk,\bk_1,\bk_2) 
g_{cc'}(\eta')\phi^{(0)}_{c'}(\bk_1)
g_{dd'}(\eta')\phi^{(0)}_{d'}(\bk_2)\ .
\eeq
Repeating the procedure of above, we consider a small change $\delta
\phi^{(0)}_{a'}(\bk)$ to the initial conditions $\phi^{(0)}_{a'}(\bk)$,
whereupon we have
\ba 
\rd\Psi_a^{(1)}(\bk,\eta)[\phi^{(0)}_{c'},\phi^{(0)}_{d'}]
& = & 
\int_0^{\eta} \rd\eta' g_{ab}(\eta-\eta')\int \rd^3 \bk_1\rd^3 \bk_2
\gamma^{(s)}_{bcd}(\bk,\bk_1,\bk_2) g_{cc'}(\eta')g_{dd'}(\eta')\nn\\
& \times & 
\left[
\delta\phi^{(0)}_{c'}(\bk_1)\phi^{(0)}_{d'}(\bk_2)+
\phi^{(0)}_{c'}(\bk_1)\delta\phi^{(0)}_{d'}(\bk_2)
\right]\ \nn \\
& = & 2\int_0^{\eta} \rd\eta' g_{ab}(\eta-\eta')\int \rd^3 \bk_1\rd^3 \bk_2
\gamma^{(s)}_{bcd}(\bk,\bk_1,\bk_2) g_{cc'}(\eta')
\phi^{(0)}_{c'}(\bk_1)g_{dd'}(\eta')\delta\phi^{(0)}_{d'}(\bk_2)\ , \label{eq:funcderiv-n1}\ea
where the last equality follows from the fact that owing to the
symmetry of the vertex matrix $\gamma^{(s)}_{bcd}(\bk,\bk_1,\bk_2)$,
the solutions are symmetric to changes $\{\bk_1,\bk_2\}\rightarrow
\{\bk_2,\bk_1\}$ and $\{c,d\}\rightarrow\{d,c\}$.

We would like to take the functional derivative with respect to the
arbitrary initial condition $\delta \phi^{(0)}_{a'}(\bk')$, and so we
use the relation
\beq \delta\phi^{(0)}_{a}(\bk_i)=\int \rd^3 \bk_j \delta^D(\bk_i-\bk_j)\delta^{K}_{ab}
\delta\phi^{(0)}_{b}(\bk_j)\ , \label{eq:deltareln} \eeq
to rewrite the initial conditions in terms of the field that we
vary. Hence, \eqn{eq:funcderiv-n1}, becomes
\ba 
\rd\Psi_a^{(1)}(\bk,\eta)[\phi^{(0)}_{c'}(\bk_1),\phi^{(0)}_{d'}(\bk_2)] 
& = & \int_0^{\eta} \rd\eta' g_{ab}(\eta-\eta')\int \rd^3 \bk_1\rd^3 \bk_2
\gamma^{(s)}_{bcd}(\bk,\bk_1,\bk_2) g_{cc'}(\eta')\phi^{(0)}_{c'}(\bk_1)g_{dd'}(\eta')\nn\\
& \times & \int \rd^3 \bk' 
\delta^D(\bk_1-\bk')\delta^{K}_{d'a'}\delta\phi^{(0)}_{a'}(\bk')\ .
\ea
On comparing the above expression with \Eqn{eq:funderiv}, we identify
the functional derivative as 
\beq 
\frac{\delta \Psi_{a}^{(1)}(\bk,\eta)}{\delta\phi^{(0)}_{a'}(\bk')}
= \delta^D(\bk_1-\bk') \int_0^{\eta} \rd\eta' g_{ab}(\eta-\eta')\int \rd^3 \bk_1\rd^3 \bk_2
\gamma^{(s)}_{bcd}(\bk,\bk_1,\bk_2) g_{cc'}(\eta')\phi^{(0)}_{c'}(\bk_1)g_{da'}(\eta')
\ .\eeq
On taking the ensemble average we have
$\left<\phi^{(0)}_{c'}(\bk_1)\right>=0$, and hence
\beq 
\left<\frac{\delta \Psi_{a}^{(1)}(\bk,\eta)}{\delta\phi^{(0)}_{a'}(\bk')}\right>=0\ .
\eeq
One direct corollary of this result is that, the propagator for the
perturbative solutions which involve an even number of initial
conditions is zero. Hence, only odd powers of $\phi^{(0)}$ contribute.
%


\begin{itemize}
\item {Solution for $n=2$:}
\end{itemize}
Based on the previous calculation, we expect that the first non-zero
contribution to the non-linear propagator is $n=2$, since it involves
three initial conditions. This can be computed as follows.  The
solution is a functional of the form,
\ba
\Psi_a^{(2)}(\bk,\eta)[\phi^{(0)}_{c'}(\bk_1),\phi^{(0)}_{f'}(\bk_3),\phi^{(0)}_{g'}(\bk_4)]
& = & 
2\int_0^{\eta} \rd\eta' g_{ab}(\eta-\eta')\int \rd^3 \bk_1\rd^3 \bk_2 \gamma^{(s)}_{bcd}(\bk,\bk_1,\bk_2) 
\Psi^{(0)}_{c}(\bk_1,\eta')\Psi^{(1)}_{d}(\bk_2,\eta') \ \nn\\
& = & 
2\int_0^{\eta} \rd\eta' g_{ab}(\eta-\eta')\int \rd^3 \bk_1\rd^3 \bk_2 \gamma^{(s)}_{bcd}(\bk,\bk_1,\bk_2) 
g_{cc'}(\eta')\phi^{(0)}_{c'}(\bk_1)\nn \\
& \times &  
\int_0^{\eta'} \rd\eta'' g_{de}(\eta'-\eta'')
\int \rd^3 \bk_3\rd^3 \bk_4 \gamma^{(s)}_{efg}(\bk_2,\bk_3,\bk_4) \nn\\
& \times & 
g_{ff'}(\eta'')\phi^{(0)}_{f'}(\bk_3)
g_{gg'}(\eta'')\phi^{(0)}_{g'}(\bk_4)\ ,
\ea
where the factor of two comes from the fact that the first equation is
symmetric to interchanging, $\{c,d\}\rightarrow\{d,c\}$ and
$\{\bk_1,\bk_2\}\rightarrow\{\bk_2,\bk_1\}$.  Repeating the procedure
of above, we now consider a small change $\delta
\phi^{(0)}_{a'}(\bk')$ to the initial conditions
$\phi^{(0)}_{a'}(\bk')$, whereupon
\ba
\rd\Psi_a^{(2)}(\bk,\eta)[\phi^{(0)}_{c'}(\bk_1),\phi^{(0)}_{f'}(\bk_3),\phi^{(0)}_{g'}(\bk_4)]
& = & 
2\int_0^{\eta} \rd\eta' g_{ab}(\eta-\eta')\int \rd^3 \bk_1\rd^3 \bk_2 \gamma^{(s)}_{bcd}(\bk,\bk_1,\bk_2) 
g_{cc'}(\eta')\int_0^{\eta'} \rd\eta'' g_{de}(\eta'-\eta'')\nn \\
& \times &  
\int \rd^3 \bk_3\rd^3 \bk_4 \gamma^{(s)}_{efg}(\bk_2,\bk_3,\bk_4) 
g_{ff'}(\eta'')g_{gg'}(\eta'') 
\left[\delta\phi^{(0)}_{c'}(\bk_1)\phi^{(0)}_{f'}(\bk_3)\phi^{(0)}_{g'}(\bk_4)\right. \nn \\
& + & \left. \phi^{(0)}_{c'}(\bk_1)\delta\phi^{(0)}_{f'}(\bk_3)\phi^{(0)}_{g'}(\bk_4)+
\phi^{(0)}_{c'}(\bk_1)\phi^{(0)}_{f'}(\bk_3)\delta\phi^{(0)}_{g'}(\bk_4)
\right]\ .
\ea
We notice that the last two terms are symmetric to changes
$\{\bk_3,\bk_4\}\rightarrow \{\bk_4,\bk_3\}$ and
$\{f,g\}\rightarrow\{g,f\}$ and this gives us a factor of 2. Next we
use \eqn{eq:deltareln} to repeatedly rewrite the perturbations
$\delta\phi^{(0)}$ in terms of the the arbitrary initial field
$\delta\phi^{(0)}_{a'}(\bk')$. Hence,
\ba
\rd\Psi_a^{(2)}(\bk,\eta)
& = & 
2\int_0^{\eta} \rd\eta' g_{ab}(\eta-\eta')\int \rd^3 \bk_1\rd^3 \bk_2 \gamma^{(s)}_{bcd}(\bk,\bk_1,\bk_2) 
g_{cc'}(\eta')\int_0^{\eta'} \rd\eta'' g_{de}(\eta'-\eta'')\nn \\
& \times &  
\int \rd^3 \bk_3\rd^3 \bk_4 \gamma^{(s)}_{efg}(\bk_2,\bk_3,\bk_4) 
g_{ff'}(\eta'')g_{gg'}(\eta'') \nn \\
& \times & 
\int \rd^3 \bk' \left[\delta^{D}(\bk_1-\bk')\delta^{K}_{c'a'}\phi^{(0)}_{f'}(\bk_3)\phi^{(0)}_{g'}(\bk_4)
+ 2 \phi^{(0)}_{c'}(\bk_1)\phi^{(0)}_{f'}(\bk_3)\delta^{D}(\bk_4-\bk')\delta^{K}_{g'a'}
\right]\delta\phi^{(0)}_{a'}(\bk')\ .
\ea
On comparing the above with our definition for the functional
derivative as given by \eqn{eq:funderiv}, we may make the following
identification:
\ba 
\frac{\delta \Psi_{a}^{(2)}(\bk,\eta)}{\delta\phi^{(0)}_{a'}(\bk')}
& = & 
2\int_0^{\eta} \rd\eta' g_{ab}(\eta-\eta')\int \rd^3 \bk_1\rd^3 \bk_2 \gamma^{(s)}_{bcd}(\bk,\bk_1,\bk_2) 
g_{cc'}(\eta')\int_0^{\eta'} \rd\eta'' g_{de}(\eta'-\eta'')\nn \\
& \times &  
\int \rd^3 \bk_3\rd^3 \bk_4 \gamma^{(s)}_{efg}(\bk_2,\bk_3,\bk_4) 
g_{ff'}(\eta'')g_{gg'}(\eta'') \nn \\
& \times & 
\left[\delta^{D}(\bk_1-\bk')\delta^{K}_{c'a'}\phi^{(0)}_{f'}(\bk_3)\phi^{(0)}_{g'}(\bk_4)
+ 2 \phi^{(0)}_{c'}(\bk_1)\phi^{(0)}_{f'}(\bk_3)\delta^{D}(\bk_4-\bk')\delta^{K}_{g'a'}
\right]\ .
\ea
Finally, we take the ensemble average and find that the term in square
brackets becomes,
\ba 
\rightarrow & & 
\left[\delta^{D}(\bk_1-\bk')\delta^{K}_{c'a'}\left<\phi^{(0)}_{f'}(\bk_3)\phi^{(0)}_{g'}(\bk_4)\right>
+ 2 \left<\phi^{(0)}_{c'}(\bk_1)\phi^{(0)}_{f'}(\bk_3)\right>\delta^{D}(\bk_4-\bk')\delta^{K}_{g'a'}
\right]\ ;\nn \\
\rightarrow & & 
\left[\delta^{D}(\bk_1-\bk')\delta^{K}_{c'a'}\cP^{(0)}_{f'g'}(\bk_3)\delta^{D}(\bk_3+\bk_4)
+ 2 \cP^{(0)}_{c'f'}(\bk_1)\delta^{D}(\bk_1+\bk_3) \delta^{D}(\bk_4-\bk')\delta^{K}_{g'a'}\right]\ .
\ea
Consider the first term in this bracket, we note that it contains a
$\delta^{D}(\bk_3+\bk_4)$. If we integrate this first term over
$\bk_4$ then the resulting expression involves
$\gamma^{(s)}_{efg}(\bk_2,\bk_3,-\bk_3)$, which from momentum
conservation means that waves with equal and opposite momenta enter
the interaction and so completely cancel and no wave exits. Thus, the
above expression reduces to
\ba 
\left<\frac{\delta \Psi_{a}^{(2)}(\bk,\eta)}{\delta\phi^{(0)}_{a'}(\bk')}\right>
& = & 
4\int_0^{\eta} \rd\eta' g_{ab}(\eta-\eta')\int \rd^3 \bk_1\rd^3 \bk_2 \bar{\gamma}^{(s)}_{bcd}(\bk_1,\bk_2)
\delta^{D}(\bk-\bk_1-\bk_2) 
g_{cc'}(\eta')\int_0^{\eta'} \rd\eta'' g_{de}(\eta'-\eta'')\nn \\
&  &  \hspace{-1.5cm}\times
\int \rd^3 \bk_3\rd^3 \bk_4  \bar{\gamma}^{(s)}_{efg}(\bk_3,\bk_4) \delta^{D}(\bk_2-\bk_3-\bk_4)
g_{ff'}(\eta'')g_{ga'}(\eta'') 
\cP^{(0)}_{c'f'}(\bk_1)\delta^{D}(\bk_1+\bk_3)\delta^{D}(\bk_4-\bk')\ .
\ea
If we integrate over $\bk_4$, then $\bk_3$, and finally over $\bk_2$,
then we obtain the desired expression,
\ba 
\left<\frac{\delta \Psi_{a}^{(2)}(\bk,\eta)}{\delta\phi^{(0)}_{a'}(\bk')}\right>
& = & 
\delta^{D}(\bk-\bk') \times 4\int_0^{\eta} \rd\eta' g_{ab}(\eta-\eta')
\int \rd^3 \bk_1 \bar{\gamma}^{(s)}_{bcd}(\bk_1,\bk-\bk_1)
g_{cc'}(\eta')\int_0^{\eta'} \rd\eta'' g_{de}(\eta'-\eta'')\nn \\
& \times &  
 \bar{\gamma}^{(s)}_{efg}(-\bk_1,\bk') 
g_{ff'}(\eta'')g_{ga'}(\eta'') 
\cP^{(0)}_{c'f'}(\bk_1)\ .
\ea
Finally, if we assume large-scale growing mode initial conditions
$u_a=(1,1,\dots,1,1)$, then we find the first non-zero perturbative
correction to the propagator is given by:
\ba 
\delta G^{(1)}_{aa'}(\bk,\eta)
& = & 
4\int_0^{\eta} \rd\eta' g_{ab}(\eta-\eta')
\int \rd^3 \bk_1 \bar{\gamma}^{(s)}_{bcd}(\bk_1,\bk-\bk_1)
g_{cc'}(\eta')T_{c'}(k_1)\int_0^{\eta'} \rd\eta'' g_{de}(\eta'-\eta'')\nn \\
& \times &  
 \bar{\gamma}^{(s)}_{efg}(-\bk_1,\bk') 
g_{ff'}(\eta'')T_{f'}(k_1)g_{ga'}(\eta'') 
{\mathcal P}_0(\bk_1)\ .
\ea


\section{Calculating the mode-coupling power spectrum}

In this appendix, we present details of the computation of the
mode-coupling contribution to the one-loop power spectrum given by
\eqn{eq:dP1mc}. In \eqn{eq:P1loop} we identified this term as,
\ba
\left<\Psi^{(1)}_a(\bk)\Psi^{(1)}_{a'}(\bk')\right> 
& = &  
	\Bigg\langle \int_0^{\eta} \rd\eta_1 \,g_{ab}(\eta-\eta_1)
	\int \rd^3 \bk_1\, \rd^3 \bk_2\,  \gas_{bcd}(\kv,\kv_1,\kv_2)
	g_{ce}(\eta_1) \, g_{df}(\eta_1) \, \phi^{(0)}_{e}(\kv_1) \phi^{(0)}_{f}(\kv_2)
\nn \\ & \times &  
        \int_0^{\eta} \rd\eta_2 \,g_{a'b'}(\eta-\eta_2)
	\int \rd^3 \bk_3\, \rd^3 \bk_4\,  \gas_{b'c'd'}(\kv',\kv_3,\kv_4)
	g_{c'e'}(\eta_2) \, g_{d'f'}(\eta_2)\, \phi^{(0)}_{e'}(\kv_3) \phi^{(0)}_{f'}(\kv_4)\Bigg\rangle\ 
\nn\\ & = &
	\int \prod_{i=1}^{4}\rd^3 \bk_i
	\int_0^{\eta} \rd\eta_2 \,g_{a'b'}(\eta-\eta_2) \gas_{b'c'd'}(\kv',\kv_3,\kv_4)
	g_{c'e'}(\eta_2) \, g_{d'f'}(\eta_2) \, 
\nn\\ & \times &  	
	\int_0^{\eta} \rd\eta_1 \,g_{ab}(\eta-\eta_1) \gas_{bcd}(\kv,\kv_1,\kv_2)
	g_{ce}(\eta_1) \, g_{df}(\eta_1) \, 
	\left<\phi^{(0)}_{e}(\kv_1) \phi^{(0)}_{f}(\kv_2)
	\phi^{(0)}_{e'}(\kv_3) \phi^{(0)}_{f'}(\kv_4)\right>\ .
\ea
On assuming Gaussian initial conditions we may use Wick's theorem to
rewrite the product of initial fields,
\ba \left<\phi^{(0)}_{e}(\kv_1) \phi^{(0)}_{f}(\kv_2)
	\phi^{(0)}_{e'}(\kv_3) \phi^{(0)}_{f'}(\kv_4)\right>
& = &
\left[
{\mathcal P}_{ef}(\bk_1)\delta^D(\bk_1+\bk_2){\mathcal P}_{e'f'}(\bk_3)\delta^D(\bk_3+\bk_4) \right. 
\nn\\ & + &
{\mathcal P}_{ee'}(\bk_1)\delta^D(\bk_1+\bk_3){\mathcal P}_{ff'}(\bk_2)\delta^D(\bk_2+\bk_4)
\nn\\ & + & 
\left.
{\mathcal P}_{ef'}(\bk_1)\delta^D(\bk_1+\bk_4){\mathcal P}_{fe'}(\bk_2)\delta^D(\bk_2+\bk_3)
\right] \ .
\ea
We notice that the first term in this structure involves Dirac delta
functions with $\delta^{D}(\bk_1+\bk_2)$ and
$\delta^{D}(\bk_3+\bk_4)$, if we were to perform any of the
$k$-integrals, this would lead to a vertex matrix
$\gas_{bcd}(\kv,\kv_1,-\kv_1)=0$. Thus from momentum conservation the
first term vanishes. The next two terms are non-zero, but are
symmetric to relabelings, and so give a factor of 2. Thus we have,
\ba
\left<\Psi^{(1)}_a(\bk)\Psi^{(1)}_{a'}(\bk')\right> & = & 
        2 \int_0^{\eta} \rd\eta_1 \,g_{ab}(\eta-\eta_1) \, g_{ce}(\eta_1) \, g_{df}(\eta_1)
	  \int_0^{\eta} \rd\eta_2 \,g_{a'b'}(\eta-\eta_2) \, g_{c'e'}(\eta_2) \, g_{d'f'}(\eta_2)
\nn\\ & \times & 	
	\int \prod_{i=1}^{4}\rd^3 \bk_i \gas_{bcd}(\kv,\kv_1,\kv_2) \gas_{b'c'd'}(\kv',\kv_3,\kv_4) 
	{\mathcal P}_{ee'}(\bk_1)\delta^D(\bk_1+\bk_3){\mathcal P}_{ff'}(\bk_2)\delta^D(\bk_2+\bk_4)\ .
\ea
Considering the $k$-space integrals, if we perform the $\bk_4$ and
$\bk_3$ integrals then we arrive at,
\beq 
\rightarrow 
\int \rd^3 \bk_1 \rd^3 \bk_2 \gas_{bcd}(\kv,\kv_1,\kv_2) \gas_{b'c'd'}(\kv',-\kv_1,-\kv_2) 
	     {\mathcal P}_{ee'}(\bk_1){\mathcal P}_{ff'}(\bk_2)\ .
\eeq
If we now use $\gas_{abc}(\kv,\kv_1,\kv_2) =
\bgas_{abc}(\kv_1,\kv_2)\Ddel(\kv-\kv_1-\kv_2)$ to factor out the
Dirac delta functions from the vertex matrices, then we find
\ba
\left<\Psi^{(1)}_a(\bk)\Psi^{(1)}_{a'}(\bk')\right> & = & 
        2 \int_0^{\eta} \rd\eta_1 \,g_{ab}(\eta-\eta_1) \, g_{ce}(\eta_1) \, g_{df}(\eta_1)
	  \int_0^{\eta} \rd\eta_2 \,g_{a'b'}(\eta-\eta_2) \, g_{c'e'}(\eta_2) \, g_{d'f'}(\eta_2)
\nn\\ & & \hspace{-2.0cm}\times 	
	\int \rd^3 \bk_1 \rd^3 \bk_2 
	\bar{\gamma}^{(s)}_{bcd}(\kv_1,\kv_2) \delta^{D}(\bk-\bk_1-\bk_2)
	\bar{\gamma}^{(s)}_{b'c'd'}(-\kv_1,-\kv_2) \delta^{D}(\bk'+\bk_1+\bk_2)
	     {\mathcal P}_{ee'}(\bk_1){\mathcal P}_{ff'}(\bk_2) \ .
\ea
On computing the integral over $\bk_2$ and factoring our the delta
function $\delta^{D}(\bk+\bk')$, we find that our general expression
for the one-loop mode-coupling contribution to the power spectrum is
given by, (c.f.~\eqn{eq:dP1mc}) and see also
\citep{CrocceScoccimarro2006a})
\ba 
\delta P^{(1)}_{aa',\mathrm{MC}}(\kv,\eta) & = & 
        2\int \rd^3 \bk_1  \int_0^{\eta} \rd\eta_1  \int_0^{\eta} \rd\eta_2 
	\,g_{ab}(\eta-\eta_1)  \bar{\gamma}^{(s)}_{bcd}(\kv_1,\kv-\kv_1) \, 
	g_{ce}(\eta_1)   \, g_{df}(\eta_1)	 
\nn\\ & \times & 	       
       \,g_{a'b'}(\eta-\eta_2)\bar{\gamma}^{(s)}_{b'c'd'}(-\kv_1,\kv_1-\kv) \, 
       g_{c'e'}(\eta_2) \, g_{d'f'}(\eta_2)
       {\mathcal P}_{ee'}(\bk_1){\mathcal P}_{ff'}(\left|\kv-\kv_1\right|)\ .
\ea
Finally, on choosing large-scale growing mode initial conditions,
$u^{(1)}_a=(1,\dots,1)$, the above expression reduces to
\ba 
\delta P^{(1)}_{aa',\mathrm{MC}}(\kv,\eta) & = & 
        2\int \rd^3 \bk_1 \cP_{0}(k_1)\cP_{0}(\left|\kv-\kv_1\right|)
	\nn\\ & \times &  
	\int_0^{\eta} \rd\eta_1  
	\,g_{ab}(\eta-\eta_1)  \bar{\gamma}^{(s)}_{bcd}(\kv_1,\kv-\kv_1) \, 
	g_{ce}(\eta_1) T_{e}(k_1)  \, g_{df}(\eta_1)T_f(\left|\kv-\kv_1\right|)	 
\nn\\ & \times & 	       
        \int_0^{\eta} \rd\eta_2 \,g_{a'b'}(\eta-\eta_2)\bar{\gamma}^{(s)}_{b'c'd'}(-\kv_1,\kv_1-\kv) \, 
	g_{c'e'}(\eta_2) T_{e'}(k_1) \, g_{d'f'}(\eta_2)T_{f'}(\left|\kv-\kv_1\right|)\ .      
\ea


\section{Some momentum integrals}\label{app:momentum}


\subsection{Momentum integrals arising in computing $\dGj{1}_{ab}(\kv,\eta)$}
\label{app:Gkints}

When computing $\dGj{1}_{ab}(\kv,\eta)$, we encounter the momentum
integral
\beq
\int\rd^3\kv_1\,\bgas_{bcd}(\kv_1,\kv-\kv_1)\bgas_{efg}(-\kv_1,\kv)
	\cP^{(0)}_{c'f'}(k_1)\,.
\eeq
Since all non-zero matrix elements of $\bgas_{abc}(\kv_1,\kv_2)$ are
either $\al(\kv_1,\kv_2)/2$, $\al(\kv_2,\kv_1)/2$ or
$\be(\kv_1,\kv_2)$, we see that only nine distinct types of integrals
can arise in the computation. These are
\bal
I^{\del^i \del^j}_{1}(k) &=
	\int\rd^3\kv_1\,\al(\kv_1,\kv-\kv_1)\al(-\kv_1,\kv)\cP^{(0)}_{ij}(k_1)
=
	-\frac{1}{3}k^2\int\rd^3\kv_1\,\frac{1}{k_1^2}\cP^{(0)}_{ij}(k_1)\,,
\\
I^{\del^i \del^j}_{2}(k) &=
	\int\rd^3\kv_1\,\al(\kv_1,\kv-\kv_1)\al(\kv,-\kv_1)\cP^{(0)}_{ij}(k_1)
=
	-\frac{1}{3}\int\rd^3\kv_1\, \cP^{(0)}_{ij}(k_1)\,,
\\
I^{\del^i \del^j}_{3}(k) &=
	\int\rd^3\kv_1\,\al(\kv-\kv_1,\kv_1)\al(-\kv_1,\kv)\cP^{(0)}_{ij}(k_1)
=
	\int\rd^3\kv_1\,\left[\frac{k^2+k_1^2}{4k_1^2}
	+\frac{(k^2-k_1^2)^2}{16k k_1^3}\ln\frac{|k-k_1|^2}{|k+k_1|^2}\right]
	\cP^{(0)}_{ij}(k_1)\,,
\\
I^{\del^i \del^j}_{4}(k) &=
	\int\rd^3\kv_1\,\al(\kv-\kv_1,\kv_1)\al(\kv,-\kv_1)\cP^{(0)}_{ij}(k_1)
=
	\int\rd^3\kv_1\,\left[\frac{3k^2-k_1^2}{4k^2}
	-\frac{(k^2-k_1^2)^2}{16k^3 k_1}\ln\frac{|k-k_1|^2}{|k+k_1|^2}\right]
	\cP^{(0)}_{ij}(k_1)\,,
\\
I^{\del^i \del^j}_{5}(k) &=
	\int\rd^3\kv_1\,\be(\kv_1,\kv-\kv_1)\al(-\kv_1,\kv)\cP^{(0)}_{ij}(k_1)
=
	\int\rd^3\kv_1\,\left[\frac{k^2(k^2-3k_1^2)}{8k_1^4}
	+\frac{k(k^2-k_1^2)^2}{32k_1^5}\ln\frac{|k-k_1|^2}{|k+k_1|^2}\right]
	\cP^{(0)}_{ij}(k_1)\,,
\\
I^{\del^i \del^j}_{6}(k) &=
	\int\rd^3\kv_1\,\al(\kv_1,\kv-\kv_1)\be(-\kv_1,\kv)\cP^{(0)}_{ij}(k_1)
=
	-\frac{1}{6}\int\rd^3\kv_1\, \left[1+\frac{k^2}{k_1^2}\right]\cP^{(0)}_{ij}(k_1)\,,
\\
I^{\del^i \del^j}_{7}(k) &=
	\int\rd^3\kv_1\,\al(\kv-\kv_1,\kv_1)\be(-\kv_1,\kv)\cP^{(0)}_{ij}(k_1)
=
	\frac{1}{6}\int\rd^3\kv_1\, \cP^{(0)}_{ij}(k_1)\,,
\\
I^{\del^i \del^j}_{8}(k) &=
	\int\rd^3\kv_1\,\be(\kv_1,\kv-\kv_1)\al(\kv,-\kv_1)\cP^{(0)}_{ij}(k_1)
=
	-\int\rd^3\kv_1\,\left[\frac{k^2+k_1^2}{8k_1^2}
	+\frac{(k^2-k_1^2)^2}{32k k_1^3}\ln\frac{|k-k_1|^2}{|k+k_1|^2}\right]
	\cP^{(0)}_{ij}(k_1)\,,
\\
I^{\del^i \del^j}_{9}(k) &=
	\int\rd^3\kv_1\,\be(\kv_1,\kv-\kv_1)\be(-\kv_1,\kv)\cP^{(0)}_{ij}(k_1)
=
	-\frac{1}{12}k^2\int\rd^3\kv_1\,\frac{1}{k_1^2}\cP^{(0)}_{ij}(k_1)\,.
\eal
However, only five of the above integrals are independent (for any
given $i$ and $j$), and we have the following relations
\beq
I^{\del^i \del^j}_{6}(k) = \frac{1}{2}I^{\del^i \del^j}_{1}(k) 
	+\frac{1}{2}I^{\del^i \del^j}_{2}(k)\,,
\quad
I^{\del^i \del^j}_{7}(k) = -\frac{1}{2}I^{\del^i \del^j}_{2}(k)\,,
\quad
I^{\del^i \del^j}_{8}(k) = -\frac{1}{2}I^{\del^i \del^j}_{3}(k)\,,
\quad
I^{\del^i \del^j}_{9}(k) = \frac{1}{4}I^{\del^i \del^j}_{1}(k)\,.
\eeq
Then we may choose any non-degenerate linear combination of
e.g.~$\{I^{\del^i \del^j}_{1}(k),\dots,I^{\del^i \del^j}_{5}(k)\}$ as
our set of independent integrals. In order to facilitate comparison
with the known expressions for the case of a single component fluid
\citep{CrocceScoccimarro2006b}, we define:
\ba
f^{\del^i \del^j}_{1}(k) & = &
	\frac{19}{42} I^{\del^i \del^j}_{1}(k)
	+\frac{1}{6} I^{\del^i \del^j}_{2}(k)
	+\frac{5}{42} I^{\del^i \del^j}_{3}(k)
	+\frac{1}{6} I^{\del^i \del^j}_{4}(k)
	+\frac{2}{21} I^{\del^i \del^j}_{5}(k)
\\ & = &
	\int \frac{\rd^3\kv_1}{504 k^3 k_1^5}
	\left[6 k^7 k_1 -79  k^5 k_1^3 +50
   k^3 k_1^5 -21 k k_1^7 +\frac{3}{4}
   (k^2-k_1^2)^3 (2 k^2+7 k_1^2) \ln
   \frac{|k-k_1|^2}{|k+k_1|^2}\right]
   \cP^{(0)}_{ij}(k_1)
\nt\\
f^{\del^i \del^j}_{2}(k) & = &
	\frac{5}{14} I^{\del^i \del^j}_{1}(k)
	+\frac{1}{14} I^{\del^i \del^j}_{2}(k)
	-\frac{1}{14} I^{\del^i \del^j}_{3}(k)
	+\frac{1}{14} I^{\del^i \del^j}_{4}(k)
	+\frac{2}{7} I^{\del^i \del^j}_{5}(k)
\\ & = &
	\int \frac{\rd^3\kv_1}{168 k^3 k_1^5}
	\left[6 k^7 k_1 - 41  k^5 k_1^3 + 2
   k^3 k_1^5 - 3 k k_1^7 +\frac{3}{4}
   (k^2-k_1^2)^3 (2 k^2+k_1^2) \ln
   \frac{|k-k_1|^2}{|k+k_1|^2}\right]
   \cP^{(0)}_{ij}(k_1)
\nt\\
f^{\del^i \del^j}_{3}(k) & = &
	-\frac{1}{2} I^{\del^i \del^j}_{1}(k)
	-\frac{3}{2} I^{\del^i \del^j}_{2}(k)
	+\frac{5}{2} I^{\del^i \del^j}_{3}(k)
	-\frac{3}{2} I^{\del^i \del^j}_{4}(k)
	+2 I^{\del^i \del^j}_{5}(k)
\\ & = &
	\int \frac{\rd^3\kv_1}{24 k^3 k_1^5}
	\left[6 k^7 k_1 +  k^5 k_1^3 + 9 k k_1^7 +\frac{3}{4}
   (k^2-k_1^2)^2 (2 k^4+5 k^2 k_1^2+3 k_1^2) \ln
   \frac{|k-k_1|^2}{|k+k_1|^2}\right]
   \cP^{(0)}_{ij}(k_1)
\nt\\
f^{\del^i \del^j}_{4}(k) & = &
	\frac{5}{6} I^{\del^i \del^j}_{1}(k)
	+\frac{3}{2} I^{\del^i \del^j}_{2}(k)
	-\frac{3}{2} I^{\del^i \del^j}_{3}(k)
	+\frac{3}{2} I^{\del^i \del^j}_{4}(k)
	-\frac{2}{7} I^{\del^i \del^j}_{5}(k)
\\ & = &
	\int \frac{-\rd^3\kv_1}{72 k^3 k_1^5}
	\left[6 k^7 k_1 +  29 k^5 k_1^3 - 18 k^3 k_1^5 + 27 
	k k_1^7 +\frac{3}{4}
   (k^2-k_1^2)^2 (2 k^4+9 k^2 k_1^2+9 k_1^2) \ln
   \frac{|k-k_1|^2}{|k+k_1|^2}\right]
   \cP^{(0)}_{ij}(k_1)
\nt\\
f^{\del^i \del^j}_{5}(k) &= 
	I^{\del^i \del^j}_{2}(k) 
\\ & = &
	\int \frac{-\rd^3\kv_1}{3} 
	\cP^{(0)}_{ij}(k_1)\,,
\nt
\ea
as the set of independent integrals appearing in
\eqn{eq:dG1final}. Clearly the functions $\{f_1, f_2, f_3, f_4\}$ are
just the $\{f, g, h, i\}$ functions of \citep{CrocceScoccimarro2006b},
while $f_5$ is simply a constant. This constant is formally divergent
for initial power spectra that fall no faster than $k^{-3}$ at large
$k$. However, we regard this to be an artifact of using unsmoothed
(i.e.~infinitely ``choppy'') initial fields to arbitrarily low spatial
scales.  Clearly this is unphysical. Hence, in obtaining the numerical
results presented in this paper, we applied a Gaussian smoothing to
all initial power spectra, with smoothing scale
$R_{\mathrm{smooth}}=0.1 \ h^{-1} \ \mathrm{Mpc}$, i.e.~we set
$\cP^{(0)}_{ij}(k) \to \cP^{(0)}_{ij}(k) \re^{-k^2
  R_{\mathrm{smooth}}^2}$.

Finally, for the sake of completeness, we note that the $f(k)$ and
$g(k)$ functions of \Sect{ssec:approxR} may be expressed with the
basis integrals as
\ba 
f(k) & = & w_1^2 f_1^{\dc\dc}(k) + 2w_1 w_2 f_1^{\dc\db}(k) + w_2^2 f_1^{\db\db}(k)\,, \\ 
g(k) &=  & w_1^2 f_2^{\dc\dc}(k) + 2w_1 w_2 f_2^{\dc\db}(k) + w_2^2 f_2^{\db\db}(k)\,.
\ea


\subsection{Momentum integrals arising in computing $\del P^{(1)}_{ab,\mathrm{MC}}(\kv,\eta)$}
\label{app:Pkints}

When computing the mode-coupling piece of the one-loop power spectrum,
$\del P^{(1)}_{ab,\mathrm{MC}}(\kv,\eta)$, we encounter the momentum
integral
\beq
\int\rd^3\kv_1\,\bgas_{bcd}(\kv_1,\kv-\kv_1)\bgas_{efg}(-\kv_1,\kv_1-\kv)
	\cP^{(0)}_{ee'}(k_1)\cP^{(0)}_{ff'}(|\kv-\kv_1|)\,.
\eeq
Since $\al(-\kv_1,-\kv_2)=\al(\kv_1,\kv_2)$ and
$\be(-\kv_1,-\kv_2)=\be(\kv_1,\kv_2)$, we see that only six distinct
types of integrals can arise in the computation. These read
\ba
J^{\del^i \del^j \del^k \del^l}_1(k) & = &
	\int \rd^3 \kv_1\, \left[\al(\kv_1,\kv-\kv_1)\right]^2 \cP^{(0)}_{ij}(k_1)
		\cP^{(0)}_{kl}(|\kv-\kv_1|)\ ,
\nn\\ & = &
	\int \rd^3 \kv_1\, 
	\frac{k^2 x^2}{k_1^2} 
	\cP^{(0)}_{ij}(k_1) \cP^{(0)}_{kl}\left(\sqrt{k^2-2k k_1 x+k_1^2}\right)\,,
\label{eq:J1}
\\
J^{\del^i \del^j \del^k \del^l}_2(k) & = &
	\int \rd^3 \kv_1\, \al(\kv_1,\kv-\kv_1)\al(\kv-\kv_1,\kv_1) \cP^{(0)}_{ij}(k_1)
		\cP^{(0)}_{kl}(|\kv-\kv_1|)\ ,
\nn\\ & = &
	\int \rd^3 \kv_1\, 
	\frac{k^2 x (k-k_1 x)}{k_1 \left(k^2-2 k k_1 x+k_1^2\right)}
	\cP^{(0)}_{ij}(k_1) \cP^{(0)}_{kl}\left(\sqrt{k^2-2k k_1 x+k_1^2}\right)\,,
\\
J^{\del^i \del^j \del^k \del^l}_3(k) & = &
	\int \rd^3 \kv_1\, \al(\kv_1,\kv-\kv_1)\be(\kv_1,\kv-\kv_1) \cP^{(0)}_{ij}(k_1)
		\cP^{(0)}_{kl}(|\kv-\kv_1|)\ ,
\nn \\ & = &
	\int \rd^3 \kv_1\, 
	\frac{k^3 x (k x-k_1)}{2 k_1^2 \left(k^2-2 k  k_1 x+k_1^2\right)}
	\cP^{(0)}_{ij}(k_1) \cP^{(0)}_{kl}\left(\sqrt{k^2-2k k_1 x+k_1^2}\right)\,,
\\
J^{\del^i \del^j \del^k \del^l}_4(k) & = &
	\int \rd^3 \kv_1\, \left[\be(\kv_1,\kv-\kv_1)\right]^2 \cP^{(0)}_{ij}(k_1)
		\cP^{(0)}_{kl}(|\kv-\kv_1|)\ ,
\nn\\ & = &
	\int \rd^3 \kv_1\, 
	\frac{k^4 (k_1-k x)^2}{4 k_1^2 \left(k^2-2 k  k_1 x+k_1^2\right)^2}
	\cP^{(0)}_{ij}(k_1) \cP^{(0)}_{kl}\left(\sqrt{k^2-2k k_1 x+k_1^2}\right)\,,
\\
J^{\del^i \del^j \del^k \del^l}_5(k) & = &
	\int \rd^3 \kv_1\, \left[\al(\kv-\kv_1,\kv_1)\right]^2 \cP^{(0)}_{ij}(k_1)
		\cP^{(0)}_{kl}(|\kv-\kv_1|)\ ,
\nt\\ & = &
	\int \rd^3 \kv_1\, 
	\frac{k^2 (k-k_1 x)^2}{\left(k^2-2 k  k_1 x+k_1^2\right)^2}
	\cP^{(0)}_{ij}(k_1) \cP^{(0)}_{kl}\left(\sqrt{k^2-2k k_1 x+k_1^2}\right)\,,
\\
J^{\del^i \del^j \del^k \del^l}_6(k) & = &
	\int \rd^3 \kv_1\, \al(\kv-\kv_1,\kv_1)\be(\kv_1,\kv-\kv_1) \cP^{(0)}_{ij}(k_1)
		\cP^{(0)}_{kl}(|\kv-\kv_1|)\ ,
\nt\\ & = &
	\int \rd^3 \kv_1\, 
	\frac{k^3 \left[x k^2-k_1 \left(x^2+1\right) k+k_1^2x\right]}
	{2 k_1 \left(k^2-2 k  k_1 x+k_1^2\right)^2}
	\cP^{(0)}_{ij}(k_1) \cP^{(0)}_{kl}\left(\sqrt{k^2-2k k_1 x+k_1^2}\right)\,.
\label{eq:J6}
\ea
However, not all of these integrals are independent, since changing
the variable of integration as $\kv_1\to\kv_1'=\kv-\kv_1$ merely
exchanges $J^{\del^i \del^j \del^k \del^l}_1(k)$ with $J^{\del^k
  \del^l \del^i \del^j}_5(k)$ and $J^{\del^i \del^j \del^k
  \del^l}_3(k)$ with $J^{\del^k \del^l \del^i \del^j}_6(k)$.  Thus,
for $(ij)=(kl)$, $J_5$ and $J_6$ are not independent integrals:
\beq
J^{\del^i \del^j \del^i \del^j}_5(k) = J^{\del^i \del^j \del^i \del^j}_1(k)\,,
\quad
J^{\del^i \del^j \del^i \del^j}_6(k) = J^{\del^i \del^j \del^i \del^j}_3(k)\,.
\eeq
On the other hand, if $(ij)\ne (kl)$, then we find that we may express
each $J^{\del^k \del^l \del^i \del^j}_n(k)$ with some $J^{\del^i
  \del^j \del^k \del^l}_m(k)$:
\beq
\bsp
J^{\del^k\del^l\del^i\del^j}_1(k) = J^{\del^i\del^j\del^k\del^l}_5(k)\,,
\quad
J^{\del^k\del^l\del^i\del^j}_2(k) = J^{\del^i\del^j\del^k\del^l}_2(k)\,,
\quad
J^{\del^k\del^l\del^i\del^j}_3(k) = J^{\del^i\del^j\del^k\del^l}_6(k)
\\
J^{\del^k\del^l\del^i\del^j}_4(k) = J^{\del^i\del^j\del^k\del^l}_4(k)\,,
\quad
J^{\del^k\del^l\del^i\del^j}_5(k) = J^{\del^i\del^j\del^k\del^l}_1(k)\,.
\quad
J^{\del^k\del^l\del^i\del^j}_6(k) = J^{\del^i\del^j\del^k\del^l}_3(k)\,.
\esp
\eeq
Thus, we may define
\bal
	f^{\del^i\del^j\del^i\del^j}_n(k) &=  J^{\del^i\del^j\del^i\del^j}_n(k)\,,
	\qquad n=1,\dots,4\,,
	\quad (ij) = \{(\rc\rc),(\rc\rb),(\rb\rb)\}\,,
\\
	f^{\del^i\del^j\del^k\del^l}_n(k) &=  J^{\del^i\del^j\del^k\del^l}_n(k)\,,
	\qquad n=1,\dots,6\,,
	\quad (ij)(kl) = \{(\rc\rc)(\rc\rb),(\rc\rc)(\rb\rb),(\rc\rb)(\rb\rb)\}
\eal
as the set of independent integrals appearing in \eqn{eq:dP1MCfinal}.

Finally, for the sake of completeness, we note that the $F(k)$, $G(k)$
and $H(k)$ functions of \Sect{ssec:approxMC} may be expressed with
these basis integrals as
\bal
F(k) &= \frac{25}{196} J_1^{\bar{\delta}\bar{\delta}\bar{\delta}\bar{\delta}}(k)
	+\frac{25}{98} J_2^{\bar{\delta}\bar{\delta}\bar{\delta}\bar{\delta}}(k)
	+\frac{10}{49} J_3^{\bar{\delta}\bar{\delta}\bar{\delta}\bar{\delta}}(k)
	+\frac{4}{49} J_4^{\bar{\delta}\bar{\delta}\bar{\delta}\bar{\delta}}(k)
	+\frac{25}{196} J_5^{\bar{\delta}\bar{\delta}\bar{\delta}\bar{\delta}}(k) 
	+\frac{10}{49} J_6^{\bar{\delta}\bar{\delta}\bar{\delta}\bar{\delta}}(k)\,, 
\label{eq:FMCwithJ}
\\ 
G(k) &= \frac{9}{196} J_1^{\bar{\delta}\bar{\delta}\bar{\delta}\bar{\delta}}(k)
	+\frac{9}{98} J_2^{\bar{\delta}\bar{\delta}\bar{\delta}\bar{\delta}}(k)
	+\frac{12}{49} J_3^{\bar{\delta}\bar{\delta}\bar{\delta}\bar{\delta}}(k)
	+\frac{16}{49} J_4^{\bar{\delta}\bar{\delta}\bar{\delta}\bar{\delta}}(k)
	+\frac{9}{196} J_5^{\bar{\delta}\bar{\delta}\bar{\delta}\bar{\delta}}(k) 
	+\frac{12}{49} J_6^{\bar{\delta}\bar{\delta}\bar{\delta}\bar{\delta}}(k)\,, 
\label{eq:GMCwithJ}
\\ 
H(k) &= \frac{15}{98} J_1^{\bar{\delta}\bar{\delta}\bar{\delta}\bar{\delta}}(k)
	+\frac{15}{49} J_2^{\bar{\delta}\bar{\delta}\bar{\delta}\bar{\delta}}(k)
	+\frac{26}{49} J_3^{\bar{\delta}\bar{\delta}\bar{\delta}\bar{\delta}}(k)
	+\frac{16}{49} J_4^{\bar{\delta}\bar{\delta}\bar{\delta}\bar{\delta}}(k)
	+\frac{15}{98} J_5^{\bar{\delta}\bar{\delta}\bar{\delta}\bar{\delta}}(k) 
	+\frac{26}{49} J_6^{\bar{\delta}\bar{\delta}\bar{\delta}\bar{\delta}}(k)\,, 
\label{eq:HMCwithJ}
\eal
where
\beq
\bsp
 J_{n}^{\bar{\delta}\bar{\delta}\bar{\delta}\bar{\delta}}(k) &=
	w_1^4 J^{\dc\dc\dc\dc}_{n}(k) 
	+ 2 w_1^3 w_2 J^{\dc\dc\dc\db}_{n}(k) 
	+ w_1^2 w_2^2 J^{\dc\dc\db\db}_{n}(k) 
\\&	+ 2 w_1^3 w_2 J^{\dc\db\dc\dc}_{n}(k) 
	+ 4 w_1^2 w_2^2 J^{\dc\db\dc\db}_{n}(k) 
	+ 2 w_1 w_2^3 J^{\dc\db\db\db}_{n}(k) 
\\&	+ w_1^2 w_2^2 J^{\db\db\dc\dc}_{n}(k) 
	+ 2 w_1 w_2^3 J^{\db\db\dc\db}_{n}(k) 
	+ w_2^4 J^{\db\db\db\db}_{n}(k)\,.
\esp
\eeq
In \eqnss{eq:FMCwithJ}{eq:HMCwithJ} above, we clearly have
$J_5^{\bar{\delta}\bar{\delta}\bar{\delta}\bar{\delta}}(k) =
J_1^{\bar{\delta}\bar{\delta}\bar{\delta}\bar{\delta}}(k)$ and
$J_6^{\bar{\delta}\bar{\delta}\bar{\delta}\bar{\delta}}(k) =
J_3^{\bar{\delta}\bar{\delta}\bar{\delta}\bar{\delta}}(k)$. However,
to obtain the compact expressions for $F(k)$, $G(k)$ and $H(k)$ as in
\eqnss{eq:FMC}{eq:HMC}, one must substitute \eqnss{eq:J1}{eq:J6} as
they stand into the general formulae,
\eqnss{eq:FMCwithJ}{eq:HMCwithJ}, disregarding the above identities.


%
%

\section{The linear propagator for $N=3$}\label{app:linpropN3}

For $N=3$, the propagator takes the form as in \eqn{eq:gab}:
\beq
g_{ab}(\eta) = \sum_{l} \re^{l\eta}g_{ab,l}\,,
\label{eq:gabagain}
\eeq
with $l \in \{1,0,-1/2,-3/2\}$.
The $g_{ab,l}$ are constant matrices:
\bal
g_{ab,1}& = \frac{1}{5}
	\left[\begin{array}{rrrrrr}
	3w_1 & 2w_1 & 3w_2 & 2w_2 & 3w_3 & 2w_3 \\
	3w_1 & 2w_1 & 3w_2 & 2w_2 & 3w_3 & 2w_3 \\
	3w_1 & 2w_1 & 3w_2 & 2w_2 & 3w_3 & 2w_3 \\
	3w_1 & 2w_1 & 3w_2 & 2w_2 & 3w_3 & 2w_3 \\
	3w_1 & 2w_1 & 3w_2 & 2w_2 & 3w_3 & 2w_3 \\
	3w_1 & 2w_1 & 3w_2 & 2w_2 & 3w_3 & 2w_3
	\end{array}\right]\,,
\\
g_{ab,0}& =
	\left[\begin{array}{rrrrrr}
	1-w_1 & 2(1-w_1) & -w_2 & -2w_2 & -w_3 & -2w_3 \\
	0 & 0 & 0 & 0 & 0 & 0\\
	-w_1 & -2w_1 & 1-w_2 & 2(1-w_2) & -w_3 & -2w_3 \\
	0 & 0 & 0 & 0 & 0 & 0\\
	-w_1 & -2w_1 & -w_2 & -2w_2 & 1-w_3 & 2(1-w_3) \\
	0 & 0 & 0 & 0 & 0 & 0
	\end{array}\right]\,,
\\
g_{ab,-1/2}& = 
	\left[\begin{array}{rrrrrr}
	0 & -2(1-w_1) & 0 & 2w_2 & 0 & 2w_3\\
	0 & 1-w_1 & 0 & -w_2 & 0 & -w_3\\
	0 & 2w_1 & 0 & -2(1-w_2) & 0 & 2w_3 \\
	0 & -w_1 & 0 & 1-w_2 & 0 & -w_3\\
	0 & 2w_1 & 0 & 2w_2 & 0 & -2(1-w_3) \\
	0 & -w_1 & 0 & -w_2 & 0 & 1-w_3
	\end{array}\right]\,,
\\
g_{ab,-3/2}& = \frac{1}{5}
	\left[\begin{array}{rrrrrr}
	2w_1 & -2w_1 & 2w_2 & -2w_2 & 2w_3 & -2w_3\\
	-3w_1 & 3w_1 & -3w_2 & 3w_2 & -3w_3 & 3w_3\\
	2w_1 & -2w_1 & 2w_2 & -2w_2 & 2w_3 & -2w_3\\
	-3w_1 & 3w_1 & -3w_2 & 3w_2 & -3w_3 & 3w_3\\
	2w_1 & -2w_1 & 2w_2 & -2w_2 & 2w_3 & -2w_3\\
	-3w_1 & 3w_1 & -3w_2 & 3w_2 & -3w_3 & 3w_3
	\end{array}\right]\,.
\eal
The eigenvectors of the linear propagator $g_{ab}$ are:
\beq 
u^{(1)}_a=
\left(
\begin{array}{c}
  1\\
  1\\
  1\\
  1\\
  1\\
  1
\end{array}
\right)
\ ;\
u^{(2)}_a=
\left(
\begin{array}{c}
  2/3\\
  -1\\
  2/3\\
  -1\\
  2/3\\
  -1
\end{array}
\right)
\ ;\ 
u^{(3,1)}_a=
\left(
\begin{array}{c}
  w_2\\
  0\\
  -w_1\\
  0\\
  0\\
  0
\end{array}
\right)
\ ;\ 
u^{(3,2)}_a=
\left(
\begin{array}{c}
  w_3\\
  0\\
  0\\
  0\\
  -w_1\\
  0
\end{array}
\right)
\ ;\
u^{(4,1)}_a=
\left(
\begin{array}{c}
  2w_2\\
  -w_2\\
  -2w_1\\
  w_1\\
  0\\
  0
\end{array}
\right)
\ ;\
u^{(4,2)}_a=
\left(
\begin{array}{c}
  2w_3\\
  -w_3\\
  0\\
  0\\
  -2w_1\\
  w_1
\end{array}
\right)
\ . 
\eeq
The eigenvectors correspond to the following eigenvalues:
\beq
\bom{u}^{(1)} \, : \, \lambda_1 = \re^\eta\,,
\qquad
\bom{u}^{(2)} \, : \, \lambda_2 = \re^{-3\eta/2}\,,
\qquad
\bom{u}^{(3,m)} \, : \, \lambda_3 = 1\,,
\qquad
\bom{u}^{(4,m)} \, : \, \lambda_4 = \re^{-\eta/2}\,,
\qquad
m=1,2\,.
\eeq

%
%

\section{The linear propagator for general $N$}
\label{app:propagator_N}

Here we compute the linear propagator, $g_{ab}(\eta)={\cal
  L}^{-1}\left[\sigma_{ab}(s),s,\eta\right]$, for general $N$, where
${\cal L}^{-1}[f(s),s,\eta]$ denotes the inverse Laplace transform of
the function $f(s)$ from the variable $s$ to the variable $\eta$ and
$\sigma_{ab}(s)=(s \del_{ab}+\Omega_{ab})^{-1}$.

We begin by noting that
\bal
(s \bom{1}+\Omega)_{(2j-1)(2k-1)} &= s\del_{jk}\,,
&
(s \bom{1}+\Omega)_{(2j)(2k-1)} &= -\frac{3}{2}w_k\,,
\\
(s \bom{1}+\Omega)_{(2j-1)(2k)} &= -\del_{jk}\,,
&
(s \bom{1}+\Omega)_{(2j)(2k)} &= \left(\frac{1}{2}+s\right)\del_{jk}\,,
\eal
with $j,k=1,2,\ldots,N$.  Now we claim that $\sigma_{ab}(s)$ may be
written in the following form for general $N$:
\ba
\sigma_{(2j-1)(2k-1)}(s) & = &
	\frac{3w_k}{s(s-1)(2s+3)}+\frac{\del_{jk}}{s}\,, \nn
\\
\sigma_{(2j)(2k-1)}(s) & = &
	\frac{3w_k}{(s-1)(2s+3)}\,, \nn
\\
\sigma_{(2j-1)(2k)}(s) & = &
	\frac{6w_k}{s(s-1)(2s+1)(2s+3)}+\frac{2\del_{jk}}{s(2s+1)}\,, \nn
\\
\sigma_{(2j)(2k)}(s) & = & 
	\frac{6w_k}{(s-1)(2s+1)(2s+3)}+\frac{2\del_{jk}}{(2s+1)}\,, 
\label{eq:sigmaall}
\ea
($j,k=1,2,\ldots,N)$.
We prove this claim by explicit computation. We calculate
\beq
\left[\sigma(s)\cdot(s\bom{1}+\Omega)\right]_{ab} =
	\sum_{c=1}^{2N}\sigma_{ac}(s)(s\bom{1}+\Omega)_{cb} =
	\sum_{k=1}^{N}\left[\sigma_{a(2k-1)}(s)(s\bom{1}+\Omega)_{(2k-1)b}
		+\sigma_{a(2k)}(s)(s\bom{1}+\Omega)_{(2k)b}\right]\,,
\eeq
where in the last expression, we have split the sum over $c$ into two
sums, corresponding to odd and even $c$ respectively. Now we examine
the four cases corresponding to $a$ and $b$ being odd or even
separately. We find
\begin{enumerate}
\item $a=2n-1$, $b=2m-1$
\beq
\bsp
&
\sum_{k=1}^{N}\left[\sigma_{(2n-1)(2k-1)}(s)(s\bom{1}+\Omega)_{(2k-1)(2m-1)}
	+\sigma_{(2n-1)(2k)}(s)(s\bom{1}+\Omega)_{(2k)(2m-1)}\right]
\\
=&
\sum_{k=1}^{N}\left[
	\left(\frac{3w_k}{s(s-1)(2s+3)}+\frac{\del_{nk}}{s}\right)
	\left(s\del_{km}\right)
	+
	\left(\frac{6w_k}{s(s-1)(2s+1)(2s+3)}+\frac{2\del_{nk}}{s(2s+1)}\right)
	\left(-\frac{3}{2}w_m\right)
	\right]
\\
=&
	\frac{3w_m}{(s-1)(2s+3)}+\del_{nm}-\frac{3}{2}w_m
	\left(\frac{6}{s(s-1)(2s+1)(2s+3)}+\frac{2}{s(2s+1)}\right)
\\
=&
	\del_{nm}\,,
\esp
\eeq
where we have used that $\sum_{k=1}^N w_k = 1$.

\item $a=2n$, $b=2m-1$
\beq
\bsp
&
\sum_{k=1}^{N}\left[\sigma_{(2n)(2k-1)}(s)(s\bom{1}+\Omega)_{(2k-1)(2m-1)}
	+\sigma_{(2n)(2k)}(s)(s\bom{1}+\Omega)_{(2k)(2m-1)}\right]
\\
=&
\sum_{k=1}^{N}\left[
	\left(\frac{3w_k}{(s-1)(2s+3)}\right)
	\left(s\del_{km}\right)
	+
	\left(\frac{6w_k}{(s-1)(2s+1)(2s+3)}+\frac{2\del_{nk}}{(2s+1)}\right)
	\left(-\frac{3}{2}w_m\right)
	\right]
\\
=&
	\frac{3sw_m}{(s-1)(2s+3)}-\frac{3}{2}w_m
	\left(\frac{6}{(s-1)(2s+1)(2s+3)}+\frac{2}{(2s+1)}\right)
\\
=&
	0\,,
\esp
\eeq
where we have used that $\sum_{k=1}^N w_k = 1$.

\item $a=2n-1$, $b=2m$
\beq
\bsp
&
\sum_{k=1}^{N}\left[\sigma_{(2n-1)(2k-1)}(s)(s\bom{1}+\Omega)_{(2k-1)(2m)}
	+\sigma_{(2n-1)(2k)}(s)(s\bom{1}+\Omega)_{(2k)(2m)}\right]
\\
=&
\sum_{k=1}^{N}\left[
	\left(\frac{3w_k}{s(s-1)(2s+3)}+\frac{\del_{nk}}{s}\right)
	\left(-\del_{km}\right)
	+
	\left(\frac{6w_k}{s(s-1)(2s+1)(2s+3)}+\frac{2\del_{nk}}{s(2s+1)}\right)
	\left(\frac{1}{2}+s\right)\del_{km}
	\right]
\\
=&
	-\frac{3w_m}{s(s-1)(2s+3)}-\frac{\del_{nm}}{s}+\left(\frac{1}{2}+s\right)
	\left(\frac{6w_m}{s(s-1)(2s+1)(2s+3)}+\frac{2\del_{nm}}{s(2s+1)}\right)
\\
=&
	0\,.
\esp
\eeq

\item $a=2n$, $b=2m$
\beq
\bsp
&
\sum_{k=1}^{N}\left[\sigma_{(2n)(2k-1)}(s)(s\bom{1}+\Omega)_{(2k-1)(2m)}
	+\sigma_{(2n)(2k)}(s)(s\bom{1}+\Omega)_{(2k)(2m)}\right]
\\
=&
\sum_{k=1}^{N}\left[
	\left(\frac{3w_k}{(s-1)(2s+3)}\right)
	\left(-\del_{km}\right)
	+
	\left(\frac{6w_k}{(s-1)(2s+1)(2s+3)}+\frac{2\del_{nk}}{(2s+1)}\right)
	\left(\frac{1}{2}+s\right)\del_{km}
	\right]
\\
=&
	-\frac{3w_m}{(s-1)(2s+3)}+\left(\frac{1}{2}+s\right)
	\left(\frac{6w_m}{(s-1)(2s+1)(2s+3)}+\frac{2\del_{nm}}{(2s+1)}\right)
\\
=&
	\del_{nm}\,. \lhd
\esp
\eeq
\end{enumerate} 

Finally, we recognize that performing the inverse Laplace transform of
\eqn{eq:sigmaall}, we simply obtain \eqn{eq:gall}.

Similar explicit computations (which however we do not reproduce)
establish that for general $N$, the propagator has the following $2N$
eigenvalues and eigenvectors:
\begin{enumerate}
\item Eigenvalue $\lambda_1 = \re^\eta$ with multiplicity one. 
  The corresponding eigenvector is:
  \beq u_a^{(1)} = (1,1,\ldots,1,1)^T\,.  \eeq
\item Eigenvalue $\lambda_2 = \re^{-3\eta/2}$ with multiplicity one. 
  The corresponding eigenvector is:
  \beq u_a^{(2)} = (2/3,-1,\ldots,2/3,-1)^T\,.  \eeq
\item Eigenvalue $\lambda_3 = 1$ with multiplicity $N-1$. 
  The corresponding eigenvectors are:
  \beq u_a^{(3,m)} = (w_m,0,\ldots,0,-w_1,0,\ldots,0)^T
  = w_m\del_{a1}-w_1\del_{a(2m-1)}\,, \eeq
  ($m=2,\ldots,N$), i.e.~the two non-zero entries in $\bom{u}^{(3,m)}$
  are the first and the $2m-1$-th.
\item Eigenvalue $\lambda_3 = \re^{-\eta/2}$ with multiplicity $N-1$. 
  The corresponding eigenvectors are:
  \beq u_a^{(4,m)} = (2w_m,-w_m,0,\ldots,0,-2w_1,w_1,0,\ldots,0)^T
  = 2w_m\del_{a1}-w_m\del_{a2}-2w_1\del_{a(2m-1)}+w_1\del_{a(2m)}\,, \eeq
  ($m=2,\ldots,N$), i.e.~the four non-zero entries in $\bom{u}^{(3,m)}$
  are the first and second and the $2m-1$-th and $2m$-th.
\end{enumerate}
%


\end{document}